\documentclass[useAMS,usenatbib,aas_macros]{mn2e}
\usepackage{aas_macros}
\usepackage{amsmath}
\usepackage{graphicx}

\usepackage{color}

\newcommand{\equ}[1]{eq.~(\ref{eq:#1})}

\newcommand{\Equ}[1]{Eq.~(\ref{eq:#1})}

\newcommand{\equnp}[1]{eq.~\ref{eq:#1}}
\newcommand{\se}[1]{\S\ref{sec:#1}}
\newcommand{\fig}[1]{Fig.~\ref{fig:#1}}
\newcommand{\figs}[1]{Figs.~\ref{fig:#1}}

\newcommand{\Fig}[1]{Figure~\ref{fig:#1}}
\newcommand{\Figs}[1]{Figures~\ref{fig:#1}}

\newcommand{\be}{\begin{equation}}
\newcommand{\ee}{\end{equation}}
\newcommand{\bea}{\begin{eqnarray}}
\newcommand{\eea}{\end{eqnarray}}

\def\la{\langle}
\def\bul{$\bullet\,$}

\def\sdash{\!-\!}

\newcommand{\no}{\noindent}

\newcommand{\msun}{M_\odot}

\newcommand{\ifm}[1]{\relax\ifmmode#1\else$\mathsurround=0pt #1$\fi}
\newcommand{\kms}{\ifmmode\,{\rm km}\,{\rm s}^{-1}\else km$\,$s$^{-1}$\fi}

\newcommand{\kpc}{\,{\rm kpc}}
\newcommand{\pc}{\,{\rm pc}}
\newcommand{\Gyr}{\,{\rm Gyr}}

\newcommand{\Myr}{\,{\rm Myr}}

\newcommand{\ltsima}{$\; \buildrel < \over \sim \;$}
\newcommand{\lsim}{\lower.5ex\hbox{\ltsima}}
\newcommand{\gtsima}{$\; \buildrel > \over \sim \;$}
\newcommand{\gsim}{\lower.5ex\hbox{\gtsima}}
\newcommand{\prop}{\propto}

\newcommand{\rar}{\rightarrow}

\def\Rv{R_{\rm v}}

\def\Mg{M_{\rm g}}
\def\Ms{M_{\rm s}}

\def\Mci{M_{\rm ci}}
\def\Mgi{M_{\rm gi}}
\def\Msi{M_{\rm si}}
\def\fgi{f_{\rm gi}}
\def\fsgi{f_{\rm sgi}}
\def\gi{g_{\rm i}}
\def\ti{t_{\rm i}}

\def\Mb{M_{\rm b}}
\def\Md{M_{\rm d}}
\def\Rd{R_{\rm d}}
\def\Hd{H_{\rm d}}
\def\Vd{V_{\rm d}}
\def\fg{f_{\rm g}}

\def\Mc{M_{\rm c}}

\def\Rc{R_{\rm c}}

\def\Vrot{V_{\rm rot}}

\def\Rt{R_{\rm T}}
\def\Mt{M_{\rm T}}
\def\Vd{V_{\rm d}}
\def\rhod{\rho_{\rm d}}
\def\sigd{\sigma_{\rm d}}
\def\sfr{\dot{M}_{\rm sf}}
\def\SFR{\dot{M}_{\rm sf}}

\def\Mdoto{\dot{M}_{\rm out}}
\def\Mdotg{{\dot{M}_{\rm g}}}
\def\Mddotg{{\ddot{M}_{\rm g}}}
\def\Mdots{{\dot{M}_{\rm s}}}
\def\Mdotac{{\dot{M}_{\rm ac}}}
\def\Mdotso{{\dot{M}_{\rm s,loss}}}
\def\Mdotsf{{\dot{M}_{\rm sf}}}

\def\td{t_{\rm d}}

\def\tff{t_{\rm ff}}

\def\torb{t_{\rm orb}}
\def\tmig{t_{\rm mig}}

\def\tdep{t_{\rm dep}}
\def\tsf{t_{\rm sf}}
\def\tsfc{t_{\rm sfc}}
\def\tsfi{t_{\rm sfi}}
\def\epsf{{\epsilon_{\rm ff}}}
\def\epsd{{\epsilon_{\rm d}}}

\def\tac{t_{\rm ac}}
\def\tg{t_{\rm g}}
\def\tout{t_{\rm out}}
\def\fg{f_{\rm g}}
\def\fsg{f_{\rm sg}}

\def\delrho{{\delta}_{\rm {\rho}}}
\def\delmin{{\delta}_{\rm {\rho}}^{\rm min}}

\def\ga{\tau_0^{-1}}
\def\etas{\eta_{\rm s}}
\def\etag{\eta}
\def\tform{t_{\rm n}}
\newcommand{\vela}{\texttt{VELA}~}
\newcommand{\velatwo}{\texttt{VELA-2}~}
\newcommand{\velathree}{\texttt{VELA-3}~}
\newcommand{\velathreecom}{\texttt{VELA-3}}

\newcommand{\velasixcom}{\texttt{VELA-6}}
\def\Rrot{R_{\rm rot}}
\def\Vrot{V_{\rm rot}}

\title[Clump Survival and Migration]
{Clump Survival and Migration in VDI Galaxies:\\  
an Analytic Model versus Simulations and Observations}

\author[A. Dekel, N. Mandelker, F. Bournaud, et al.]
{\parbox[t]{\textwidth}
{
Avishai Dekel$^{1,2}$\thanks{dekel@huji.ac.il}, 
Nir Mandelker$^{1,3}$\thanks{nir.mandelker@mail.huji.ac.il}, 
Frederic Bournaud$^{4}$\thanks{frederic.bournaud@cea.fr},\\  
Daniel Ceverino$^{5}$,
Yicheng Guo$^{6}$,
Joel Primack$^{7}$
}
\\
\\
$^1$Centre for Astrophysics and Planetary Science, Racah Institute of Physics, 
The Hebrew University, Jerusalem 91904, Israel\\ 
$^2$SCIPP, University of California, Santa Cruz, CA 95064, USA\\ 
$^3$Kavli Institute for Theoretical Physics, University of California, 
Santa Barbara, CA 93106, USA\\
$^4$AIM, CEA, CNRS, Universit\'e Paris-Saclay, Universit\'e Paris Diderot, 
Sorbonne Paris Cit\'e, 91191 Gif-sur-Yvette, France\\ 
$^5$CIAFF, Facultad de Ciencias, Universidad Autonoma de Madrid, Madrid 28049,
Spain\\
$^6$Department of Physics and Astronomy, University of Missouri, Columbia, MO
      65211, USA\\
$^7$Department of Physics, University of California, Santa Cruz, CA 95064, USA 
}

\begin{document}

\large

\pagerange{\pageref{firstpage}--\pageref{lastpage}} \pubyear{2002}

\maketitle

\label{firstpage}

\begin{abstract}
We address the nature of the giant clumps in high-$z$ galaxies that undergo 
Violent Disc Instability, attempting to distinguish between long-lived 
clumps that migrate inward and short-lived clumps that disrupt by feedback.
We study the evolution of clumps as they migrate through the disc using an 
analytic model tested by simulations and confront theory with CANDELS 
observations. 
The clump ``bathtub" model, which considers gas and stellar gain and loss,
is characterized by four parameters: the accretion efficiency,  
the star-formation-rate (SFR) efficiency, 
and the outflow mass-loading factors for gas and stars.  
The relevant timescales are all comparable to the migration time, 
two-three orbital times.
A clump differs from a galaxy by the internal dependence of the accretion 
rate on the varying clump mass.  
The analytic solution, involving exponential growing and decaying modes,
reveals a main evolution phase during the migration,
where the SFR and gas mass are constant and the stellar mass is
rising linearly with time. This makes the inverse of the specific SFR an 
observable proxy for clump age. 
Later, the masses and SFR approach an exponential growth  
with a constant specific SFR, but this phase is hypothetical as the 
clump disappears in the galaxy center.  
The model matches simulations with different, moderate
feedback, both in isolated and cosmological settings. 
The observed clumps agree with our theoretical predictions,
indicating that the massive clumps are long-lived and migrating. 
A non-trivial challenge is to model feedback that is non-disruptive 
in massive clumps but
suppresses SFR to match the galactic stellar-to-halo mass ratio. 
\end{abstract}


\section{Introduction}
\label{sec:intro}

Most of the massive star-forming galaxies at $z\!\sim\!1\!-\!3$ have a major
component of an extended, rotating, turbulent, gas-rich disc 
\citep{genzel06,forster06,genzel08,tacconi10,genzel11,tacconi13,genzel14_rings,
forster18}
which typically shows several giant star-forming clumps
\citep{elmegreen05,genzel06,genzel08,guo12,guo15,fisher17,
guo18,huertas20,ginzburg21}.
Certain high-resolution studies using lensing and ALMA
have argued that the clump masses may be overestimated by limited resolution
\citep{cava18,dessauges19,rujopakarn19}, 
but they might have missed the gathering of small clumps as substructure of 
giant clumps \citep[as explained by][]{faure21}.

\smallskip 
The common wisdom is that most of these clumps are formed in-situ 
by gravitational disc instability \citep{toomre64}.
This has been studied in simulations of isolated discs 
\citep{noguchi99,immeli04_a,Immeli04_b,bournaud07c,genzel08}
and in cosmological simulations
\citep{dsc09,agertz09,cdb10,ceverino12},
in particular \citet[][M14]{mandelker14} and \citet[][M17]{mandelker17}.
With the high gas fraction at high redshift \citep[e.g.,][]{daddi10,tacconi18}, 
the characteristic Toomre clump mass is as large as a few percent of the disc 
mass, thus associated with ``violent" disc instability (VDI), 
where the clumps play a significant dynamical role in the galaxy evolution 
\citep[e.g.,][]{dsc09,db14}.

\smallskip 
It may be argued that 
the standard analysis of linear Toomre instability, assuming small 
perturbations in an otherwise uniform disc, may be invalid in its pure form
in high-$z$ 
cosmological discs, as they are continuously perturbed in a nonlinear manner 
and are constantly stimulated by external perturbations, such as mergers, tidal
interactions and intense instreaming. 
Indeed, we find in cosmological simulations that Lagrangian proto-clump 
regions, in their linear regime, sometimes have high Toomre-$Q$ values in the 
range $2\!-\!5$, as opposed to the $Q\!\sim\!1$ expected in the standard Toomre 
analysis \citep[][Mandelker et al., in prep.]{inoue16}. 
This calls for a modified instability theory for VDI.
Other works have called into question the validity of linear Toomre
analysis in the context of high-$z$ disc galaxies
for other reasons \citep[e.g.,][]{behrendt15,tamburello15,mayer16}. 
Nevertheless, the successes of the standard Toomre analysis in explaining 
and predicting many features of high-$z$ clumpy discs 
\citep[see also][]{fisher17} indicates that it can 
serve as a useful tool for obtaining a qualitative understanding. 
According to VDI, bound clumps that survive disruptive feedback
are predicted to form with a Toomre mass and a high gas fraction preferably
in the outer disc (\se{obs} below) 
and to migrate due to VDI-driven torques into the disc
centre in a few disc orbital times, which is a few hundred Megayears at 
$z\sim 2$
\citep{dsc09,cdb10,cacciato12,forbes12,forbes14a}.
We note, however, that 
this migration may become slower after a wet compaction event
\citep{zolotov15}, when the disc becomes an extended ring whose migration 
rate is suppressed by a massive central bulge \citep{dekel20_ring}.

\smallskip 
This paper is partly motivated by 
an ongoing controversy concerning whether the massive clumps disrupt 
on a disc dynamical timescale due to stellar feedback or survive longer
and possibly complete their migration to the central bulge in ten dynamical
times or more. 
\citet{dsc09}, based on \citet{ds86}, estimated that supernova feedback by 
itself may not have enough power to disrupt the massive clumps. 
\citet{murray10} argued that momentum-driven radiative 
stellar feedback, enhanced by infra-red photon trapping, could disrupt the 
clumps on a dynamical timescale, as it does in the local giant molecular 
clouds. However, \citet{kd10} pointed out that such an explosive 
disruption would be possible in the high-redshift giant clumps only if the 
star-formation rate (SFR) efficiency in a free-fall time is $\epsf\!\sim\!0.1$, 
significantly larger than what is implied by the observed Kennicutt-Schmidt 
relation in different types of galaxies at different redshifts,
namely $\epsf$ of about one to a few percent
\citep{tacconi10,kdm12,freundlich13}.
\citet{dk13} proposed instead that outflows from high-redshift clumps, with
mass-loading factors $\etag$ of order unity to a few, are driven by steady
momentum-driven outflows from stars over many tens of free-fall times. Their
analysis is based on the finding from high-resolution 2D simulations that
radiation trapping is negligible because it destabilizes the wind
\citep{krum_thom12,krum_thom13}. 
Each photon can therefore contribute to the wind momentum
only once, so the radiative force is limited to $\sim\!L/c$, where $L$ is the
luminosity of the clump.  Combining
radiation, protostellar plus main-sequence winds, and supernovae, \citet{dk13}
estimated the total direct injection rate of momentum into the outflow to be
$\sim\!2.5L/c$. The adiabatic phase of clustered supernovae and main-sequence 
winds may double this rate \citep{Gentry17}. The predicted values are thus
$\etag\!\sim\! 1\!-\!5$, shown in \citet{dk13} to be consistent with the
values deduced from the pioneering observations of outflows from giant clumps
\citep{genzel11,newman12}. They concluded that most massive clumps are
expected to complete their migration prior to gas depletion. With the
additional gas accretion onto the clumps, they argued that the clumps are
actually expected to grow in mass as they migrate inward. 
 
\smallskip 
Simulations that put in very strong winds by hand at the sub-grid level
\citep{genel12}, or that include enhanced radiative feedback of $30\!-\!50 L/c$
due to strong trapping \citep{Hopkins12b,oklopcic17},
indeed produced very short-lived clumps (SL clumps)
that are typically unbound and disrupt on timescales in the ball park of
the disc dynamical timescales, 
ten to a few tens of million years at $z\!\sim\!2$. %
On the other hand, simulations with more modest momentum driving of 
$2\!-\!3 L/c$
\citep{moody14,bournaud14,mandelker17} reveal the formation of clumps of two
types. In particular, the \velathree
cosmological simulations that were analyzed 
in M14 and M17 reveal that a significant fraction of 
the giant clumps above $\sim\!10^{8.5}\msun$ are long-lived clumps
(LL clumps) that are compact, round and bound and they survive feedback during 
their inward migration that lasts $\sim\!300\Myr$ or more.
This is while the less massive clumps tend to be SL clumps that are
diffuse, elongated and unbound and they disrupt on  
one to a few disc dynamical timescale. These SL clumps may in certain ways be 
similar to the clumps detected in the simulations with stronger feedback,
though some of the low-mass \velathree
clumps do survive for a few dynamical times 
before they more gradually disrupt by the accumulated supernovae and possible
tidal stripping.
Cosmological simulations of a similar nature to \velathreecom, 
of the same suite of galaxies 
but with a stronger supernova feedback, turn out to produce more SL clumps 
at the expense of LL clumps even in the massive end 
(\velasixcom, Ceverino et al. 2021, in preparation).
In turn,
isolated galaxy simulations have emphasized the dependence of clump
formation, mass and longevity on the gas fraction in the disc where they form
\citep{fensch21,renaud21}.
Thus, one may consider two extreme general types of giant clumps, 
one where the clumps are long-lived and manage to migrate through 
a fair fraction of the disc radius,
and the other where the clumps form and die young not too far from where they 
formed. When they are very short lived, they may poetically resemble popcorn or 
the lights of a Xmass tree. 

\smallskip 
The main difference between the observable predictions of these extreme 
types of clumps is the distribution of stellar ages among the massive in-situ 
clumps, which are typically expected to be less than $\sim\!50\Myr$ if the 
clumps disrupt on a disc dynamical timescale, 
while they are expected to last for up to a few hundred Megayears 
if the clumps survive till the end of migration. 
The distribution of radial distances may be another distinguishing feature,
where it is expected to be broader for migrating climps.
The time evolution of LL clumps is expected to be observable as
correlations between clump properties and clump age. Due to the migration,
this evolution could be translated to gradients of clump properties as a 
function of radial distance from the galaxy center. 
In contrast, the SL clumps are not expected to show
a wide span of ages and migration-driven radial gradients.
For both types of clumps, certain radial gradients 
might have been implemented at clump formation, reflecting possible radial
gradients in disc conditions (see \se{LLvsSL} below).
The radial gradients in clump properties for LL and SL clumps, as well as for 
clumps that were formed ex-situ to the disc and for the background disc itself,
were measured in cosmological simulations (M14,M17).
They confirmed and partly quantified the above expectations for the LL clumps.
These theoretical expectations for the two scenarios should be sharpened 
and physically understood better in order to allow a meaningful comparison with 
observations.

\smallskip 
We focus here on a theoretical understanding of the evolution of clump 
properties during migration through the disc,
as a function of the physical ingredients involved.
For this we first follow the
evolution of the giant clumps using a simple analytic model. This idealized
model is in the general spirit of the bathtub model used extensively for the
whole galaxy or the disc \citep{bouche10,dave12,cacciato12,kd12,lilly13, 
dekel13,dm14,krum18}, except that it applies here to a single giant clump. 
The feature unique to clumps is the proportionality of the gas accretion rate 
onto a clump to its total clump mass, which can vary on a short timescale.
For a whole galaxy,
the gas (baryonic) accretion rate onto the galaxy can be crudely related to 
the gas (baryonic) mass,
as the cosmological specific accretion rate of gas onto the central
galaxy is expected to be roughly comparable to the total specific accretion 
rate, \citep[e.g.,][]{dekel13}.
The accretion onto a clump, unlike a whole galaxy,
is not directly constrained by the evolution of the cosmological accretion 
rate \citep{dm14}. 

\smallskip 
During clump migration, clump gas turns into stars and the clump exchanges
gas and stars with the disc through accretion, feedback-driven outflows and
tidal stripping \citep{bournaud07c,bournaud14}. The evolution of clump
properties is governed by the balance between these processes under
conservation of gas and stellar mass.
The evolution of clump properties is obtained here by analytically solving the 
two corresponding continuity equations, and this evolution in time can be 
translated to radial gradients within the disc.
The analytic model allows a study of how the clump properties and evolution 
depend on the assumed efficiency of accretion, star formation and  
outflows. It also points at connections between the different clump properties,
and suggests alternative observable proxies to quantities that suffer from
large observational uncertainties, such as stellar age.  

\smallskip 
The model predictions are compared to the results of hydrodynamical AMR 
simulations, both of isolated
gas-rich galaxies mimicking high-redshift discs at high resolution and of
zoomed-in high-redshift galaxies in their cosmological setting, 
spanning a range of moderate feedback strengths. 

\smallskip 
Preliminary observational estimates of clump ages and radial gradients using 
a relatively small number of clumps 
\citep{forster11b,guo12,guo15,zanella19}
seem to be consistent with the scenario of LL migrating clumps.
The same is true for low-redshift gas-rich analogs of high-redshift discs
\citep{fisher19,lenkic21}.
However, the individual estimates of clump properties carry large
uncertainties, especially the age estimates but also the mass
and SFR estimates \citep[e.g.,][]{huertas20,ginzburg21}.  
It is therefore important to use large samples of clumps,
and to consider systematic errors in their properties 
\citep[e.g., as done in][]{huertas20}.
A major step in the direction of a large sample has been made by
\citet{guo18}, who assembled and studied a very large sample consisting of
thousands of clumps in massive galaxies from the CANDELS survey at 
$z\!=\!0.5\!-\!3$. 
In a first test of our model against observations, attempting to 
distinguish between the competing scenarios for clump
survival, we perform below a comparison of our theoretical 
predictions to the clump properties and their radial gradients 
in a large subsample of the observed CANDELS clumps,
focusing on star-forming clumps at $z\!=\!1.5\!-\!3$.
In this comparison, we allow for systematic errors in the clump masses.
Comparisons of our model to the larger sample of CANDELS clumps, with improved
mass estimates via machine learning \citep{huertas20},
are deferred to a future study.
Using a machine-learning method for comparing theory to observations,
The two types of clumps as identified in the \velathree simulations, LL and SL
clumps, were found in the CANDELS sample, 
with consitent relative distributions of clump masses, radial positions and 
host-galaxy masses \citep{ginzburg21}.

A potential practical difficulty in the comparison to observations
may arise due to contamination of the observed
clumps by stars from the underlying disc.
This may smear out the differences in clump properties, such as their age,
and possibly generate false radial gradients even in the SL scenario.
If the density contrast in clumps is significant, the contamination effects are
weak (M17). 
If, however, the contrast is low, these effects could be
significant \citep{oklopcic17}. One should therefore seek distinguishing
observable clump properties that are less sensitive to contamination by the 
disc, or that can be corrected for it, and attempt to estimate the possible
effects of contamination both in the simulations and in the observations. 


\smallskip 
As pointed out, 
the distribution of clump ages is the key distinguishing feature between the
competing scenarios for clump survival. However, the observational age
estimates are very crude and potentially biased, e.g., by an assumed SFR that
exponentially decays in time in the SED fitting. We use our model and
simulations to evaluate the actual star-formation history (SFH) in clumps, and
attempt to derive an observable proxy for clump age based on the actual SFH. 
Motivated by the results of our toy model, we
test the hypothesis that the SFR in a clump is rather constant with time, 
such that the inverse of the sSFR is a proxy for time since clump formation,
capable of distinguishing between the competing scenarios. 
We also evaluate, using isolated-galaxy simulations with low gas fraction and
strong feedback as well as short-lived clumps in the cosmological simulations, 
to what extent the inverse of the sSFR is a proxy for age also for SL clumps.

\smallskip 
The paper is organized as follows. 
The following two sections present the bathtub toy model for clumps.
In \se{toy_ingredients} we summarize and parameterize the processes of clump 
mass gain and loss, and in \se{toy_results} we put the ingredients together 
in the continuity equations for gas and stars and solve them analytically for 
the evolution of clump properties.
In passing, we propose a method for observationally estimating clump
ages based on their sSFR.  
The subsequent two sections present the results from the simulations in
comparison to the model predictions.
In \se{sims} we describe the isolated-galaxy and cosmological simulations used
and the way we identify clumps in them and evaluate their properties.
In \se{evolution} we present the simulation results for clump evolution
and compare them to the toy-model predictions in order to test the validity of 
the model and interpret the simulation results.
In \se{obs} we summarize the observational results reported in
\citet{guo18} from CANDELS and compare them to the theory predictions.
In \se{LLvsSL} we use the observations and theoretical predictions
to constrain the nature of the giant clumps. 
In \se{conc} we summarize our conclusions and discuss them.

\section{Analytic model: mass gain \& loss}
\label{sec:toy_ingredients}

In order to spell out a bathtub model for clump evolution,  
we first address here each of the main processes of mass gain and loss 
in the gas and stellar components of the clumps,
and derive the corresponding rates of mass exchange and the
associated timescales. These ingredients will be inserted in the conservation
equations for gas and stars in \se{toy_results}.

\subsection{Clump migration}

The giant clumps in VDI discs migrate toward the disc centre following
angular-momentum and energy loss by several processes such as 
torques from the perturbed disc,
clump-clump interactions and dynamical friction. 
The migration time can be estimated in several different ways
\citep[e.g.,][]{dsc09,krum_burkert10,dekel13}, all leading to
estimates in the same ball park of ten to twenty disc dynamical times.
Assuming that the clump maintains its initial Toomre mass,
we adopt here the estimate based on clump encounters by \citep{dsc09},
\be
\tmig \simeq 2.1 Q^2 \delta^{-2} \td \sim (10-20)\,\td \, ,
\label{eq:tmig}
\ee
{\no}where $\td \!=\! \Rd/\Vd$
is the disc crossing time, $\Rd$ is the characteristic disc radius and $\Vd$
is the characteristic disc circular velocity.
The quantity $\delta$ is the mass
fraction in the cold disc within the disc radius, which at the cosmological 
steady state has been argued to be $\delta\sim 0.3$ \citep{dsc09}. 
The Toomre parameter is $Q \!\sim\! 1$\footnote{In a 
thick disc the Toomre $Q$ is expected to be slightly smaller than unity
\citep{goldreich65_thick}, while in cosmological simulations we sometimes find
$Q$ values that are somewhat larger than unity \citep{inoue16}. The migration
time in our simulations therefore ranges from ten to a few tens of dynamical 
times.}. 
The migration time is thus comparable to $2\sdash 3$ orbital
times at the outer disc, $\torb\!=\!2\pi\td$.

\smallskip
With a higher value of $Q$ \citep[e.g.,][]{inoue16}, the migration time would 
be somewhat longer. On
the other hand, if the clump changes its mass during migration, the migration
time scales inversely proportional to the clump mass $\Mc(t)$ [as in dynamical
friction, or in clump encounters, equations 16-18 of \citet{dsc09}], so
\be
\tmig \sim 10\,\td\,(\Mt/\Mc) \, ,
\label{eq:tmig2}
\ee
where $\Mt$ is the initial mass of the clump, assumed to be the Toomre mass.
A corrected estimate for $\tmig$ can be obtained once $\Mc(t)$ is known.
The inward radial migration velocity when the clump mass is $\Mc$
can be crudely estimated from \equ{tmig} and \equ{tmig2} as
\be
V_{\rm mig} \sim \frac{1}{10} \frac{\Rd}{\td} \frac{\Mc}{\Mt} \, .
\label{eq:vmig}
\ee
Assuming that the clump forms at $r\sim\Rd$ (leading to an upper limit for the
migration time),
one can integrate $V_{\rm mig}$ in time to estimate the clump position $r(t)$,
and evaluate $\tmig$ when $r(\tmig)\!=\!0$.
We find that the change in total mass during migration is typically not large, 
so the initial estimate of $\tmig$, which is crude anyway, is a fair 
approximation for our purpose here.
However, \equ{vmig} turns out not to be a good approximation for translating
the time dependence of the clump mass to radial gradients within the disc.

\subsection{Gas accretion onto clumps}

As the clump spirals in toward the disc centre, it accretes matter from the
surrounding disc. An estimate of the accretion rate is provided by the
entry rate into the tidal (Hill) sphere of the clump in the galaxy, $\Rt$,
\be
\dot{M}_{\rm ac} \simeq \alpha\, \rho_{\rm d}\,(\pi \Rt^2)\,\sigma_{\rm d} \, .
\label{eq:alpha}
\ee
Here $\rho_{\rm d}$ is the density in the cold disc (gas or young
stars), $\pi\Rt^2$ is the cross section for entry into the tidal sphere,
and $\sigma_{\rm d}$ is the velocity dispersion in the disc representing
here the relative velocity of the clump with respect to the rest of the
ring within which the clump is orbiting.
We crudely assume here that $\rho_{\rm d}$ and $\sigma_{\rm d}$ as well as 
$\Rt$ are cosntant within the disc throughout the clump migration.
The parameter $\alpha$ represents the fraction of the mass entering
the tidal radius that is actually bound to the clump, and is crudely expected 
to be on the order of 0.2-0.4 \citep{dk13,mandelker14}. 
If only gas that is inflowing with respect to the clump center becomes
bound to the clump, the value of $\alpha$ would be on the low side.
We assume here that $\alpha$ is constant throughout the clump lifetime,
and test this assumption using simulations in \se{evolution}.

\smallskip
The tidal or Hill radius $\Rt$ about the clump is where the self-gravity force
by the clump balances the tidal force exerted by the total mass distribution in
the galaxy along the galactic radial direction. If the disc is in marginal
Toomre instability with $Q \!\sim\! 1$, this is the same as the Toomre radius 
of 
the proto-clump patch that eventually contracts to form the clump \citep{dsc09},
\be
\Rt \simeq 0.5\,\delta \Rd\, ,
\label{eq:Rt}
\ee
where the clump mass is given by
\be
\frac{\Mc}{\Md} \simeq \left( \frac{\Rt}{\Rd} \right)^2 \, ,
\label{eq:Mc}
\ee
with $\Md$ referring to the mass of the cold disc. Also when
$Q \!\sim\! 1$, the disc half-thickness $\Hd$ is comparable to $\Rt$,
\be
\frac{\Hd}{\Rd} \simeq \frac{\sigma_{\rm d}}{\Vd} \, ,
\label{eq:hr}
\ee
and
\be
\delta \simeq \sqrt{2}\frac{\sigd}{\Vd} \, ,
\label{eq:delta-sv}
\ee
with the numerical factor corresponding to a flat rotation curve.
Note that \equ{alpha} is valid only when $\Rt$ is not larger than $\Hd$, such
that the clump accretes from a 3D distribution around it.

\smallskip
We can now evaluate the rate of clump growth by accretion
using \equ{alpha}. We insert $\Rt$ from \equ{Rt},
write $\rhod \!=\! \Md/(2\pi\Rd^2 \Hd)$, and use \equ{hr} for $\Hd$ to obtain
the desired expression for the accretion rate,
\be
{\dot{M}_{\rm ac}} \simeq \frac{\alpha}{2} \frac{\Mc}{\td} \, .
\label{eq:ac}
\ee
The timescale for accretion is thus,
\be
\tac = \frac{\Mc}{\dot{M}_{\rm ac}} \simeq \frac{2}{\alpha} \td \, .
\label{eq:tac}
\ee
This turns out to be about half the migration time of \equ{tmig},
already indicating that the clump mass may grow by a factor of a few during 
the migration.

\subsection{Star Formation Rate}
\label{sec:sfr}

For a clump of gas mass $\Mg$ the SFR can be modeled as
\be
{\rm SFR} = \SFR = \epsf \frac{\Mg}{\tff}
= \epsd \frac{\Mg}{\td} \, ,
\label{eq:sfr}
\ee
where $\tff$ is the clump free-fall time and $\epsf$ the corresponding
SFR efficiency.
The free-fall time is
\be
\tff \simeq \sqrt{\frac{\Rc^3}{G\Mc}} \, ,
\label{eq:tff}
\ee
where $\Mc$ and $\Rc$ are the clump mass and radius. 
The SFR efficiency per free-fall time has to be of order 
$\epsf \!\sim\! 0.02$ in order to match the observed Kennicutt-Schmidt law 
at different environments and redshifts \citep[e.g.,][]{kdm12}. 
We thus assume $\epsf$ to be a constant during the clump lifetime.
The effective value of $\epsf$, as defined for the clump mass and free-fall
time, may be larger if star formation actually
occurs in dense sub-regions within the clump where the free-fall timescale is
shorter.
For our purpose here, it is convenient to replace $\epsf$ by $\epsd$, where 
$\tff$ in the clump is replaced by the disc $\td$, as in \equ{sfr},
and where we assume a constant density contrast between the clump interior
and the disc over all. 
The constancy of $\epsf$, and therefore of $\epsd$, is expected in the Toomre 
regime of star formation, that is relevant
for VDI high-$z$ discs \citep{kdm12}, so we adopt it here. 
The ratio $\td/\tff$ is given by the
square root of the density contrast between the star-forming region in the
clump and the disc, namely $\td/\tff\!\sim\!3\!-\!10$ if the density contrast 
is $10\!-\!100$ \citep[e.g.,][]{ceverino12}. 
We thus expect $\epsd\!\sim\!0.1$ or higher. 

\smallskip
We assume that a fraction $\mu$ of the mass in forming stars remains in
stars, the rest assumed to be instantaneously lost from the stars due to
supernovae and stellar winds and deposited back in the inter-stellar medium
within the clump, namely in $\Mg$. The value of $\mu$ is estimated to be 
between 0.5 and 0.8,
depending on the duration over which it is computed \citep{kd12}.
For the relevant timescales for clump evolution, on the order of 
$\sim 100\Myr$, we adopt hereafter $\mu\!=\!0.8$.

\smallskip
The two observable timescales of interest are the depletion time and
star-formation time.
The depletion time is
\be
\tdep \equiv \frac{\Mg}{\SFR} = \epsd^{-1} \td \, ,
\label{eq:tdep}
\ee
where the actual timescale for gas consumption by star formation is
only slightly different, $\mu^{-1}\tdep$.
The star-formation time is defined as the inverse of the specific SFR (sSFR),
\be
\tsf \equiv \frac{\Ms}{\SFR} = \fsg \epsd^{-1} \td \, ,
\label{eq:tsf}
\ee
where $\fsg$ is the stellar-to-gas mass ratio, related to the conventional
gas fraction $\fg$ by
\be
\fsg\equiv \frac{\Ms}{\Mg}, \quad
\fg \equiv \frac{\Mg}{\Ms+\Mg} = (1+\fsg)^{-1} \, .
\label{eq:fg}
\ee

\subsection{Gas Outflow}

The star formation in the clump produces gas outflow from the clump via
various stellar feedback processes, including supernovae feedback and radiative
pressure feedback \citep[e.g.,][]{dk13}. One can approximate the outflow rate as
\be
\Mdoto = \etag \SFR \, ,
\label{eq:etag}
\ee
where $\etag$ is the mass-loading factor. 
We assume here that when averaged over a dynamical time (or more) $\etag$
is a constant throughout the clump lifetime. 
It is estimated to be of order unity, or a few, both from observations 
\citep[e.g.,][]{genzel11,newman12} 
and based on theoretical considerations \citep[e.g.,][]{dk13}.

\smallskip
The timescale for gas outflow is thus
\be
\tout = \frac{\Mg}{\Mdoto} = \frac{\tff}{\etag\epsf}
\simeq \frac{\td}{\etag\epsd} \gsim \etag^{-1}\, \tdep .
\label{eq:tout}
\ee
For $\etag \!\sim\! 1$, this is comparable to the depletion timescale,
which for $\epsd\!\sim\! 0.1$ is comparable to the migration timescale.

\subsection{Stellar mass exchange}

Accretion of stars from the disc into clumps is expected
to be less efficient than the gas accretion because even when they enter the
Hill sphere they are less likely to become bound to the clump, as they
tend to have a larger velocity dispersion and they do not dissipate.
On the other hand, tidal stripping of older stars
from the clump is expected to be more efficient than gas stripping because
these stars tend to be on less bound orbits. 
We may therefore expect a net effect of stellar mass loss, 
that may partly compensate for the stellar mass gain by star formation.
The rate of tidal stripping after clump formation is not expected to be large,
and is likely to be smaller than both the rate of gas accretion and the rate 
of gas outflow driven by feedback.
For a net stellar mass loss from the clump
$\Mdotso$, stripping minus accretion, we define
\be
\etas = \frac{\Mdotso}{\SFR} \, .
\label{eq:etas}
\ee
Negative values would refer to a net accretion rather than stripping.
Since $\etas$ is expected to be relatively unimportant, we allow
it to be constant during the clump migration, and use its value only for 
estimating the qualitative effect of stellar exchange between the clump and the
disc.\footnote{For cases where the stellar mass exchange is more significant 
one may try more accurate modeling of stellar accretion 
(e.g., with an efficiency $\alpha_{\rm s}$) and
tidal stripping (e.g., as an increasing function of $\Ms$ and a decreasing
function of $\Mc$). We do not attempt this here.}

\begin{table*}
\centering
  \begin{tabular}{llllc}
      \hline
  Quantity & Meaning & Definition & Reference &Fiducial value   \\
      \hline
 Basic properties  &         & & & \\
 $\Mg$  & clump gas mass  & & \equnp{Mgt} & \\
 $\Ms$  & clump stellar mass  & & \equnp{Mst}  & \\
 $\Mc$  & clump total baryonic mass & $\Mc=\Ms+\Mg$  & \equnp{Mc} & \\
 SFR, $\Mdotsf$  & star-formation rate in the clump  &  &   \equnp{sfr}  &  \\
 $\td$ & dynamical time of the disc   & $\td=\Rd/\Vd$ & \fig{td}  & 
        $\sim 30\Myr\, (1+z)_3^{-3/2}$ \\
 $\tmig$ & typical clump migration time  & $\tmig \sim (10-20)\,\td$ & 
        \equnp{tmig2} & $\sim (10-20)\, \td$ \\
      \hline
 Parameters  &         & & & \\
 $\alpha$ & gas accretion efficiency   & $\Mdotac = 0.5 \alpha \Mc/\td$ &
             \equnp{ac} & 0.2-0.4  \\
 $\epsd$  & SFR efficiency per disc dynamical time & 
          SFR = $\epsf\Mg/\tff = \epsd \Mg/\td$ & \equnp{sfr} & 0.1-0.4 \\
 $\etag$  & gas outflow mass-loading factor & $\Mdoto=\etag$ SFR & 
             \equnp{etag} & 1 \\
 $\etas$ & stellar exchange rate factor & $\Mdotso=\etas$ SFR & 
             \equnp{etas} & $\ll 1$ \\
 $\mu$  & fraction of formed stellar mass left in stars & & \se{sfr}
              & 0.8 \\
 $\fsgi$ or $\fgi$ & initial $\fsg$ or $\fg$ at clump formation  & & \se{fsgi} 
        & $\fgi \lsim 1$ \\
      \hline
 Model solution  &         & & & \\
 $t_1$, $t_2$ & timescales of growing and decaying modes &
 $t_1=\vert \beta_1\vert^{-1}$, $t_2=\vert\beta_2\vert^{-1}$ & \equnp{beta_i} & 
  $\sim 10\,\td$ \\ 
      \hline
Other Properties  &         & & & \\
 $\fg$  & gas fraction          & $\fg=\Mg/\Mb$  & \equnp{fg} & $\sim 0.5$ \\
 $\fsg$ & stellar-to-gas ratio  & $\fsg=\Ms/\Mg =\fg^{-1}-1 $ & 
    \equnp{fg}, \ref{eq:fsg} & $\sim 1$ \\
 $\tac$ & accretion time & $\tac=\Mc/\Mdotac=2\alpha \td$ & \equnp{tac} &
      $\sim 10\,\td$ \\
 sSFR  & specific star-formation rate    & sSFR$=$SFR$/\Ms$ &           & \\
 $\tsf$  & star-formation time & $\tsf=\,$sSFR$^{-1} =\fsg \epsd^{-1} \td$  &
            \equnp{tsf} & $\sim 10\,\td$ \\
 $\tdep$ & depletion time & $\tdep=\Mg/$SFR$\,=\epsd^{-1}\td$  & \equnp{tdep} & 
      $\sim 10\, \td$ \\
 $\tout$ & outflow time & $\tout=\Mg/\Mdoto=\eta^{-1} \tdep$ & \equnp{tout} & 
      $\sim 10\,\td$ \\
     \hline
   \end{tabular}
  \caption{List of clump properties and model parameters.} 
\label{tab:par}
\end{table*}

\section{Analytic clump bathtub model}
\label{sec:toy_results}

\smallskip
We note that the timescales for changes in clump properties as
evaluated in \se{toy_ingredients} 
($\tac$, $\tsf$, $\tdep$, $\tout$), 
with the fiducial values of the model parameters 
($\alpha \!\sim\! 0.2$, $\epsd \!\sim\! 0.1$, and $\etag \!\sim\! 1$,
with $\fsg \!\sim\! 1$),
are all in the ball park of $\sim\! 10 \td$, which is on the order of the
migration time $\tmig$.
This implies that the clump properties are expected to evolve during the
migration, significantly but not by orders of magnitude.
The ingredients discussed in \se{toy_ingredients} are now put together
in mass conservation equations for each clump, to be solved analytically
and to be compared to simulations (\se{evolution}) and observations (\se{obs}).

\subsection{Gas mass conservation}

\smallskip
The gas mass in the clump varies due to the net effect of growth by
accretion and decrease by gas consumption into stars and gas loss by
outflows,
\be
\Mdotg = \Mdotac - \mu\,\sfr - \Mdoto \, .
\label{eq:cont_gas_1}
\ee
Using the expressions of the previous section, this becomes
\be
\Mdotg = 0.5\,\alpha\, \td^{-1} \Mc - (\mu+\etag)\, \epsd\, \td^{-1} \Mg \, .
\label{eq:cont_gas_2}
\ee
\Equ{cont_gas_2} could be re-written in short as
\be
\Mdotg = g \Mg \, , \quad
g(t) \equiv \td^{-1} [0.5\, \alpha\, \fg^{-1} - (\mu+\etag)\,\epsd] \, .
\label{eq:cont_gas_3}
\ee
However, the timescale $\tg=g^{-1}$ is not necessarily constant, 
as $\fg$ may evolve
in time even if $\alpha$, $\etag$ and $\epsd$ are roughly constant.
In fact, $g(t)$ may change sign as $\fg$ evolves, so the associated timescale
$g^{-1}$ may diverge at this moment.
Thus, unfortunately,
$\Mg(t)$ cannot be obtained by a simple integration of \equ{cont_gas_3}
with a constant $g$, and one has to work harder on solving
\equ{cont_gas_2}.

\smallskip
\Equ{cont_gas_2} is reminiscent of the analogous continuity equation for
the gas mass in the whole galaxy, with source and draining terms on the
right hand side \citep[eq.~1 of][]{dm14}.
For a whole galaxy, while the draining term is similar,
the source term is external, driven by the cosmological gas 
accretion rate and its penetration into the central galaxy.
In the case of a constant source term, 
the draining term being proportional to negative $\Mg$
drives the system to a quasi-steady state, where $\Mg$ is constant on a
timescale shorter than the long timescales for variations in the accretion
time and in the disc dynamical time, and therefore the SFR follows the 
accretion rate, as it varies slowly with cosmological time.
For a clump, however, the source term depends
on the total clump mass, which involves both $\Mg$ and $\Ms$ 
and could vary on a short timescale. This leads to a different solution.

\smallskip
In order to solve \equ{cont_gas_2}, it is convenient to separate $\Mc$ into
$\Ms+\Mg$ and rearrange \equ{cont_gas_2}
into terms that are proportional separately to $\Ms$ and $\Mg$,
\be
\Mdotg = 0.5\,\alpha\,\td^{-1}\Ms +[0.5\,\alpha-(\mu+\etag)\,\epsd]\,
\td^{-1}\Mg \, .
\label{eq:Mdotg}
\ee
The first term is proportional to $\Ms$, which in general, like $\Mg$,
varies in time.
This implies that in order to follow the evolution of $\Mg$ one has 
to also take into account the variation in stellar mass.
This reflects the main difference between the bathtub model for
a clump and for a whole galaxy.

\subsection{Stellar mass conservation}

The stellar mass evolves at a rate
\be
\Mdots = (\mu-\etas)\,\SFR
= (\mu-\etas)\, \epsd\, \td^{-1} \Mg \, .
\label{eq:Mdots}
\ee
A slight overestimate of the clump stellar mass can be obtained for the simple 
case of $\etas\!=\!0$, where the stellar mass grows primarily by star
formation, ignoring stellar mass exchange with the disc.
An underestimate of the clump mass would be obtained with $\etas\!=\!\mu$,
where the stellar mass loss exactly compensates for the star formation,
such that the stellar mass remains constant in time.

\subsection{Analytic solution for clump evolution}

\Equ{Mdotg} and \equ{Mdots} can be solved analytically for the evolution of
$\Mg(t)$, $\Ms(t)$ and the associated quantities of interest.
We assume that the model parameters are constant during the migration of the
clump, including $\alpha$, $\epsd$, $\etag$ and $\etas$ 
($\mu\!=\!0.8$ is naturally assumed throughout), as well as $\td$ of the disc.
We set the mass units such that the initial total clump mass at $t\!=\!0$,
assumed to be the Toomre mass, is $\Mci\!=\!1$.
The initial condition is characterized by the initial gas fraction
$\fgi$, giving rise to the initial gas mass and stellar mass
$\Mgi$ and $\Msi$.

\smallskip
Taking a time derivative of \equ{Mdotg} we obtain
\be
\Mddotg=0.5\,\alpha\,\td^{-1}\Mdots 
+[0.5\,\alpha-(\mu+\etag)\,\epsd]\,\td^{-1}\Mdotg \, .
\ee
Inserting $\Mdots$ from \equ{Mdots} we obtain a differential equation for
$\Mg(t)$,
\be
\Mddotg+b\Mdotg+c\Mg = 0 \, ,
\label{eq:quad}
\ee
\be
b\equiv [(\mu+\etag)\,\epsd - 0.5\,\alpha]\,\td^{-1}\, ,
\ee
\be
c\equiv - 0.5\,\alpha\,(\mu-\etas)\,\epsd\,\td^{-2} \, .
\ee

\smallskip
We try a solution of the sort
\be
\Mg(t) = m_1 e^{\beta_1 t} + m_2 e^{\beta_2 t} \, .
\label{eq:Mgt}
\ee
Inserting each term of \equ{Mgt} in \equ{quad},
we obtain a quadratic equation for each $\beta$, independent of the
corresponding $m$,
\be
\beta^2 + b \beta + c = 0 .
\ee
The two roots (which always exist as $c\!<\!0$) are
\be
\beta_{1,2} = 0.5 (-b \pm d) \, , \quad
d\equiv (b^2-4c)^{1/2} \, .
\label{eq:beta_i}
\ee
These represent the two characteristic timescales
\be
t_1 \equiv \beta_1^{-1} , \quad t_2 \equiv \beta_2^{-1} \, .
\ee
The large root, $\beta_1 \!=\! \beta_2+d$,
is commonly positive, representing an exponentially
growing mode that dominates at $t\!>\!t_1$.
The small root, $\beta_2$, could be positive as well, such that $t_1\!<\!t_2$.
Alternatively it could be negative, thus representing an exponentially 
decaying mode, with $t_2 \!=\! \vert \beta_2 \vert ^{-1}$ either smaller or 
larger than $t_1$, which may in principle dominate at 
$t\!<\!t_1$ and $t\!<\!t_2$.

\smallskip
The coefficients $m_1$ and $m_2$ are determined by the initial condition
$\Mgi$, via
\be
\Mg(t=0) = \Mgi = m_1+m_2
\label{eq:icondition1}
\ee
and
\be
\Mdotg(t\!=\!0) \!=\! \gi\Mgi \!=\! m_1\beta_1 + m_2\beta_2 , \quad
\gi \equiv g(t\!=\!0) .
\label{eq:icondition2}
\ee
The first equality in \equ{icondition2}
is based on the definition of $g$ in \equ{cont_gas_3}, and 
the second equality is based on a time derivative of $\Mg(t)$ from \equ{Mgt}.
Solving \equ{icondition1} and \equ{icondition2} yields
\be
m_1= \frac{(\gi-\beta_2)}{d} \Mgi \, , \quad
m_2= \frac{(\beta_1-\gi)}{d} \Mgi \, .
\ee
The dependence of these coefficients on the initial condition is
both through the explicit $\Mgi$ and through $\gi$.

\smallskip
The stellar mass is obtained by a time integration of the SFR,
as given in \equ{Mdots},
namely integrating $\Mg(t)$ from \equ{Mgt}. We obtain
\be
\begin{aligned}
\Ms(t) =& \Msi + \epsd (\mu-\etas) \\
&\times \left[\frac{m_1}{\beta_1\td}(e^{\beta_1 t}-1)
      +\frac{m_2}{\beta_2\td}(e^{\beta_2 t}-1) \right] \, .
\label{eq:Mst}
\end{aligned}
\ee
\Equ{Mgt} and \equ{Mst} provide the evolution of the stellar-to-gas mass ratio,
\be
\fsg(t)\!=\!
 \frac{\Msi+\epsd (\mu\!-\!\etas) [\frac{m_1}{\beta_1\td} (e^{\beta_1 t}\!-\!1)
                                +\frac{m_2}{\beta_2\td} (e^{\beta_2 t}\!-\!1)]
}
      {m_1 e^{\beta_1 t} + m_2 e^{\beta_2 t} } \, .
\label{eq:fsg}
\ee
The conventional gas fraction is given by $\fg(t) \!=\! [1+\fsg(t)]^{-1}$,
and the timescale for star formation is
$\tsf(t)\!=\!\epsd^{-1}\fsg(t)\td$ from \equ{tsf}.

\begin{figure*} 
\centering
\includegraphics[width=0.49\textwidth,trim={0.6cm 0.4cm 0.6cm 1.0cm},clip]
{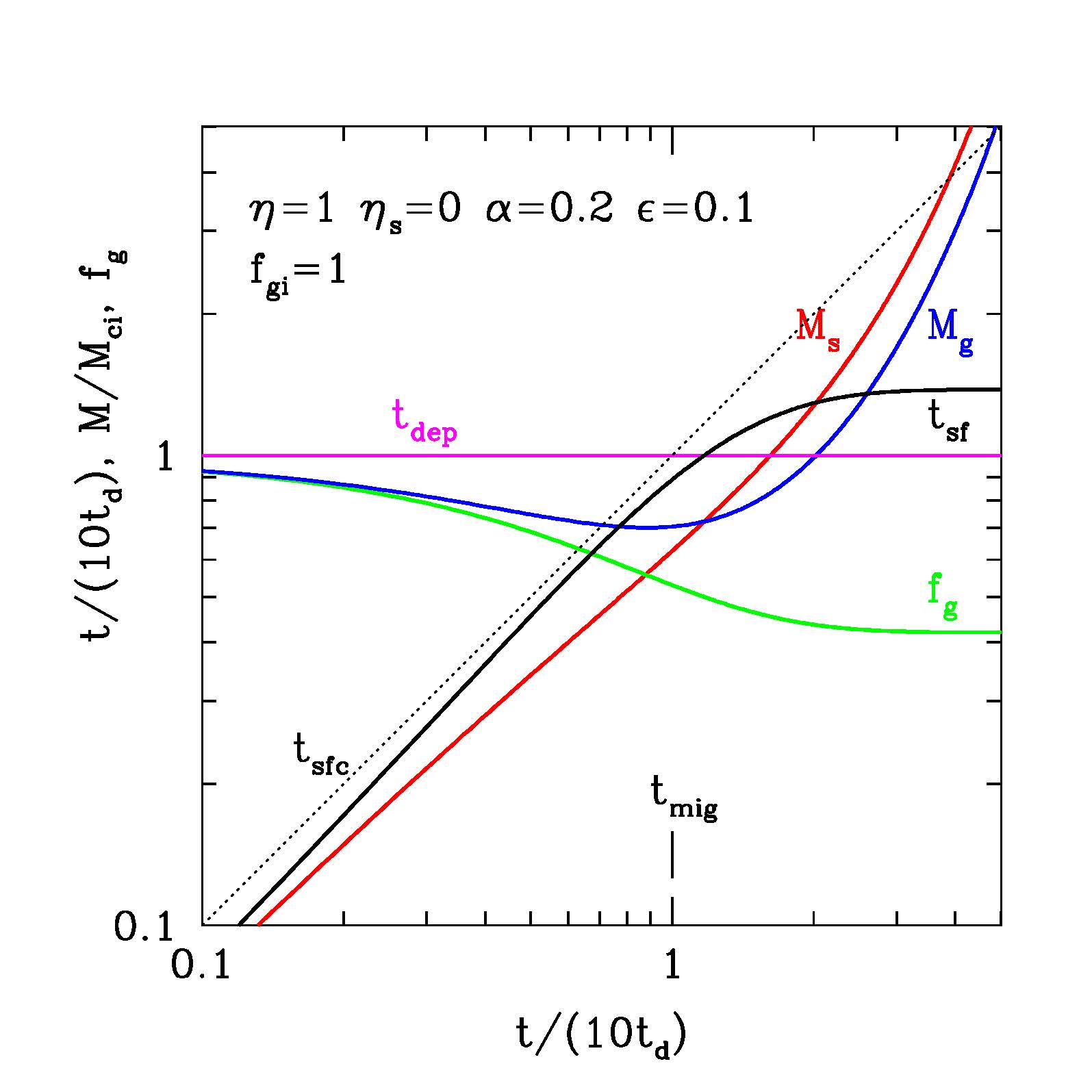}
\includegraphics[width=0.49\textwidth,trim={0.6cm 0.4cm 0.6cm 1.0cm},clip]
{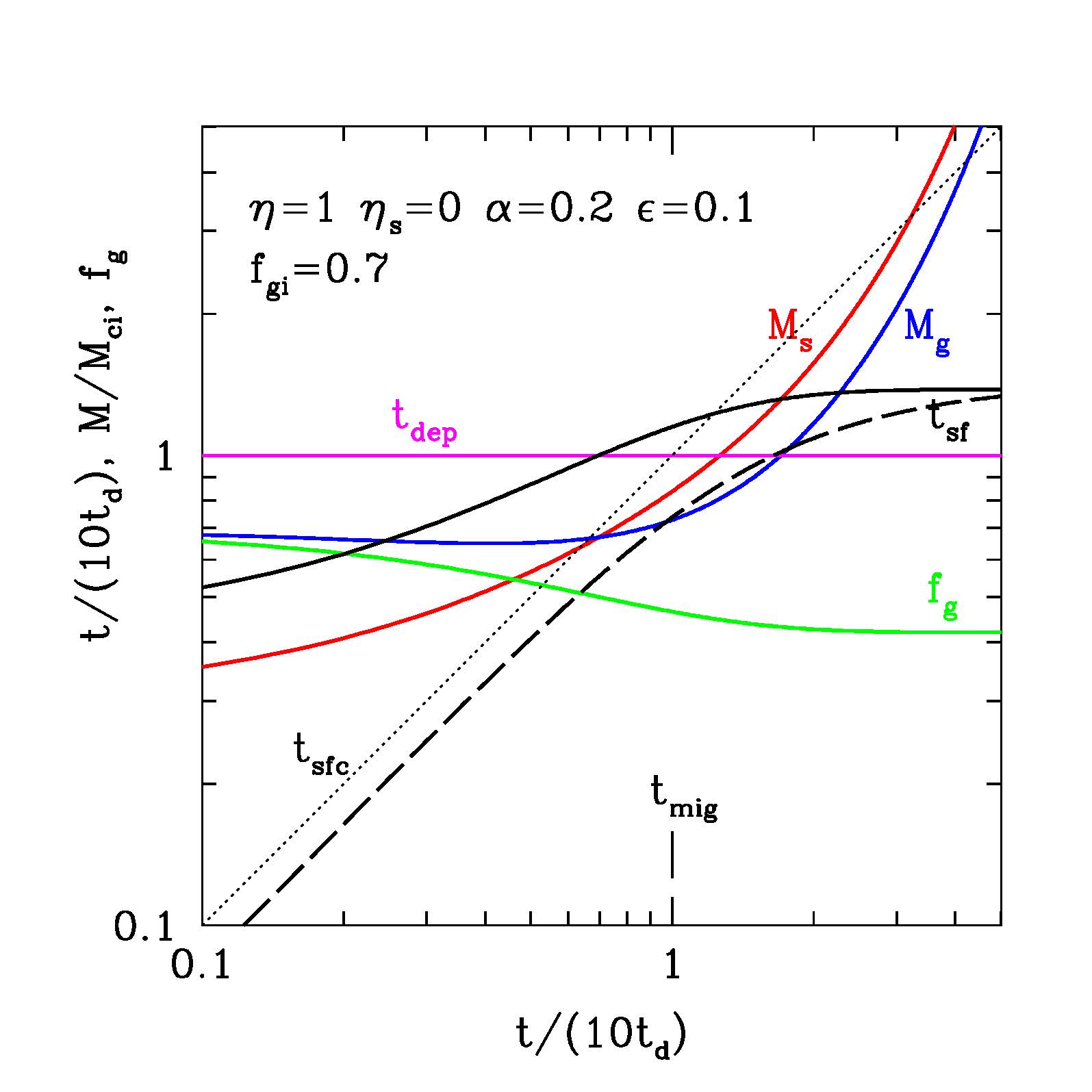}
\caption{
Toy-model evolution of clump properties, for the choice of parameters
indicated, and with initial gas fraction $\fgi\!=\!1$ (left) and 0.7 (right).
Shown are clump gas mass $\Mg$ (blue), stellar mass $\Ms$ (red),
gas fraction $\fg$ (green),
depletion time $\tdep$ (magenta), star-formation time $\tsf\!=$sSFR$^{-1}$
(black)
and its corrected version $\tsfc$ for $\fgi\!<\!1$, \equ{tsfc} (dashed black),
which differs from $\tsf$ only when $\fgi\!<\!1$.
We see the phase during migration where the SFR, proportional to
$\Mg$, is roughly constant, such that $\tsfc$ is a proxy for the clump
lifetime (dotted black). 
At late times the clump converges to a quasi-steady state with an
exponential mass growth and a constant $\tsf$,
but this is supposed to `occur' only after the migration is complete.
For $\fgi\!<\!1$ there is an early phase dominated by the initial gas
fraction, where $\tsf$ is an overestimate of the clump time,
but this is properly corrected for by $\tsfc$,
which is a close lower bound for the clump time.
}
\label{fig:massa}
\end{figure*}

\smallskip
\subsection{Three stages of evolution}

\Fig{massa} shows the time evolution of the different clump quantities of 
interest for a given choice of the model parameters.
Time is plotted with respect to $10\td$, assumed to represent $\tmig$.
Mass is in units of the initial clump mass $\Mci$.
Shown are $\Mg$ (blue), $\Ms$ (red), $\fg$ (green), $\tsf$ (black)
and $\tdep$ (magenta).
The parameters chosen for this figure are
$\alpha=0.2$, $\epsd=0.1$, $\etag=1$ and $\etas=0$.
The two panels show the solutions for two different initial conditions,
$\fgi=1$ (left) and $\fgi=0.7$ with an initial stellar component (right).
The exponents in this case, independent of $\fgi$,
are $(\beta_1,\beta_2)=(0.058, -0.138)$ in units of $\td^{-1}$,
thus representing a growing and a decaying mode.
The corresponding times $t=\vert\beta\vert^{-1}$, in units of $\td$,
are $(t_1,t_2)=(17.3, 7.25)$, namely both on the order of $\tmig$ and with 
$t_1$ somewhat larger.
We identify three stages of evolution, as follows.

\subsubsection{Exponential growth at late times}

\no
At $t>t_1$ (and $t>t_2$) the system approaches a quasi steady state.
In this stage the growing mode dominates, so both $\Mg$ (and the SFR)
and $\Ms$ grow exponentially, but their ratio converges to an asymptotic value,
\be
\fsg \rar \frac{\epsd(\mu-\etas)}{\beta_1 \td} \, , \quad (t>t_1) \, .
\ee
This makes $\tsf$ saturate at a maximum level of
\be
\tsf \rar \frac{\mu-\etas}{\beta_1} \, , \quad (t>t_1) \, .
\ee
For the choice of parameters in \fig{massa}, the asymptotic values 
are $\fsg=1.38$, corresponding to $\fg=0.42$, and $\tsf=13.8 \td$.

\smallskip
The exponential growth of $\Mg$ (and therefore $\Ms$) in this regime is
because the accretion of gas from the disc into the clump is proportional to
the total clump mass itself, which is driven by the growing stellar mass that
has become comparable to and slightly larger than the gas mass.
This results in a quasi steady state with a coherent exponential growth of 
both $\Mg$ and $\Ms$ at a fixed $\fg$.

\smallskip
Another way to understand this asymptotic exponential growth regime 
is by noting that with a constant $\fg$ the
time derivative of $g(t)$ in \equ{cont_gas_3} vanishes, so $g$ is a constant
and $\Mdotg \prop \Mg$, yielding an exponential growth $\Mg \prop \exp (gt)$.

\smallskip
We note, however, that if the migration is as rapid as estimated in \equ{tmig},
this asymptotic exponential phase is largely hypothetical as it is typically
supposed to occur after the clump has completed its migration and 
merged into the bulge.

\subsubsection{The main stage of migration: a constant SFR}
\label{sec:main}

\no
Prior to $t_1$, at $t\!<\!t_1$ (and $t\!<\!t_2$), 
the migrating clump is at it's main stage of evolution.
As $t \! \rar\! 0$, the gas mass approaches a constant $\Mgi$,
\be
\Mg \simeq m_1 (1+\beta_1 t) + m_2 (1+\beta_2 t)
\rar \Mgi \, ,
\ee
and the stellar mass approaches $\Msi$ accordingly,
\be
\Ms \simeq \Msi + \epsd\, (\mu-\etas)\, \Mgi\, t/\td 
\rar \Msi \, .
\ee
In this whole regime prior to $t_1$, $\Mg$ is roughly constant,
therefore the SFR is roughly constant, so
the growing term of $\Ms$ is proportional to $t$.
From \equ{tsf} and \equ{fsg}, we obtain during this period
\be
\tsf \simeq \epsd^{-1}\fsgi\, \td +(\mu-\etas) t \, , \quad (t<t_1) \, .
\label{eq:tsf_early}
\ee
For a small enough $\Msi$ ($\fgi\!\sim\! 1$ or $\fsgi\!\ll\! 1$), 
this expression simplifies to
\be
\tsf \simeq (\mu-\etas) t \, , \quad (\fgi\sim 1,\, t<t_1) \, .
\ee
This linear growth with $t$ is seen in \fig{massa} in the case $\fgi\!=\!1$ 
throughout the whole stage $t\!<\!t_1$.

\smallskip
For the above choice of parameters and $\fgi\!=\!1$, the coefficients for
$\Mg(t)$ are $(m_1,m_2) \!\simeq\! (0.3,0.7)$
in units of $\Mgi$.
The fact that in this case $m_2\!>\!m_1>0$ means that the decaying
mode is dominant at early times, $t\!<\!t_2$, such that
$\Mg$ slowly declines with time until the growing
mode takes over after $t_1$.
However, the gas mass is roughly constant until $t\sim t_1$, as the growing and
decaying modes roughly balance each other.
For $\fgi \!\sim\! 1$, the minimum of $\Mg(t)$, obtained between $t_2$ and 
$t_1$, is at a mass value not much smaller than the initial $\Mgi \!\sim\! 1$.
It turns out that for all values of $\fgi$ the curves for $\Mg(t)$
go through a fixed point, near $t\!\sim\! 10\td$ ($t\!\simeq\! 9\td$ and 
$\Mg\!\simeq\! 0.7$ for the above choice of parameter values).

\smallskip
We note that for this choice of parameters, the total clump mass $\Mc$ 
is roughly constant until the end of the migration at $t \!\sim\! 10\td$. 
This is because the clump mass is
dominated by the gas mass that is roughly constant. The constant $\Mc$
makes the accretion rate into the clumps roughly constant, so the solution
tends toward a constant $\Mg$, with the input balanced by the output,
similar to the quasi-steady-state solution for a whole galaxy under a 
constant cosmological accretion rate \citep{dm14}.

\subsubsection{Effect of initial stellar mass at early times}
\label{sec:fsgi}

With an initial stellar component such that $\fgi\! <\! 1$, 
there is a third, early stage, 
where the star-formation time approaches an asymptotic minimal value
\be
\tsfi = \epsd^{-1}\fsgi\, \td \, , \quad (t<\ti) \, ,
\ee
valid at $t\!<\!\ti$, defined by
\be
\ti = \frac{\fsgi}{\epsd (\mu-\etas)} \td \, .
\ee
This follows from \equ{fsg}, with $\vert \beta_1 t\vert \!\ll\! 1$ and
$\vert \beta_2 t \vert \!\ll\! 1$ and demanding that the second term in the
numerator is smaller than the first.
As long as $\ti$ is significantly smaller than $\tmig$, $t_1$ and $t_2$,
there is a period between these two times where the 
evolution is of the sort discussed in \se{main}, 
namely a roughly constant SFR and $\tsf \simeq t$.
For example, if min$\{t_1,\tmig\} \!\sim\! 10\td$, this requires
$\fsgi \!\ll\! 10\epsd (\mu-\etas) \!\sim\! 1$, 
namely an initial gas fraction close to unity.

\smallskip
However, once $\Msi$ is not negligible, the early evolution of $\tsf$ deviates
from the pure linear $\tsf \!\prop\! t$.
As can be seen in \fig{massa}, already for $\fgi\!=\!0.7$, the growth of 
$\tsf(t)$ throughout the period $(\td,\tmig)$ is significantly shallower than 
a linear growth, with $\tsf$ a severe overestimate of $t$ at early times.

\smallskip 
If one wishes to use $\tsf$ as an observable proxy for the clump age $t$,
and make sure that it is not a severe overestimate of the age at early times,
one could define a corrected estimator for the star-formation time by
\be
\tsfc = \frac{\Ms-\Msi}{\sfr} \, . 
\label{eq:tsfc}
\ee
As seen in the right panel of \fig{massa}, for $\fgi\!=\!0.7$, $\tsfc$ is 
indeed growing roughly linearly with time prior to $\sim\! 10\td$, 
where it deviates from $t$ by only $\sim\! 20\%$. 
The initial stellar mass is $\Msi\! =\! (1-\fgi) \Mci$, so it can be estimated 
for a given initial (Toomre) clump mass and an initial gas fraction that can be
estimated from the gas fraction in the cold component of the clump-forming 
disc.

\subsection{Dependence on model parameters}

One can show that for the parameters within a factor of $\sim 2$ about their
fiducial values $\etag \!\sim\! 1$, $\alpha \!\sim\! 0.2$ and 
$\epsd \!\sim\! 0.1$, the value of $t_1$ is robustly of order 
$t_1 \!\gsim\! 10 \td$, quite insensitive to the actual values of the 
parameters.
If the estimated rapid migration of \equ{tmig} is valid,
this implies that $t_1 \!\sim\! \tmig$, so the asymptotic exponential growth 
phase occurs only after the end of migration. 
Thus, the main stage where $\tsf \!\sim\! t$ extends all
the way till the end of migration.  This is provided that the initial clump is
gas rich such that the onset of this regime, at $\ti$, is long before $t_1$.
The corrected $\tsfc$ is a good proxy for $t$ even prior to $\ti$.

\smallskip
Short-lived clumps may be modeled by a strong feedback.
When $\etag$ is larger than unity, say $\etag \!\sim\! 3$,
while $\beta_1$ remains similar to the case of $\etag \!\sim\! 1$ discussed 
above, the value of $\vert\beta_2\vert$ becomes significantly larger than
$\beta_1$, so the gas mass declines significantly during $t\!<\!t_1$.
Still, $\tsf \!\prop\! t$ in this regime, though it is an overestimate of $t$ 
if $\fgi\!<\!1$.
However, the corrected $\tsfc$ is an excellent approximation to the clump 
time for any value of $\fgi$, especially during the first few dynamical times 
after which the short-lived clump is expected to be disrupted.

\smallskip
We note in passing that
in some local Giant Molecular Clouds (GMC) there are certain indications 
for an early exponential 
growth while the clump is still gas dominated \citep{palla99}.
Our toy model would allow an early exponential growth when $t_1$ is 
sufficiently small compared to $10\td$, or when the migration is slower. 
This requires significant deviations from the fiducial
values of the model parameters assumed above for the high-$z$ giant clumps in
their main phase of evolution, 
in particular very efficient accretion with $\alpha$ closer to unity, 
and/or negligible outflows with $\etag \!\ll\! 1$, 
as well as a low SFR efficiency $\epsd \!<\! 0.1$.  
This may be valid in the early, collapsing phase of the GMC.
The exponential phase may alternatively be reached if the migration is 
significantly slower than assumed here. This would be the case if
the disc becomes a ring about a large central mass, which suppresses the ring's
inward mass transport rate \citep{dekel20_ring}. Such conditions are expected
after a generic wet-compaction event that typically occurs when the galaxy
becomes larger than the golden mass of $M_{\rm s,gal} \!\sim\! 10^{10}\msun$.

\section{Simulations}
\label{sec:sims}

\subsection{Isolated galaxy simulations}


\subsubsection{The simulations}

We use here simulations similar to those described in
\citet[][e.g., model G2]{bournaud14} and in \citet{perret14}. The
RAMSES AMR code \citep{teyssier02}
is utilized with a maximum resolution of $3.5\pc$ and an SFR
efficiency of $\epsf\!=\!0.02$ per free-fall time on the local cell size. Six
gas-rich disc galaxies were simulated in isolation, mimicking the evolution of
$z \!\sim\! 2$ discs for a duration of $\sim 1 \Gyr$. The galaxy baryonic 
masses are in the range $4\!\times\! 10^{10}\!-\! 4\!\times\! 10^{11}\msun$ 
and the initial gas fraction is 0.6. The typical disc crossing time at the 
radius where most clumps form is $\td\!\sim\! 24 \pm 3\Myr$.

\smallskip
These simulations include subgrid models for supernova feedback, 
and four of them also incorporate models for 
stellar feedback by photo-ionization and by radiation pressure (RP), as
described in \citet{renaud13}. The radiative pressure is assuming a momentum
supply to the outflow at a rate of $\sim\! 2.5 L/c$ for a luminosity $L$, 
namely the photons are weakly trapped with each photon
contributing 2.5 times. This 
radiative contribution is comparable to the contribution by supernova feedback
and 
is comparable to the momentum driving by all
stellar sources as predicted by \citet{dk13}. A few clumps from these
simulations were analyzed in \citet{bournaud14}, and the relevant results are
summarized in figures 6 and 8 there. 
The simulations with and without radiation-pressure feedback, both including
supernova feedback, are termed hereafter RP and NoRP respectively.

\smallskip
\Fig{image_cosmo} shows face on views of the surface cold-gas density
in NoRP and RP simulated disks during their main clumpy phase.
They both show giant star-forming clumps with masses $\Mc\!>\!10^8 \msun$,
tending toward larger masses when the RP is not implemented.

\subsubsection{Isolated clump analysis}

We briefly describe here our method for detecting clumps in the isolated galaxy
simulations, and refer to \citet{bournaud14} for further details.
The clumps are first detected visually in 2D face-on projections of the 
baryonic mass.  The center of each clump is then determined in 3D through an 
iterative procedure as follows. 
An initial guess for the clump center is obtained 
visually, assuming that it lies in the disk mid-plane. 
The center of mass of the baryons within a sphere of radius $200\pc$ about
this initial center is determined, and the procedure is repeated about this
new center until the correction is smaller than the resolution of the 
simulation.
The clump radius and mass about this centre is determined as follows.
Using a radius $R_1$, initially set to
$100\pc$, we compute the mass in a sphere of radius $R_1$ and in a sphere with 
a volume twice as large (i.e. a radius $2^{1/3}R_1$). If the mass increase is
larger than 30\%, we iterate the process with a larger 
radius $R_1 \!+\! 100\pc$ until the mass increase is by less than 30\%. 
In practice, the clump radii
range from $200\pc$ to $500\pc$. These positions and radii are used to measure
gas and stellar masses, mass flows, and SFR in the clumps.
Given the method used, clump masses are determined with an uncertainty better
than 30\%.  
Background subtraction was performed by subtracting the average density near a 
spherical contour around the clump at twice the clump radius. 
Another technique, based on the average density at the same galacto-centric 
radius as the clump center but outside of the clump itself, gave
very similar results. 
This background subtraction may be an over correction, because some of the
subtracted background may belong to the clump (see \se{contamination}). 

\smallskip
Here we analyze all the giant clumps with
total masses $\Mc \!\geq\! 10^{8}\msun$, 
which amount to 17 clumps from the two NoRP galaxies
and 31 clumps form the four galaxies with RP.

\smallskip
The SFR in each clump is directly measured from the simulations. 
%
The gas outflow rate from a clump, being mostly in the polar direction
perpendicular to the disc, is measured through parallel disc surfaces of
radius $1 \kpc$ at a height $\pm 1 \kpc$ above and below the clump center
and the galactic disc plane. 
This turned out to be a less noisy measurement than the outflow through 
the spherical boundary of the clump, which suffers from diffuse disk gas 
entering and leaving the clump \citep{bournaud14}.
The former method was found to lead on average to only a small, 
23 percent underestimate of the outflow. 
%
The gas accretion rate onto the clump, allowed to come from all directions,
is then evaluated through mass conservation, \equ{cont_gas_1} (with $\mu=0.8$),
where the net rate of change of $\Mg$ is measured from $\Mg$ in two
successive output times separated by $\sim\! 10\Myr$. 
The net stellar mass-loss rate is measured by
identifying stellar particles moving in and out from the clump.


\subsection{Cosmological simulations}
\label{sec:cosmo_sims}

\subsubsection{The simulations}

\smallskip
We use here two zoom-in cosmological simulations from the \vela suite,
out of 34 galaxies with halo masses $10^{11}-10^{12}\msun$ at $z \!\sim\! 2$, 
which have been used to explore many aspects of galaxy formation at high 
redshift
\citep[e.g.,][]{ceverino14,moody14,zolotov15,ceverino15_shape,
inoue16,tacchella16_prof,tacchella16_ms,tomassetti16,ceverino16_drops,
ceverino16_outflow,mandelker17,
dekel20_flip,dekel20_ring}. 
The simulations utilize the ART code \citep{krav97,krav03,ceverino09}, 
which follows the evolution of a gravitating $N$-body system and the
Eulerian gas dynamics with an AMR maximum resolution of $17.5-35\pc$
in physical units at all times.  The dark-matter particle mass is $8.3\times
10^4 \msun$ and the minimum mass of stellar particles is $10^3 \msun$.
The code incorporates gas and metal cooling,
UV-background photoionization and self-shielding in dense gas, 
stochastic star formation, 
stellar winds and metal enrichment, thermal feedback from 
supernovae \citep{cdb10}, and feedback from radiation pressure. 
In the current implementation of radiation pressure feedback, 
referred to as model ``RadPre" in \citet{ceverino14}, ``RP" in
M17, 
and \velathree in other places,
the momentum driving efficiency is $\sim 3 L/c$. This is comparable to the 
theoretical estimates and to the
radiative pressure contribution in the RP isolated simulations 
(which is comparable to the supernova contribution),   
but is on the low side compared to certain other strong-feedback simulations
\citep[e.g.,][]{genel12,oklopcic17}. 
Further details regarding the feedback and other sub-grid models  
can be found in \citet{ceverino14} and M17. 

\begin{figure} 
\centering
\includegraphics[width=0.48\textwidth]
{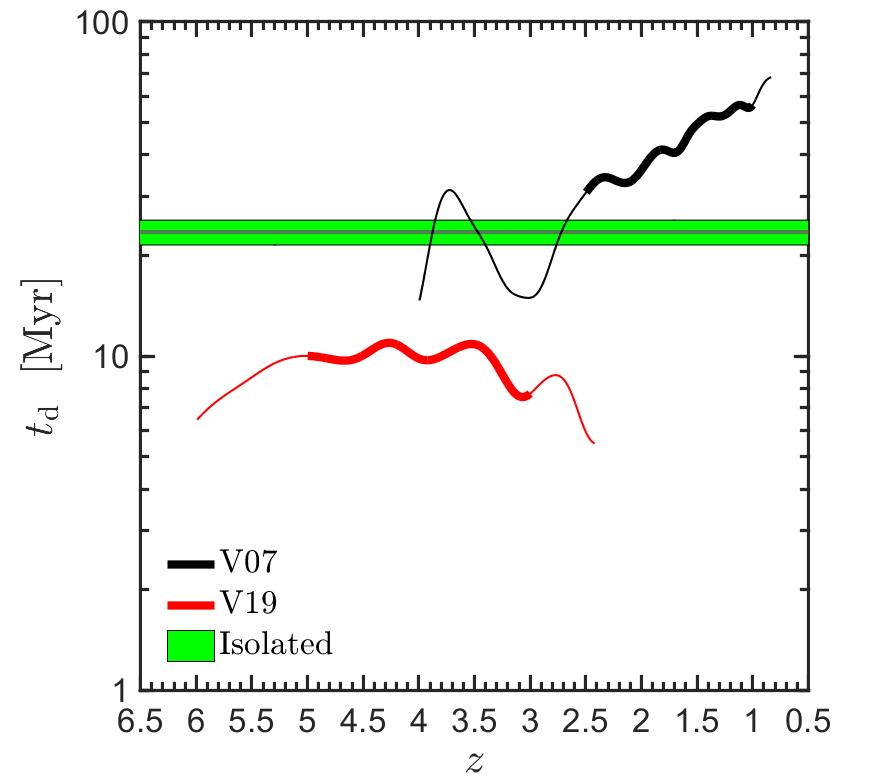}
\caption{
Evolution of disc dynamical time $\td$ in the simulations.
The curves for the cosmological simulations
are Gaussian smoothed with an FWHM that corresponds to
10\% of the Hubble time.
Marked are the relevant redshift ranges from which clumps were selected in the
cosmological simulations.
}
\label{fig:td}
\end{figure}

\smallskip
M17 have analyzed the clumps in the \velathree galaxies as
well as in counterparts that lacked the radiation pressure feedback 
(referred to as RP and NoRP, or \velathree and \velatwo, respectively).  
They found that with RP a
significant fraction of the clumps of $\Mc \!<\! 2\!\times\! 10^8\msun$ are 
disrupted on timescales in the ball park of one to a few disc dynamical times, 
while most of the more massive clumps survive 
and migrate to the disc centre. The migration inward of these long-lived clumps
produced galacto-centric radial gradients in their properties, such as mass, 
stellar age, gas fraction and sSFR, which distinguished them from
the short-lived clumps. 
However, in that study the output timesteps were $\sim\! 100\Myr$, 
which prevented a detailed study of the individual clump evolution.
In this study, we have re-simulated two of the \vela galaxies with RP 
(V07 and V19, see table 1 in M17), saving about 10 snapshots 
per disc crossing time. 
Both V07 and V19 have extended clumpy phases of VDI, in the redshift ranges
$z\!=\!2.5\!-\!1$ and $5\!-\!3$ respectively, where
massive, long-lived, star-forming clumps are continuously formed.

\smallskip
As in M17, the disc is defined as a cylinder with radius
$\Rd$ and half height $\Hd$, containing 85\% of the cold gas 
($T\!<\!1.5\!\times\! 10^4~{\rm K}$) and the young stars (ages $<\!100\Myr$)
within $0.15\Rv$. 
Thus, if the disc surface-density profile is exponential, $\Rd$ corresponds 
to twice the exponential scale radius.
The disc stars are also required to obey a kinematic criterion, 
that their specific angular momentum parallel to that of the disc, $j_{\rm z}$,
is at least $70\%$ of the maximal possible value it could have given its 
galacto-centric distance, $r$, and orbital velocity, $v$, namely 
$j_{\rm max}\!=\!vr$. 

\smallskip
The disc crossing time in the main body of the disc is computed as 
$\td\!=\!\Rrot/\Vrot$, where $\Rrot$ and $\Vrot$ are the mass-weighted radius 
and azimuthal velocity in the cylindrical ring of height $\pm\! 0.5 \kpc$
and radii $(0.5\!-\!1)\Rd$. For an exponential disc $\Rrot\!\simeq\! 0.78 \Rd$.
The evolution of $\td$ in the two simulations in the relevant redshift
ranges is shown \fig{td}.

\smallskip
In V07, the main VDI phase occurs at $z\!\sim\! 2.5\!-\!1$, following a
dramatic phase of wet compaction into a star-forming ``blue nugget"
\citep[][figure 2]{zolotov15}.  During this time, the disc radius steadily
increases from $\Rd\!\sim\! 9$ to $20\kpc$ while $\Hd/\Rd\!\sim\! 0.1\!-\!0.2$. 
The disc crossing time increases from $\td\!\sim\! 32$ to $57\Myr$,
the baryonic disc mass increases from $\Md\!\sim\! 2.7$ to 
$6.5\!\times\! 10^{10}\msun$, 
and the gas fraction in the disc decreases from $\fg\!\sim\! 0.28$ to $0.13$. 
The ``cold" mass fraction in the disc, referring to cold gas and
young stars, declines from $f_{\rm cold}\!\sim\! 0.4$ to $0.14$.

\begin{figure*} 
\centering
\includegraphics[width=0.48\textwidth,trim={0.0cm 0.8cm 0.0cm 0.2cm},clip]
{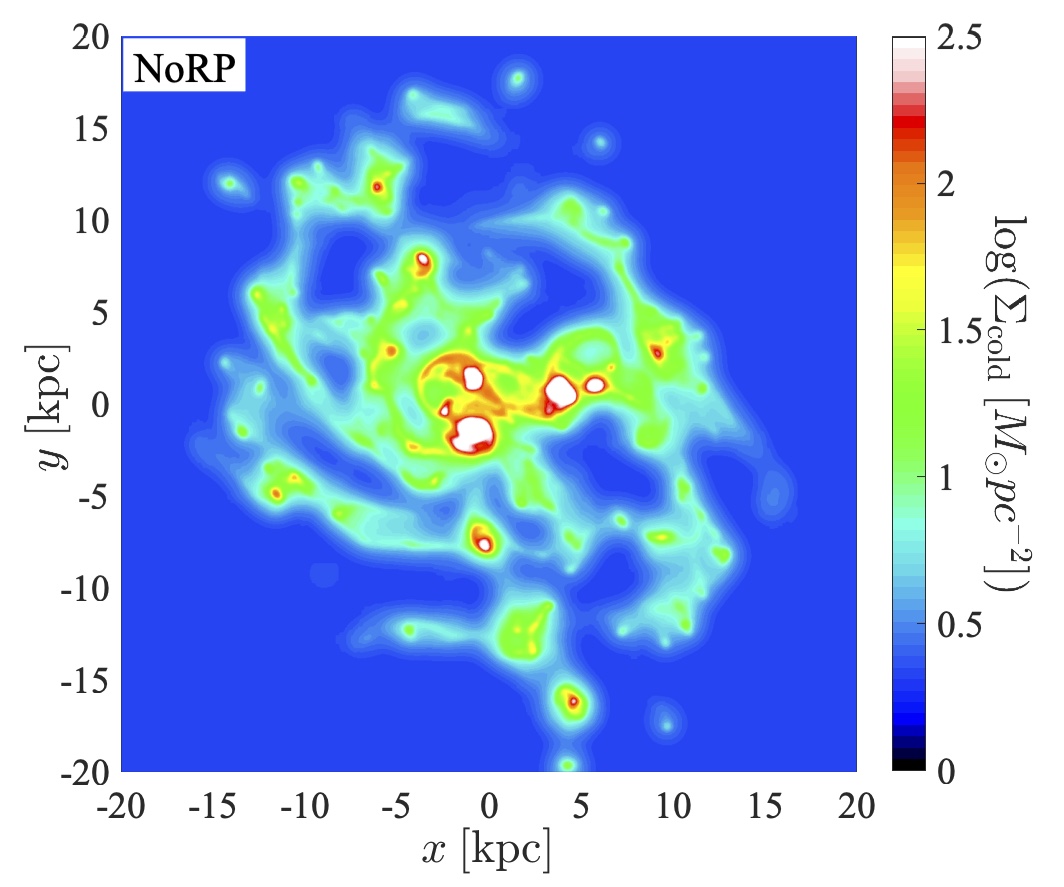}
\includegraphics[width=0.48\textwidth,trim={0.0cm 0.8cm 0.0cm 0.2cm},clip]
{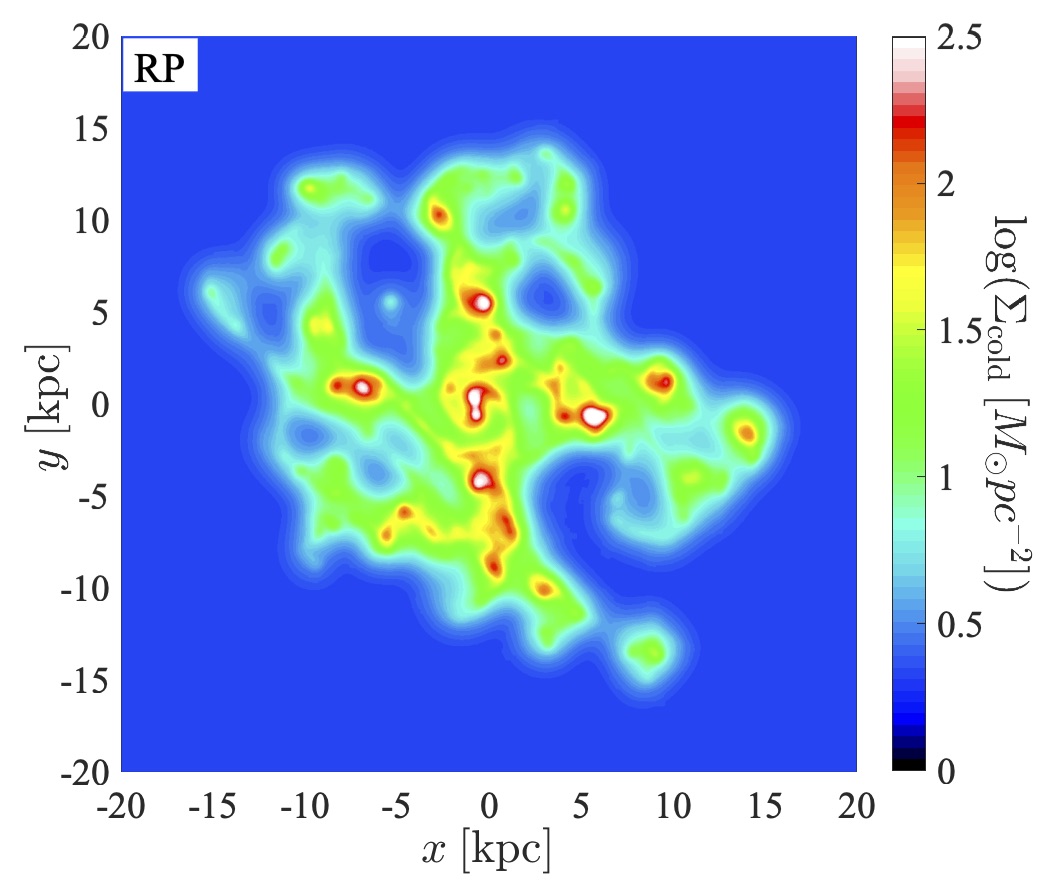}
\includegraphics[width=0.48\textwidth,trim={0.0cm 0.8cm 0.0cm 0.8cm},clip]
{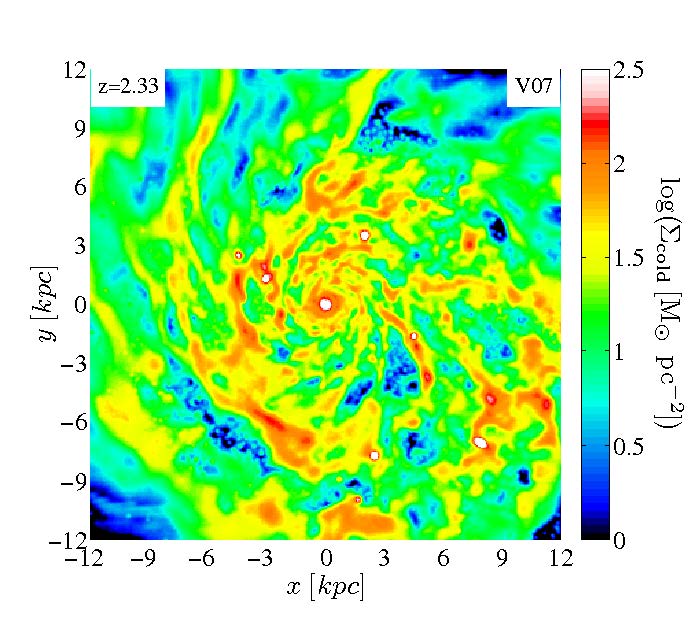}  
\includegraphics[width=0.48\textwidth,trim={0.0cm 0.8cm 0.0cm 0.8cm},clip]
{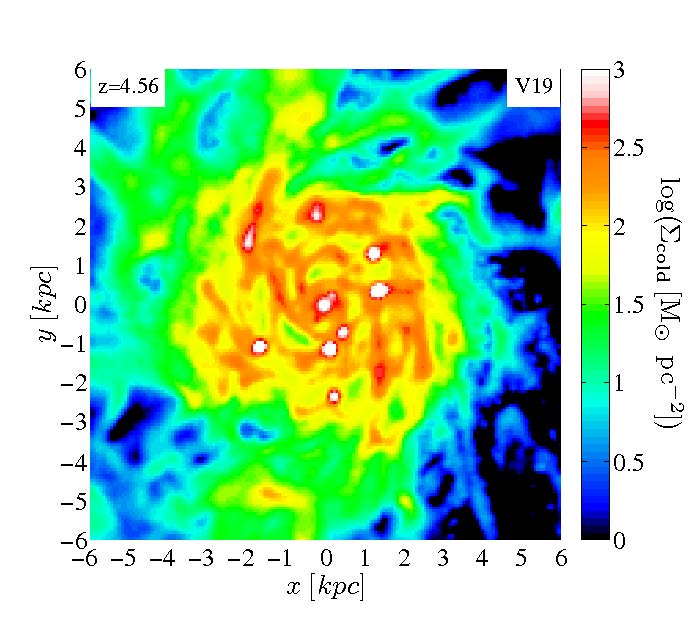}  
\caption{
Face-on views of the surface gas density in simulated discs showing giant 
clumps.
{\bf Top:} Isolated simulations, NoRP and RP of G1 \citep{bournaud14}.
Both show giant clumps, indicating less massive clumps when RP is incorporated.
{\bf Bottom:} Cosmological simulations, V07 at
$z\!=\!2.33$ (left) and V19 at $z\!=\!4.56$ (right) from the \velathree suite. 
Shown here is the
surface density of the ``cold" mass, comprised of cold gas
($T\!<\!1.5\!\times\!10^4~{\rm K}$) and young stars ($<\!100~\Myr$).
The disc radii are $\Rd \!=\! 11.8$ and $3.2\kpc$ respectively.
In both cases the disc half-thickness is $\Hd \!\sim\! \Rd / 6$.
The baryonic masses of the discs are
$\Md/(10^{10}\msun) \!\sim\! 3.50$ and $1.35$, and the fractions of cold-mass 
are $28\%$ and $33\%$ respectively. Both discs show giant clumps with
masses $\Mc\!>\!10^8\msun$.
The gas density in and around the $\sim\!1\kpc$ vicinity of the clumps  
is similar in the isolated simulations and the cosmological simulations at
$z\!\sim\! 2$, while the background density is higher in the cosmological
simulations where gas is continuously accreted.
}
\label{fig:image_cosmo}
\end{figure*}

\smallskip
In V19, the VDI phase occurs at higher redshifts, $z\!\sim\! 5\!-\!3$. 
During this phase the disc radius is $\Rd\!\sim\! 3\!-\!4\kpc$,
$\Hd/\Rd\!\sim\! 0.1\!-\!0.2$ and $\td\!\sim\! 10\Myr$.  
The disc mass increases from $\Md\!\sim\! 1.0$ to $1.5\!\times\! 10^{10}\msun$, 
the gas fraction in the disc decreases from $\fg\!\sim\! 0.23$ to $0.07$,
and the cold mass fraction is decreases from $f_{\rm cold}\!\sim\! 0.45$ to
$0.18$.
V19 thus represents a galaxy that is quite different from V07; 
being at higher redshift it is more compact and with a shorter dynamical time,
where the specific accretion rate, sSFR and gas fraction are significantly
higher.
These two galaxies allow us to explore cosmological galaxies at extremely 
different conditions, and hopefully identify robust features in the evolution 
of their clumps that are captured by our toy model.

\smallskip
The gas fractions in the discs of these two simulations are somewhat lower 
than estimated in typical observed galaxies
at similar redshifts. This has been discussed in detail in several papers which
used the \vela suite \citep[e.g.,][]{zolotov15,tacchella16_ms,mandelker17}. 
As highlighted
in those papers, while they are state-of-the-art in terms of high-resolution
AMR hydrodynamics and the treatment of key physical processes at the subgrid
level, the \vela simulations are not perfect in terms of their treatment of
star-formation and feedback, much like other simulations. Star-formation tends
to occur too early, leading to lower gas fractions later on. The stellar masses
at $z\!\sim\! 2$ are a factor of $\sim\! 1.5\!-\!2$ higher than inferred for 
similar mass
halos from abundance matching \citep[e.g.,][]{rodriguez17,moster18,behroozi19}. 
However, for the purposes of
the present study, the relatively low gas fractions during the peak VDI phase
would only underestimate the actual accretion of fresh gas onto clumps during
their migration, providing a lower limit on clump survival. The effect of gas
fraction on clump properties and survival in simulated isolated discs
is further discussed in \citet{fensch21}.
Regardless, the toy model can still be tested with the relevant parameters.

\smallskip
\Fig{image_cosmo} shows face on views of V07 at $z\!=\!2.33$ and V19 at 
$z\!=\!4.56$,
shortly after the start of their VDI phase. The figure shows the surface
density of the cold mass, integrated over $\pm\!\Rd$ perpendicular to the disc, 
where $\Rd\!\simeq\! 12$ and $3.2\kpc$ for V07 and V19 respectively. 
While these two discs are at very different redshifts, with different masses 
and sizes, both show giant star-forming clumps with masses 
$\Mc\!>\!10^8 \msun$.

\subsubsection{Cosmological clump analysis}
          
Clumps are identified in 3D and followed through time following the method 
detailed in M17. Here we briefly summarize the main features.

\smallskip
We search for clumps within a cube of side $L\!=\!3\Rd$ centered on the main 
galaxy centre.
Using a cloud-in-cell interpolation, we deposit the mass in a uniform 
grid with a cell size of $\Delta\!=\!70\pc$, $2\!-\!4$ times the maximal AMR
resolution. We then smooth the density in the grid cells, $\rho$, into a 
smoothed density $\rho_{\rm W}$, using
a spherical Gaussian filter of FWHM of ${\rm min}(2.5\kpc, 0.5\Rd)$. 
The density fluctuation at each grid point is 
$\delrho \!=\! (\rho\! -\! \rho_{\rm W})/\rho_{\rm W}$.
This is performed separately for the cold mass and the stellar mass, 
and at each point we adopt the maximum of the two residual values. 
We define clumps as connected regions containing at least 8
grid cells with a density fluctuation above $\delmin\!=\!10$. 
No attempt has been made to remove unbound mass from the clump.
We define the clump centre as the baryonic density
peak and the clump radius, $\Rc$, as the radius of a sphere with the same
volume as the clump. 
\textit{Ex situ} clumps, which joined the disc as minor
mergers, are identified by their dark matter content and the birth place of
their stellar particles and are not considered further here.

\smallskip
The SFR in the clumps is calculated using the mass in stars younger
than $30\Myr$, which is long enough for good statistics
and short enough for stellar mass loss to be negligible. 
%
Outflow rates from the clumps are measured using each of three spherical shells 
of width $\Delta r\!=\!140\pc$ centered at radii $r\!=\!\Rc$, $1.5\Rc$ and 
$2\Rc$.  The gas outflow rate is 
$\dot{M}\!=\!\Delta^{-1} \sum_i V_r \rho$, where the sum is over cells within 
the shell with $V_r\!>\!0$ and a 3D velocity larger than the escape velocity 
from the shell, $V^2\!>\!2G\Mc/r$.  
We adopt the average of the outflow rates in the three shells, 
which may vary by $\sim\! 0.1~{\rm dex}$.
This is different from how we measure outflows in the isolated
galaxy simulations, through parallel discs.
This is because the cosmological disc planes are thicker and highly perturbed 
and the outflows through slabs are noisier due to ongoing accretion and
mergers. We verified that the two methods do not lead to systematically
different results.
The gas accretion rates are then computed as in the isolated galaxies,
namely from the gas continuity equation,
\equ{cont_gas_1} (using $\mu\!=\!0.8$), after obtaining the net rate of change 
from the clump gas mass in successive output times. 
Using instead baryonic-mass conservation yields nearly identical results. 
Stellar mass loss or gain is measured by identifying stellar particles moving 
in and out from the clump, as for the isolated galaxy simulations.

\smallskip
Individual clumps, that contain at least 10 stellar particles,
are traced through time based on their stellar particles.
For each such clump at a given snapshot, we search for all
``progenitor clumps" in the preceding snapshot, 
defined as clumps that contributed at least $25\%$ of
their stellar particles to the current clump. If a given clump has more
than one progenitor, we consider the most massive one as 
the main progenitor and the others as having merged, thus creating a clump 
merger tree. 
If a clump in snapshot $i$ has no progenitors in snapshot $i\!-\!1$,
we search the previous snapshots back to two disc crossing times before 
snapshot $i$. If no progenitor is found in this period, snapshot $i$ is
declared the initial, formation time of the clump, and for that clump $t$ is
set to zero at that time.\footnote{If the mass weighted mean stellar age of the
clump at its initial snapshot is less than the timestep since the previous
snapshot, we set the initial clump time to this age rather than to zero. This
introduces an error of a few Megayears in the clump age.}

\smallskip
We select for analysis 
here %
clumps that formed at $z\!=\!2.5\!-\!1$ and $5\!-\!3$ 
for V07 and V19 respectively.
Based on the disc crossing times, the expected migration time for clumps,
which is about $10\!-\!20$ dynamical times, 
is in the range $400\!-\!1000\Myr$ for V07 and only $\sim 150\Myr$ for V19. 
In both
cases, we find this to be a good match to the actual migration times of massive
clumps. We only consider for analysis 
here %
clumps that survive for longer than $50\Myr$. 
Similar to our analysis of the isolated simulations, we also limit our sample 
of clumps based on their masses at $50\Myr$ (averaged from $40$ to $60\Myr$), 
to be more massive than $10^8\msun$. 
In order to relate to the observed clumps that are detected in UV 
\citep{guo18}, we also apply 
a lower SFR threshold of $0.1 \msun {\rm yr}^{-1}$ at $t\!=\!200\Myr$. 
This turns out to have no
effect in V19, but it removes about half the clumps that meet the mass and
lifetime criteria in V07, all in the mass range $10^8\!-\!10^{8.5}\msun$ 
(see also M17, Figure 10).  
We note that these passive clumps that were removed, most of which have formed
ex-situ, tend to move on perturbed orbits that 
deviate from the main disc, such that the gas density along their orbits
fluctuates, invalidating the toy model assumption of a constant density.
The orbits of the clumps selected by the criteria of mass and SFR above 
are all confined to the disc such that the accretion model is valid.
Our final clump sample contains 37 clumps from V07 and 12 clumps from V19.
We note that we have not applied here an explicit selection based on the 
M17 characterization of the clump as LL or SL. While most of the clumps
selected here are naturally LL in-situ clumps, 
some of them may be SL clumps, though the $50\Myr$ age threshold implies that
they live more than one-two disc dynamical times 
(SL clumps a la M17 can live for up to 20 clump free-fall times, 
which could reach a few disc dynamical times).

\smallskip
In summary, our sample consists of massive, star-forming, bound and mostly 
long-lived clumps from four types of simulations.
First, two RAMSES isolated-disc simulations, NoRP and RP, with 17 and 31 clumps
from two and four galaxies respectively.
Second, two ART-\vela cosmological simulations (with RP), 
V07 in the redshift range $z\!=\!2.5\!-\!1.0$ and V19 at $z\!=\!5\!-\!3$, 
with 37 and 12 clumps respectively.
In all these simulations the feedback is not overwhelmingly strong, such that
the massive clumps above $10^{8.5}\msun$ tend to 
survive feedback during their migration.


\section{Simulated clumps versus model}
\label{sec:evolution}

\subsection{Model validity and parameters}
\label{sec:param}

\begin{figure*} 
\centering
\includegraphics[width=0.45\textwidth]
{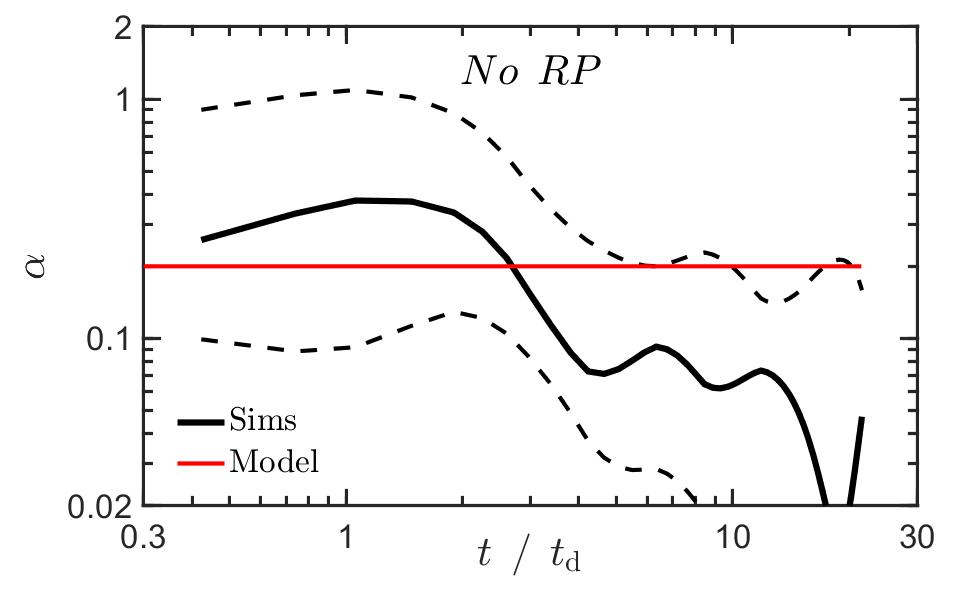}
\includegraphics[width=0.45\textwidth]
{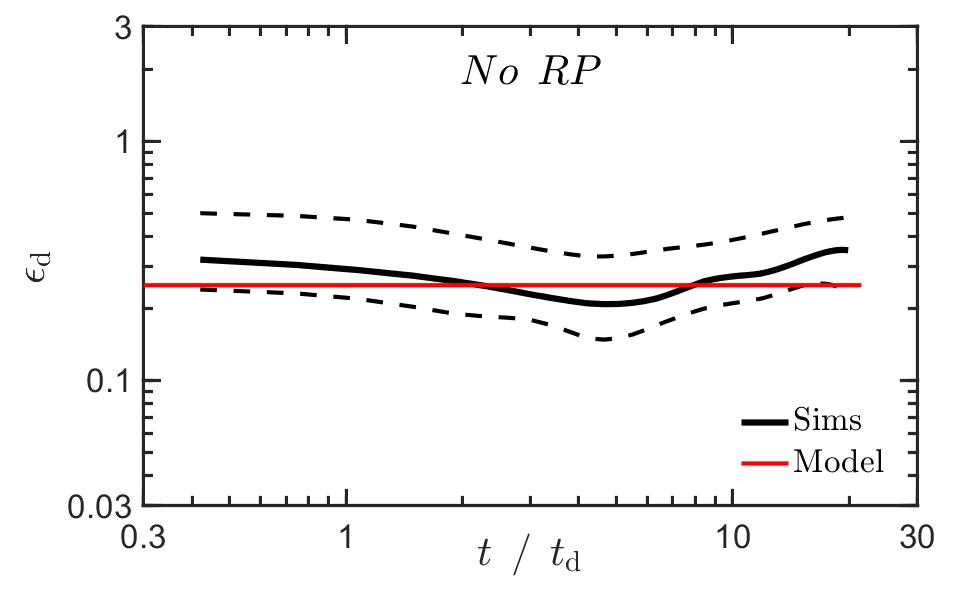}
\includegraphics[width=0.45\textwidth]
{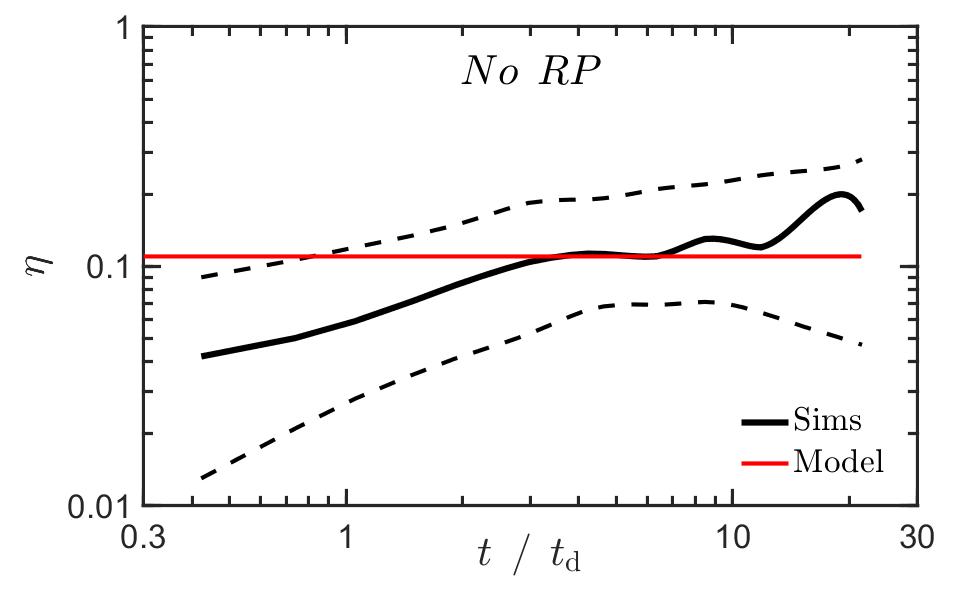}
\includegraphics[width=0.45\textwidth]
{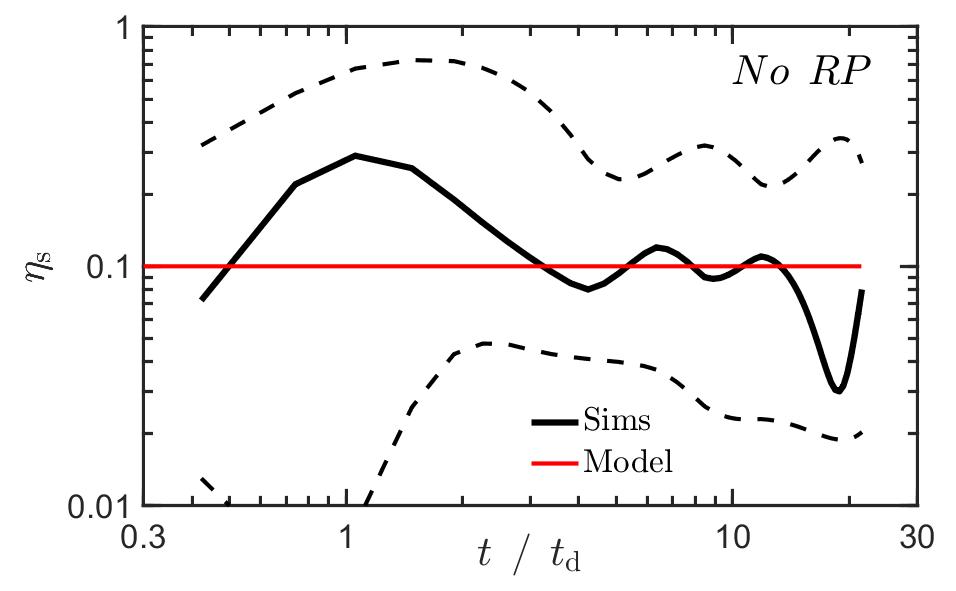}
\includegraphics[width=0.45\textwidth]
{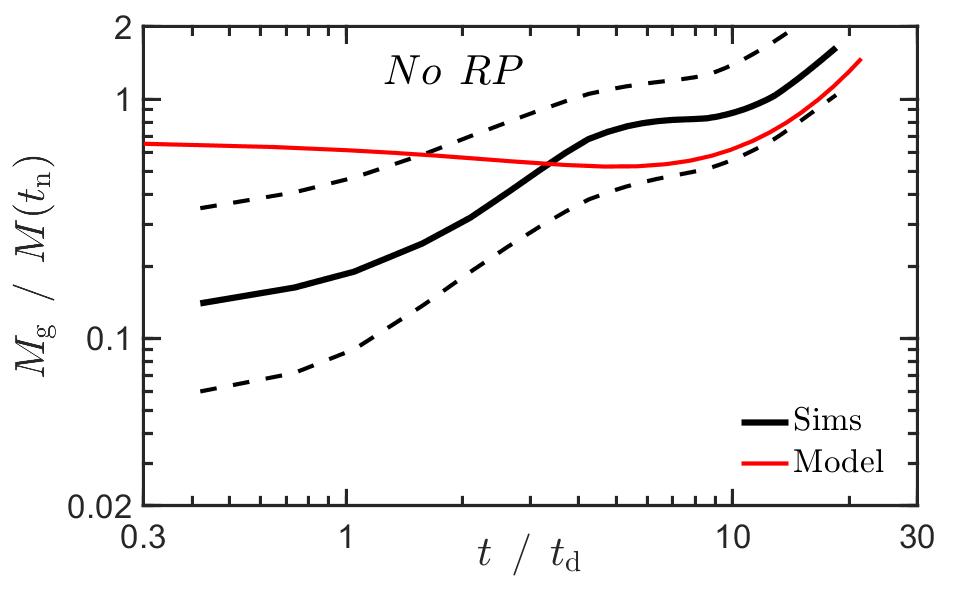}
\includegraphics[width=0.45\textwidth]
{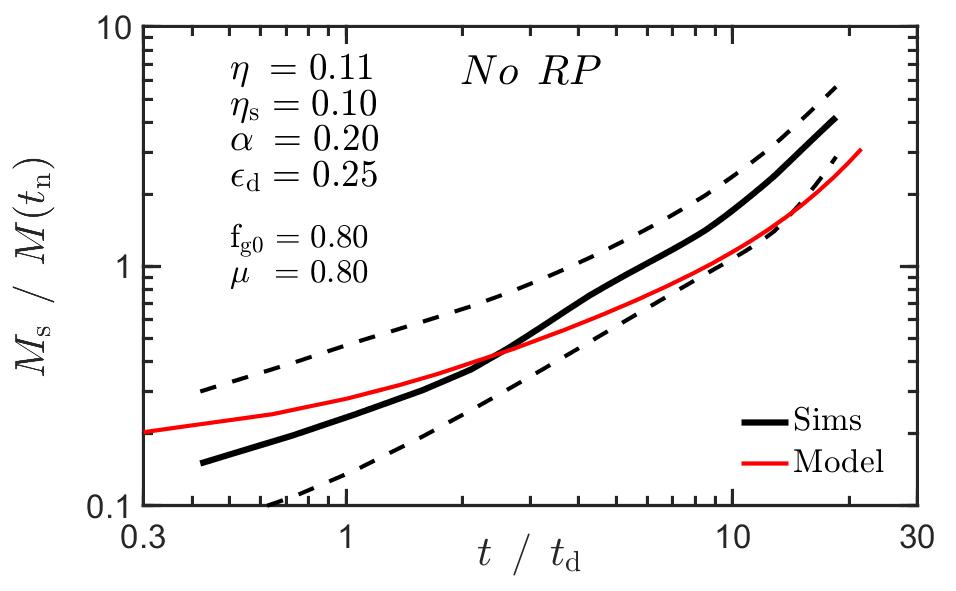}
\includegraphics[width=0.45\textwidth]
{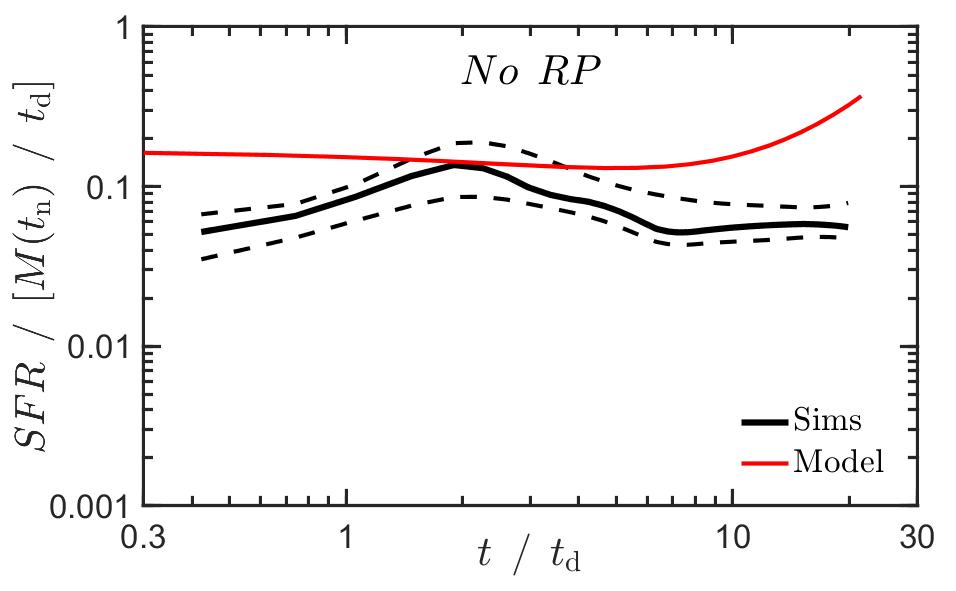}
\includegraphics[width=0.45\textwidth]
{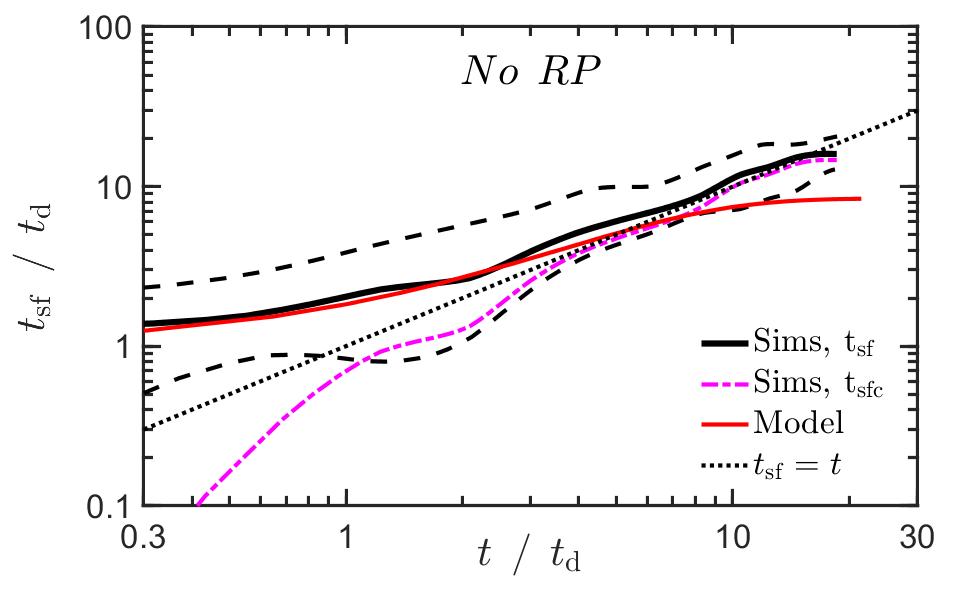}
\caption{
Evolution of simulated clump properties versus toy model,
here for the isolated galaxies with no RP.
Stacked are the clumps with $\Mc\!>\!10^8\msun$ that live for more than 
$200\Myr$, here 17 clumps.
The black curves refer to medians (solid) and 68\% scatter (dashed),
Gaussian smoothed with FWHM$\!=\!0.1$dex.
Time from clump formation is measured in units of the 
instantaneous disc dynamical time $\td$, shown in \fig{td}.
Model fits are shown (red) using the marked values of the parameters. 
The four top panels refer to the model parameters $\alpha$, $\epsd$, $\etag$
and $\etas$.
The third row from top shows the gas and stellar masses. 
Before stacking, the total mass of each clump, $\Mc$, is normalized to unity
at $\tform\!=\!3\td$, and $\Mg$ and $\Ms$ are normalized accordingly, thus 
given with respect to $\Mc(\tform)$.
The SFR in the bottom-left panel is normalized 
accordingly, so it is given in terms of $\Mc/\td$ at $\tform$.
The bottom-right panel shows $\tsf=$ sSFR$^{-1}$. 
Also shown is the corrected $\tsfc$.
The quantity of interest $\tdep/\td$ is given by $\epsd^{-1}$.
%
%
}
\label{fig:NoRP}
\end{figure*}

\begin{figure*} 
\centering
\includegraphics[width=0.45\textwidth]
{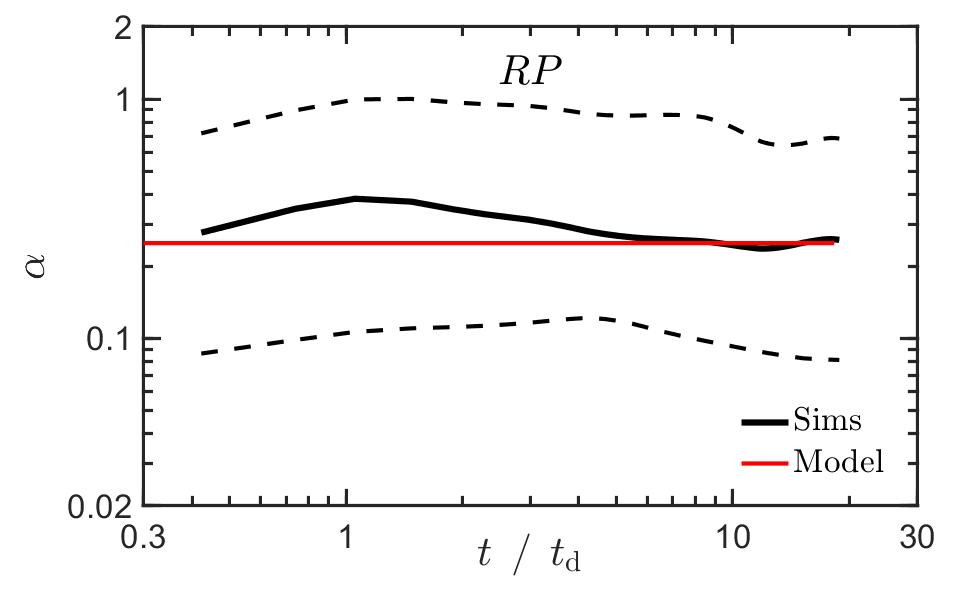}
\includegraphics[width=0.45\textwidth]
{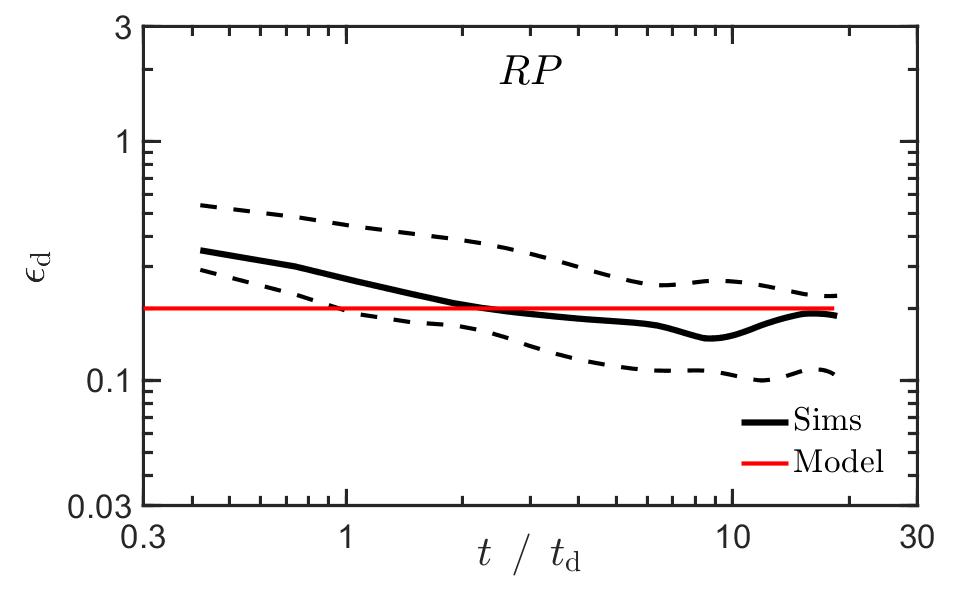}
\includegraphics[width=0.45\textwidth]
{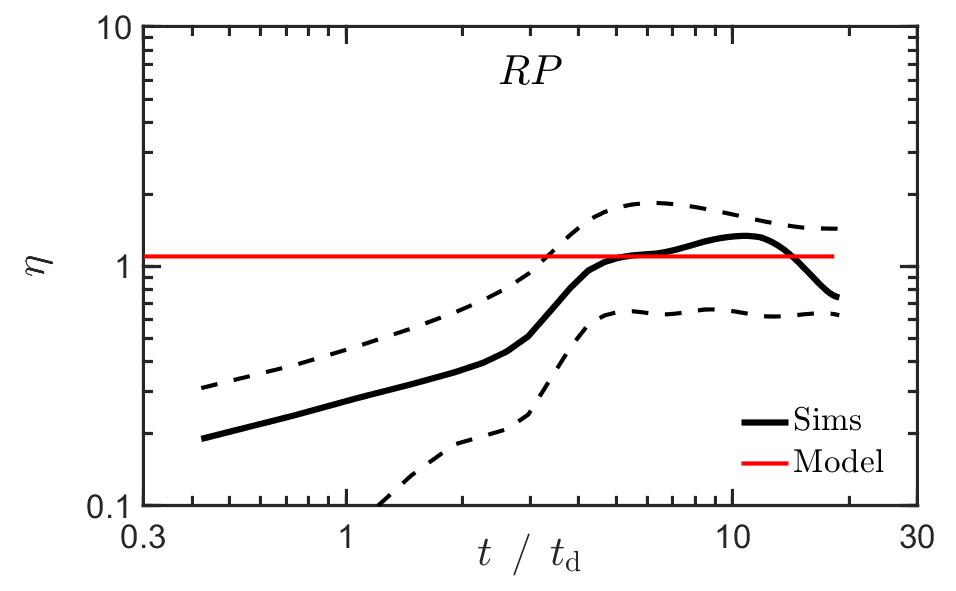}
\includegraphics[width=0.45\textwidth]
{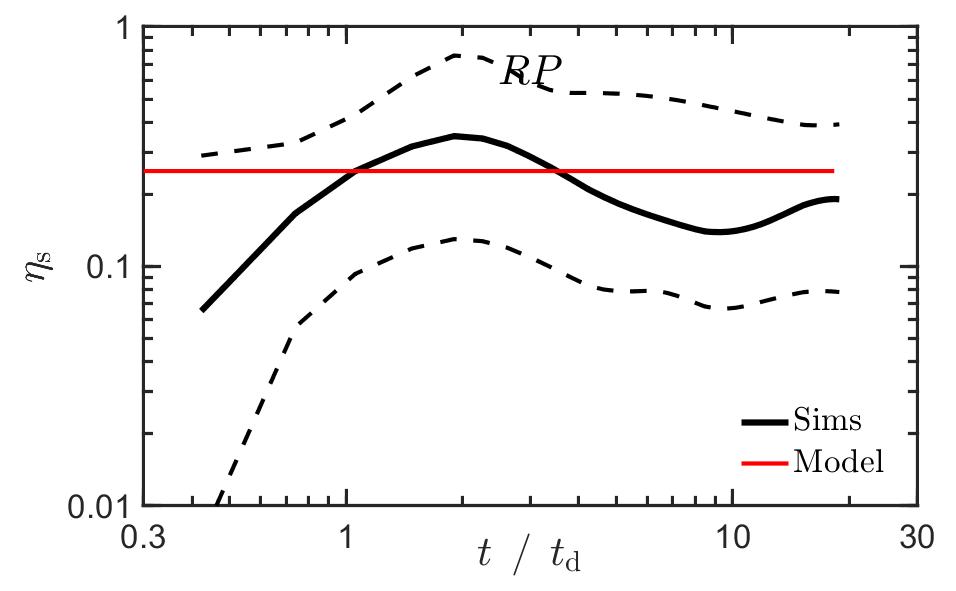}
\includegraphics[width=0.45\textwidth]
{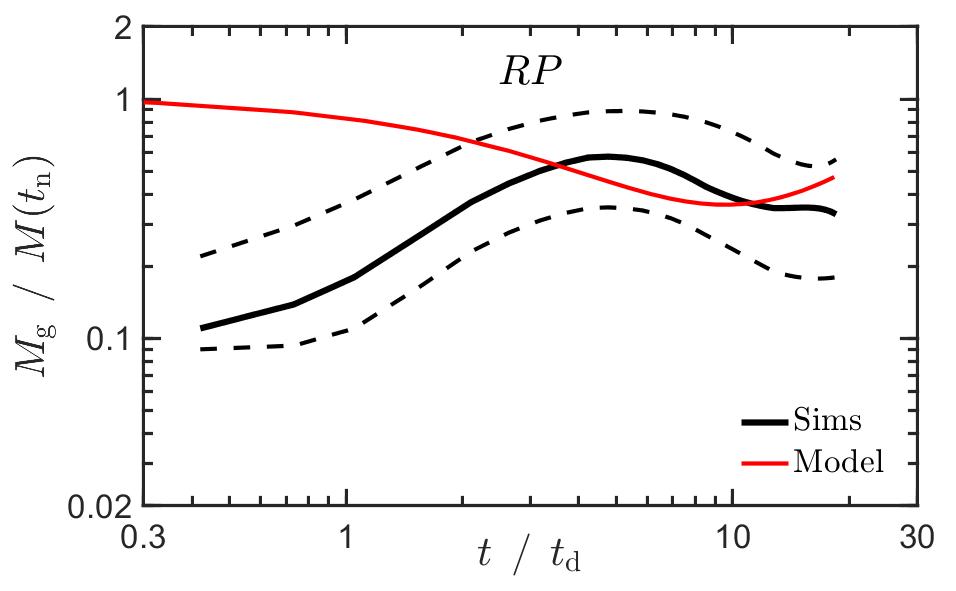}
\includegraphics[width=0.45\textwidth]
{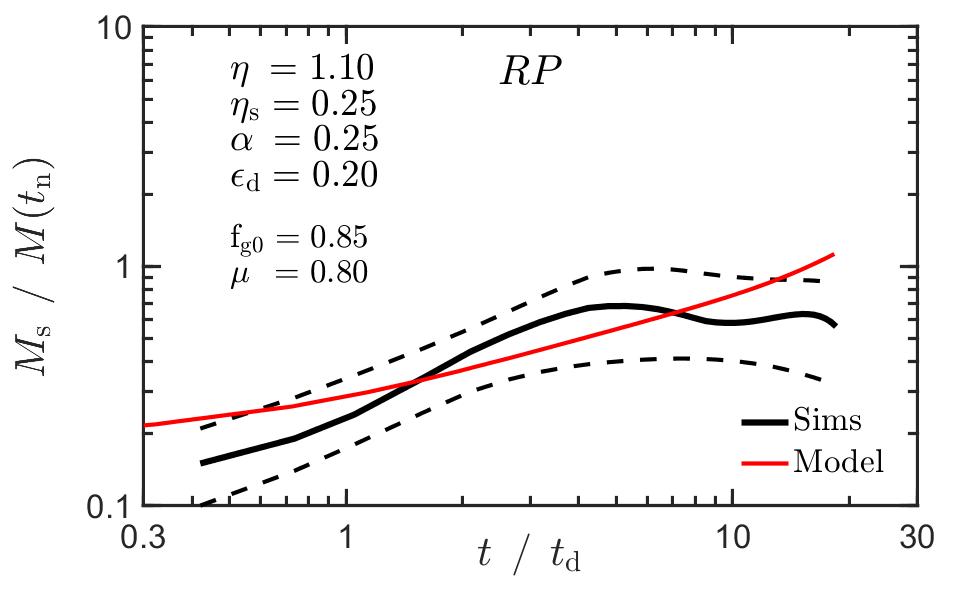}
\includegraphics[width=0.45\textwidth]
{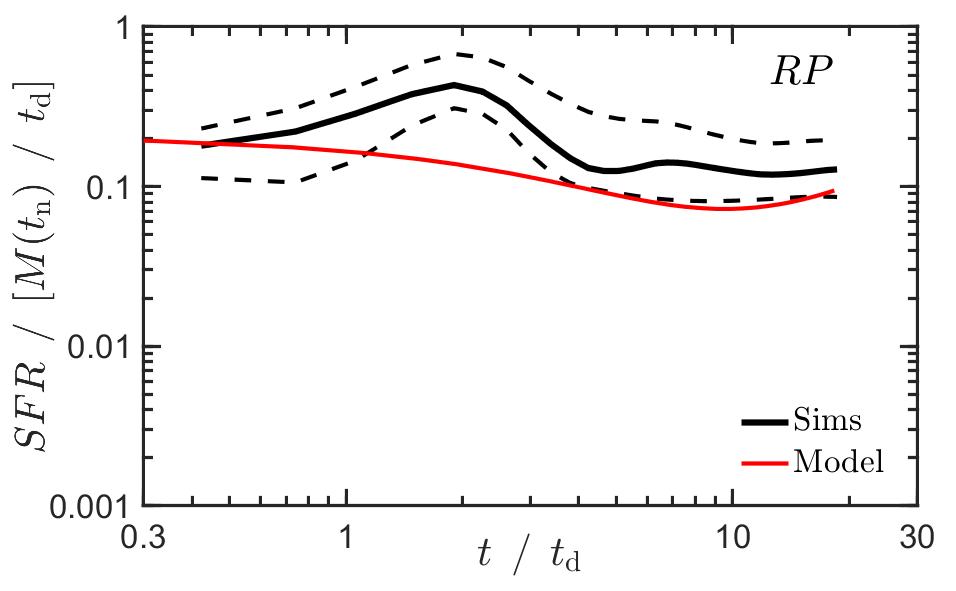}
\includegraphics[width=0.45\textwidth]
{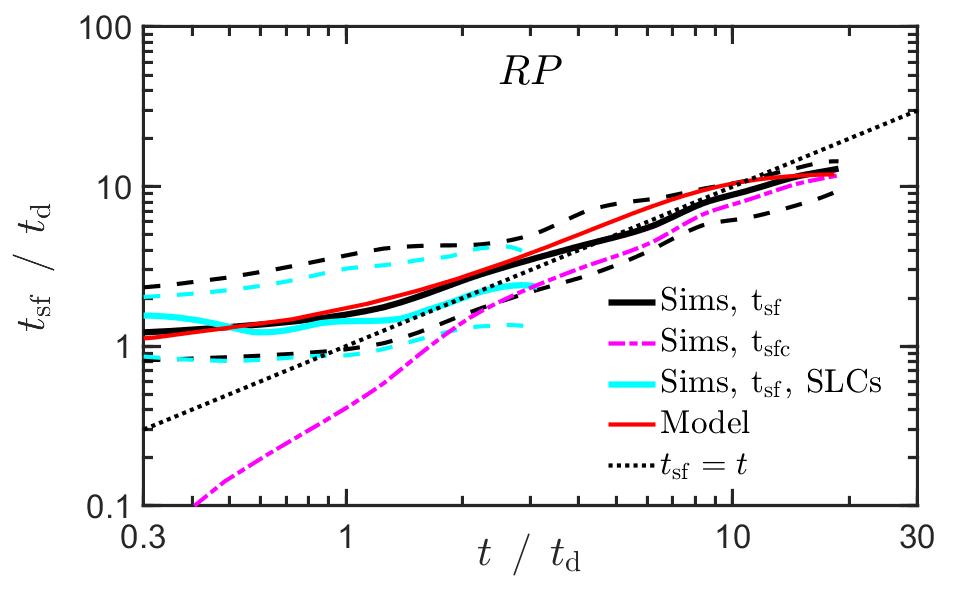}
\caption{
Evolution of simulated clump properties versus toy model,
here for the isolated galaxies with RP.
The details are the same as in \fig{NoRP}, here with 31 clumps. 
%
%
The bottom-right panel shows $\tsf$ also for short-lived clumps from
simulations of isolated galaxies with a low gas fraction and stronger feedback. 
}
\label{fig:RP}
\end{figure*}

\begin{figure*} 
\centering
\includegraphics[width=0.45\textwidth]
{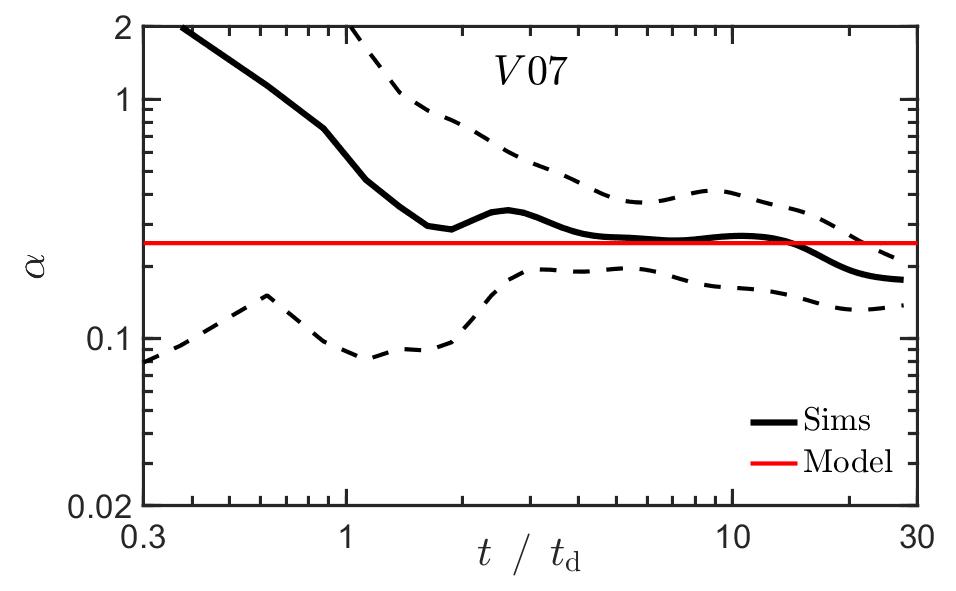}
\includegraphics[width=0.45\textwidth]
{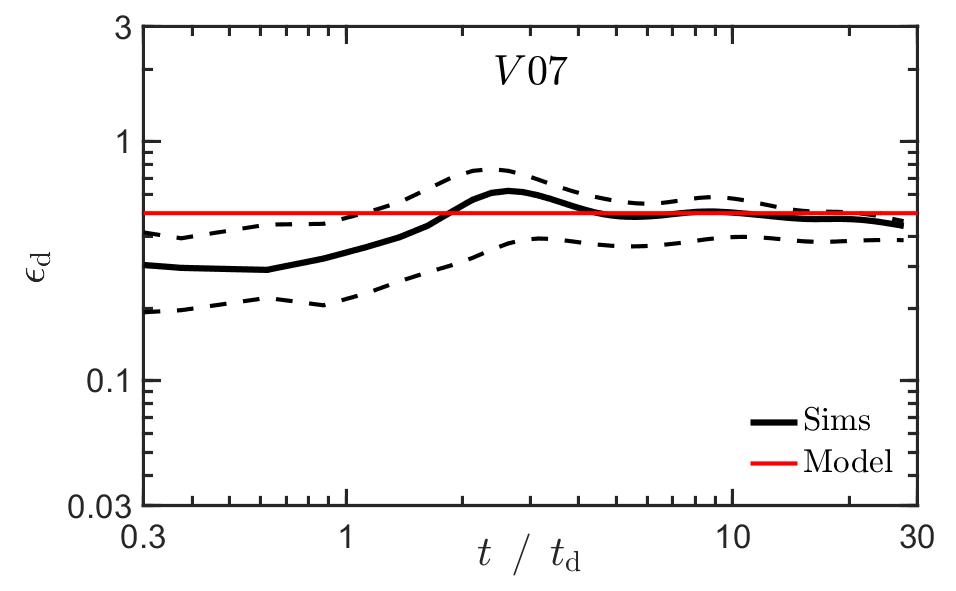}
\includegraphics[width=0.45\textwidth]
{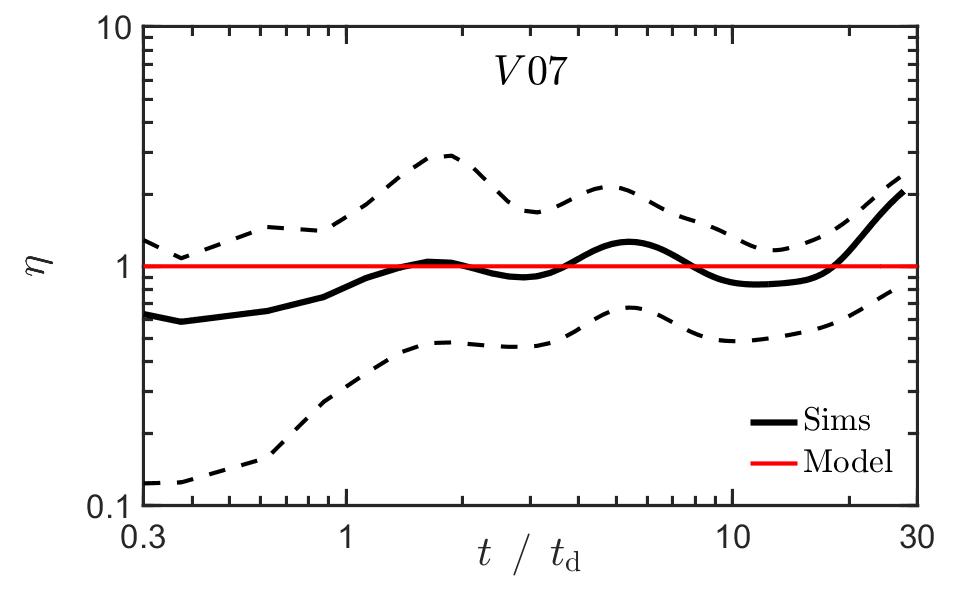}
\includegraphics[width=0.45\textwidth]
{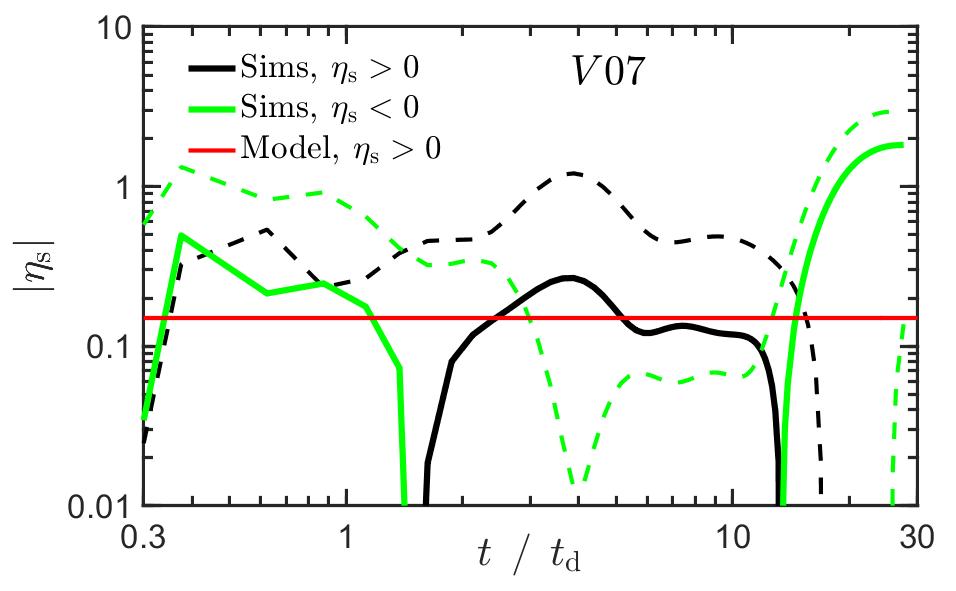}
\includegraphics[width=0.45\textwidth]
{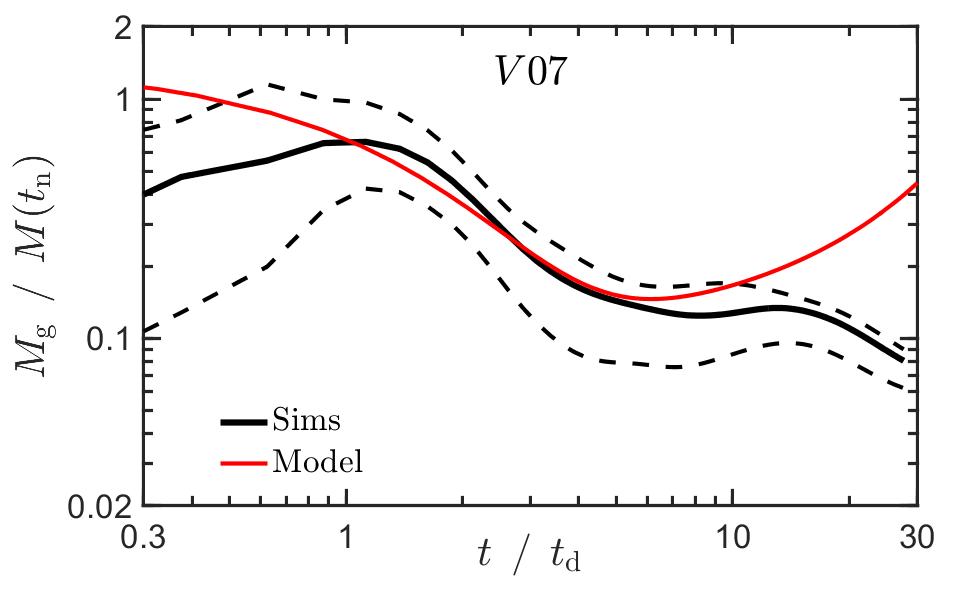}
\includegraphics[width=0.45\textwidth]
{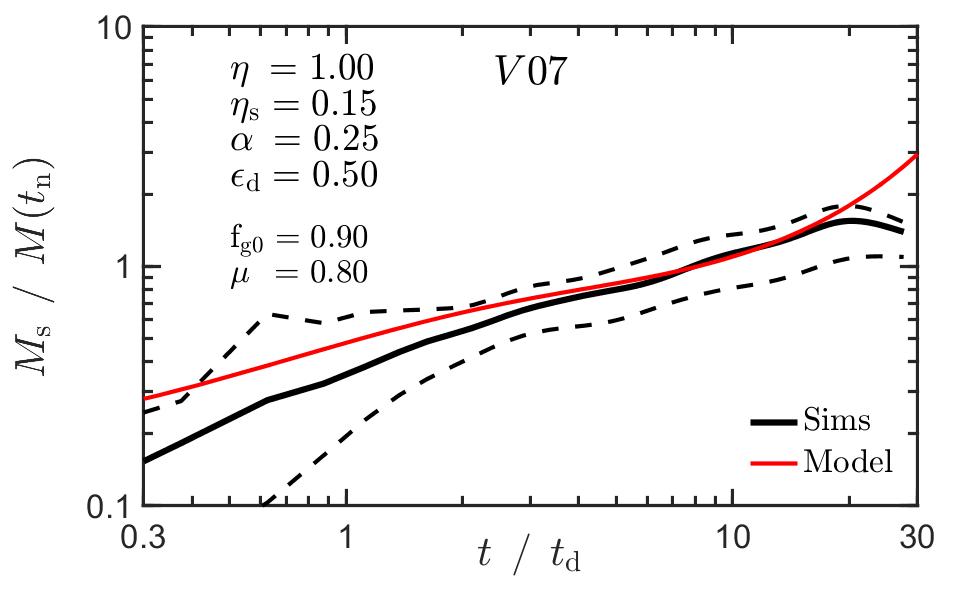}
\includegraphics[width=0.45\textwidth]
{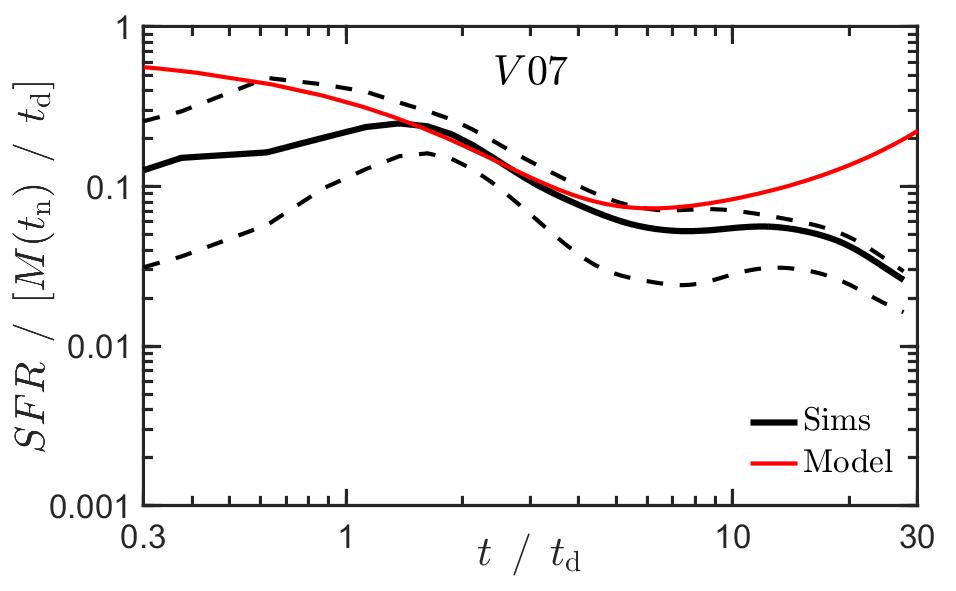}
\includegraphics[width=0.45\textwidth]
{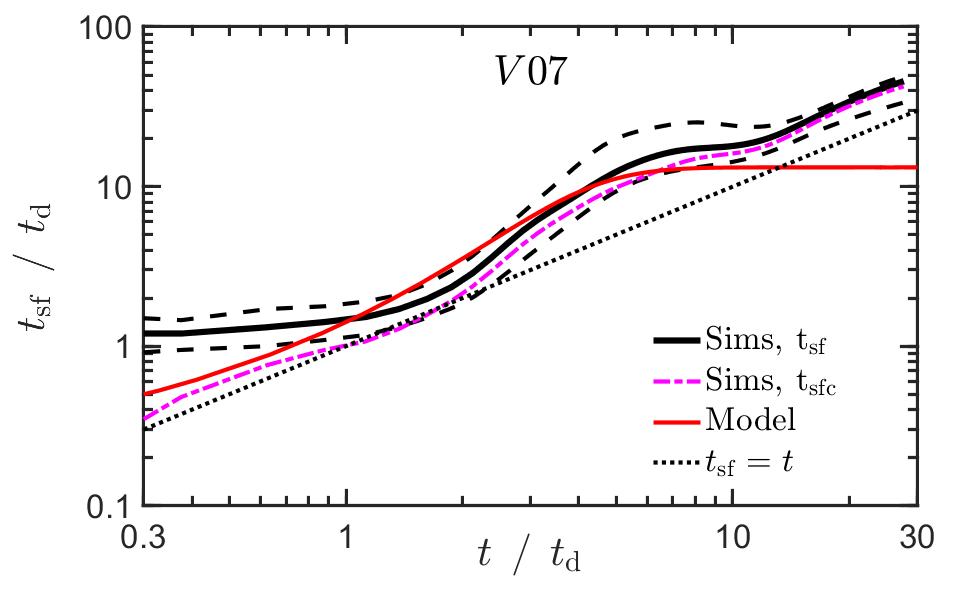}
\caption{
Evolution of simulated clump properties versus toy model,
here for the cosmological galaxy V07 at $z\!=\!2.5\!-\!1$.
The details are the same as in \fig{NoRP}, here with 37 clumps.
}
\label{fig:V07}
\end{figure*}

\begin{figure*} 
\centering
\includegraphics[width=0.45\textwidth]
{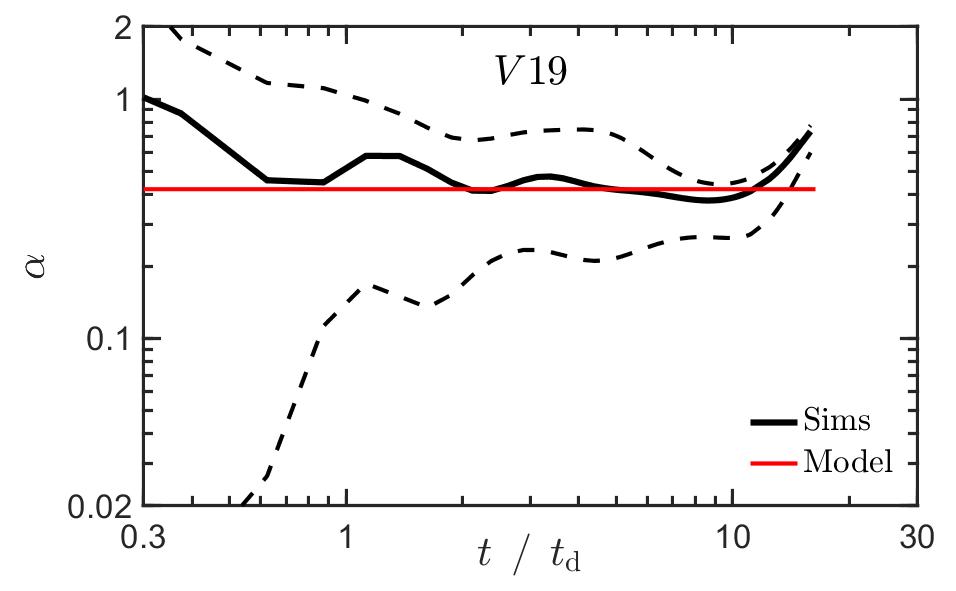}
\includegraphics[width=0.45\textwidth]
{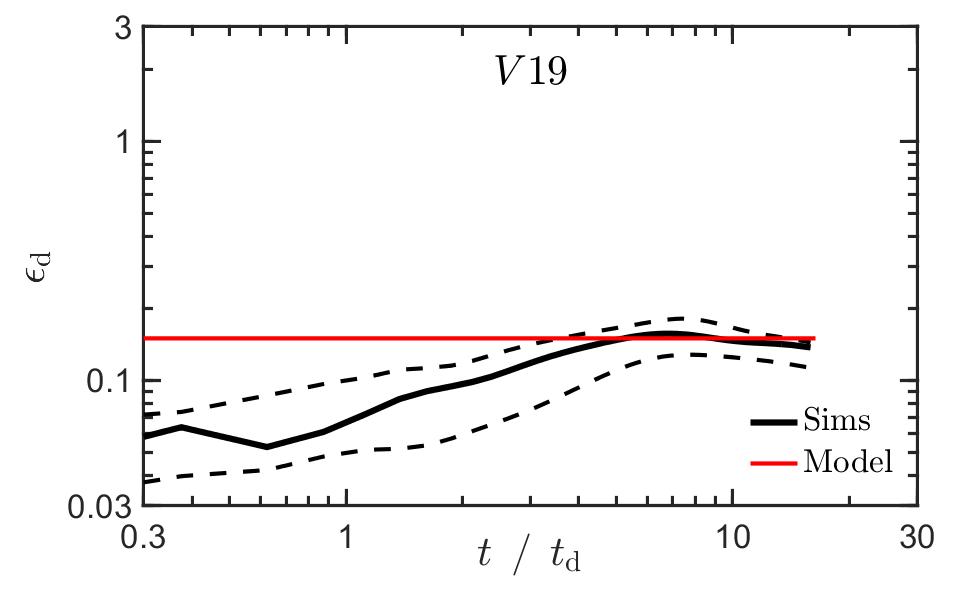}
\includegraphics[width=0.45\textwidth]
{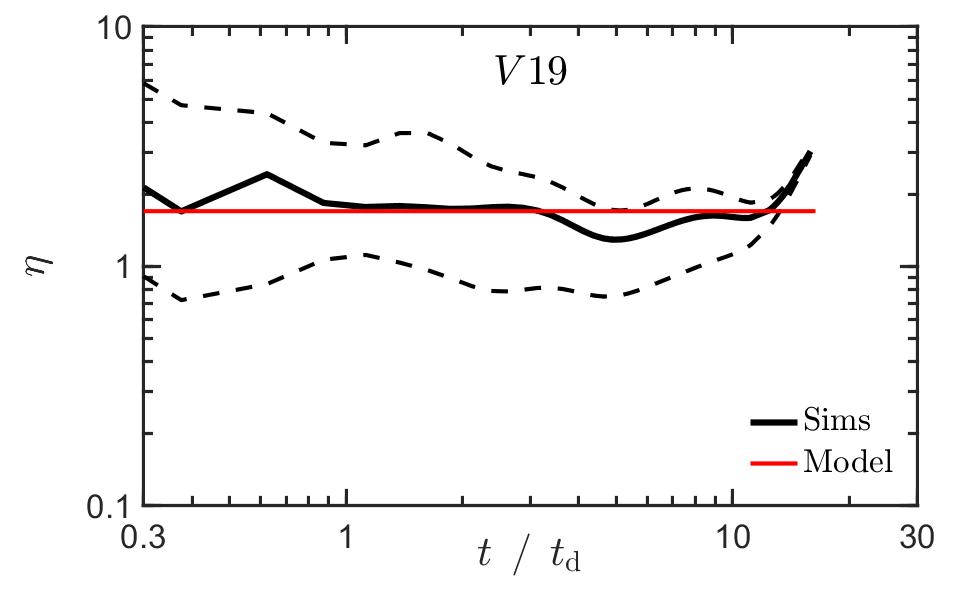}
\includegraphics[width=0.45\textwidth]
{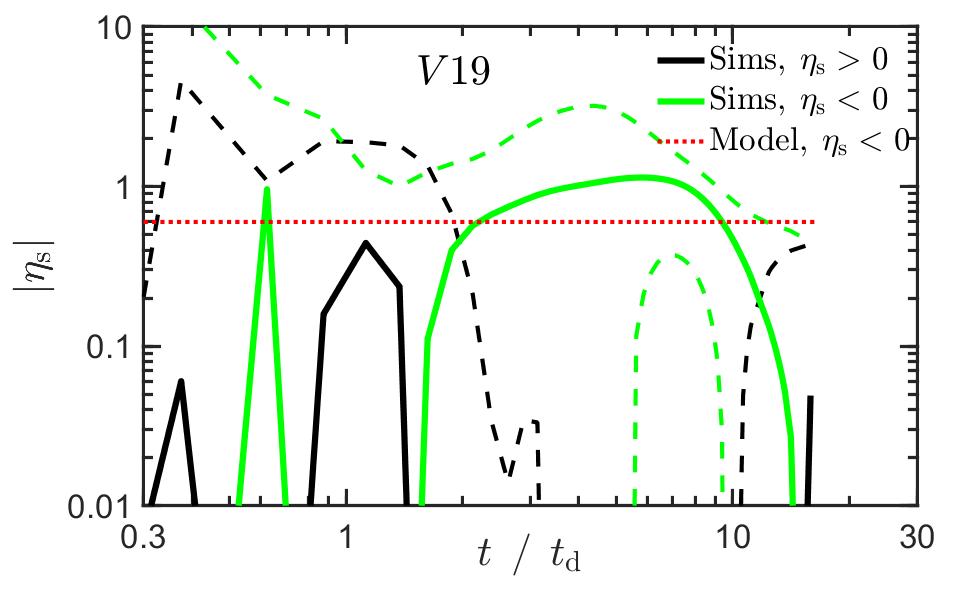}
\includegraphics[width=0.45\textwidth]
{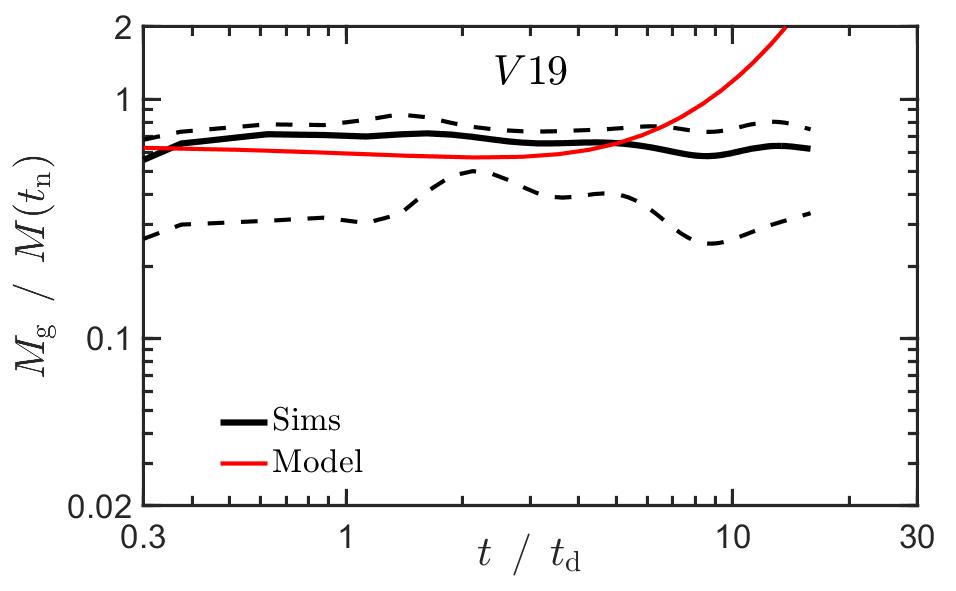}
\includegraphics[width=0.45\textwidth]
{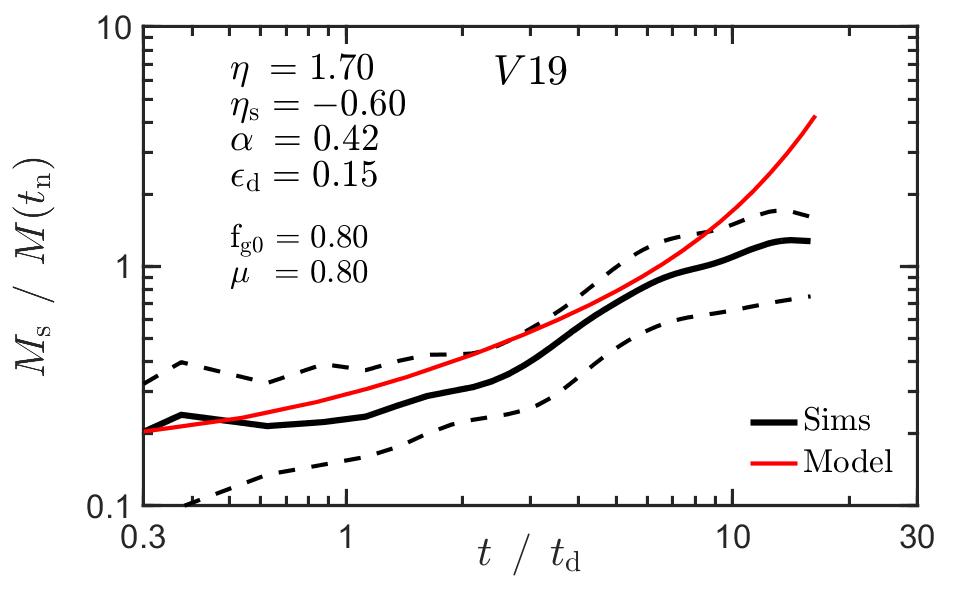}
\includegraphics[width=0.45\textwidth]
{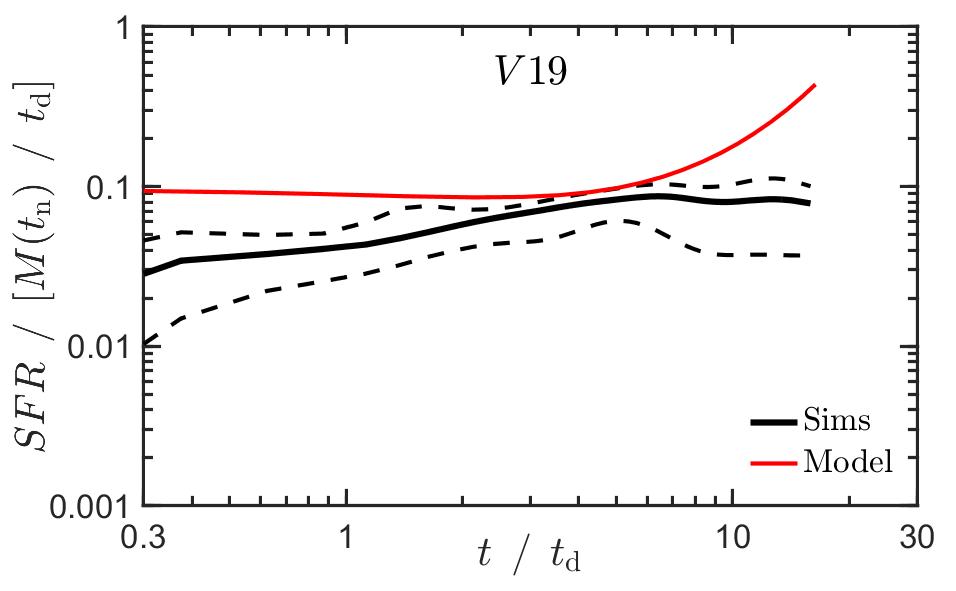}
\includegraphics[width=0.45\textwidth]
{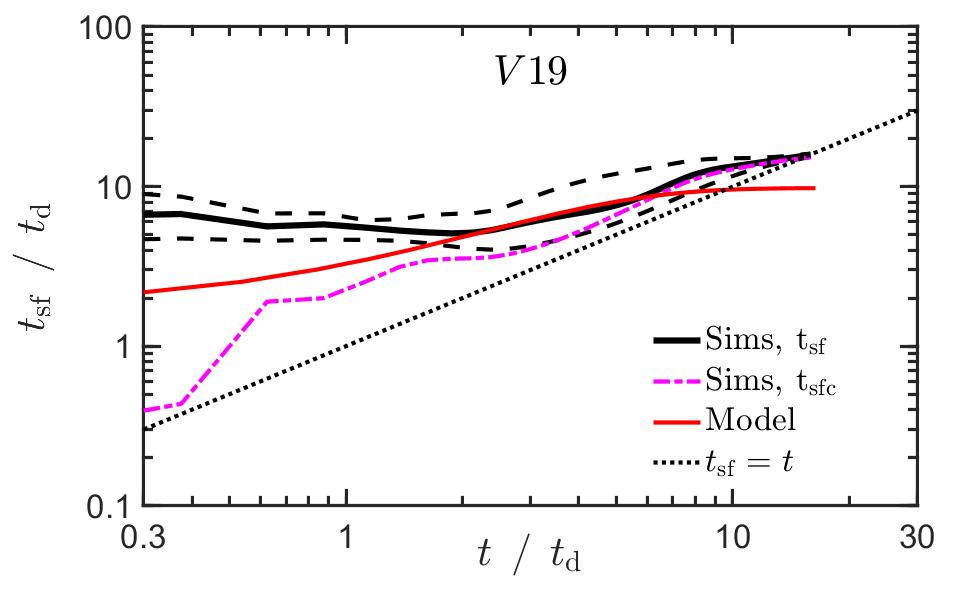}
\caption{
Evolution of simulated clump properties versus toy model,
here for the cosmological galaxy V19 at $z\!=\!5\!-\!3$.
The details are the same as in \fig{NoRP}, here with 12 clumps.
}
\label{fig:V19}
\end{figure*}

\Figs{NoRP} to \ref{fig:V19} display the evolution of clump properties in the
four types of simulations during the clump lifetimes, starting from their
formation time and evolving for tens of disc dynamical times.
The typical clumps form in the outer disc, migrate inwards 
(see, e.g., \fig{tc_d}), and merge into the bulge at
the galaxy center after $\tmig \!\sim\! (10\!-\!20) \td$, 
as predicted in \equ{tmig}.
A fraction of the clumps live longer, e.g., because they have low
masses compared to the Toomre mass, or because they form at very large radii,
or for other reasons.
Shown in the figure are the medians over the clumps of the given simulation
type, and the 68\% scatter about
them at time $t$ since clump formation, measured with respect to the disc
dynamical time, $\td$.
The values of $\td(z)$ for the different galaxies are shown in \fig{td}.
Shown in comparison are the results from the toy model with the parameter
values as derived separately from each of the four simulation types.
The four top panels refer to the model parameters $\alpha$, $\epsd$, $\etag$
and $\etas$.
The third row from top shows the clump gas and stellar masses.
Before stacking, the total mass of each clump, $\Mc$, is normalized to unity
at $\tform=3\td$, and $\Mg$ and $\Ms$ are normalized accordingly, thus
they are given with respect to $\Mc(\tform)$.
The actual median clump masses at $\tform$ are
$\Mc(\tform)\!=\! 10^{8.43},\, 10^{8.36},\, 10^{8.35},\, 10^{8.15}\msun$ for
NoRP, RP, V07 and V19 respectively.
The SFR in the bottom-left panel is normalized to $\Mc/\td$ at $\tform$.
The bottom-right panel shows $\tsf$, the inverse of sSFR.
Recall that $\tdep/\td$ is given by $\epsd^{-1}$.

\smallskip 
These figures allow an evaluation of the validity of the toy model by 
verifying the degree of constancy of the assumed model parameters and then 
estimating their values in comparison to their a priori estimates.
They then serve for exploring the validity of the robust model predictions in
the main stage of clump evolution.

\smallskip 
We note that in certain ways the clumps go through an initial adjustment
phase of a duration on the order of a disc dynamical time
after which they settle to a more stable behavior. 
This is expected in the isolated simulations, where the clumps form
out of an otherwise unperturbed disc, but it is evident also in the
cosmological simulations.
For example, the specific accretion rate tends to start high and settle into 
a rather constant lower value. Similarly, the SFR tends to climb to a peak 
value and then decline and settle into a rather constant value.
We crudely identify the time when the clump settles in as $\tform$,
which we choose to be $\tform\! =\! 3\td$.
This $\tform$ is used as the anchor of the normalization of the clump masses
before stacking the clumps and determining the median for the plot.
We focus hereafter on the more stable phase after $\tform$, and
evaluate the model parameters in this phase.

\smallskip 
We next inspect the evolution of the model parameters one by one in the four
simulation types.
The median efficiency of gas accretion, $\alpha$, as defined in \equ{alpha},
is roughly constant after the first 
disc dynamical time, except in the isolated simulations NoRP, where it is high 
till $\tform$ and lower but rather flat in the range 
$t/\td \!\simeq\! 4\!-\!15$. 
The fluctuations are on the order of $\pm 0.1$dex.
This makes the model assumption of a constant $\alpha$ crudely acceptable in
the main stage of evolution.
In the isolated simulations we estimate $\alpha\!\simeq\! 0.2$ 
and $0.25$ in NoRP and RP respectively. 
In the cosmological simulations we estimate at late times 
$\alpha \simeq 0.25$ and $0.4$ for V07 and V19 respectively.
These are in the ball park of the values assumed a priori.

\smallskip 
The median SFR efficiency per disc dynamical time, $\epsd$, as defined in 
\equ{sfr}, is rather constant, with fluctuations on the order of $\pm\!0.1$dex,
validating the model assumption of a constant $\epsd$.
This is partly by construction, as the simulations assume a local SFR law
which is roughly consistent with a constant local $\epsf$ in \equ{sfr}.
However, the fact that the global properties of $\Mg$ and $\SFR$ are related
to $\td$ via a roughly constant $\epsd$ does not follow in a trivial way.
In the isolated simulations we deduce $\epsd \simeq\! 0.25$ and $0.2$ for
NoRP and RP respectively.
In the cosmological simulations we estimate $\epsd\!\sim\! 0.5$ and $0.15$
for V07 and V19 respectively.
We note that with a higher density threshold for star formation
within the clumps, corresponding to lower values of $\tff$, one can obtain
higher values of $\epsd$ while the values of $\epsf$ are the same universally,
as $\epsd/\epsf\!=\!\td/\tff$.
The higher value of $\epsd$ for V07 compared to V19
indeed reflects a higher value of $\td/\tff\!\sim\! 10$ compared to $\sim\! 5$ 
in V19, stemming from the different density contrasts of the clumps with 
respect to the disc in the two simulated galaxies.
Similarly, the slightly higher value for NoRP versus RP reflects the tendency 
of the star-forming region to be of higher density when the feedback is weaker.

\smallskip 
The median gas outflow mass-loading factor, $\etag$, is roughly constant in the
cosmological simulations after $\sim\!\td$, but in the isolated simulations it
starts lower and flattens off only after $\sim\! \tform \!=\! 3\td$.
The fluctuations are on the order of $\pm\! 0.2$dex or smaller.
We note that the instantaneous mass-loading factor can fluctuate by 
a factor of a few over a disc dynamical time 
(e.g., Fig.~13 of M17), 
for example following a sudden boost of the SFR due to a clump merger, 
but these peaks become significantly lower when smoothed over 
$0.1$dex in $t/\td$, 
validating the assumption of a constant $\etag$ after sufficient smoothing. 
In the isolated simulations we estimate $\etag \!\simeq\! 0.1$ and $1$ 
for NoRP and RP respectively. The low $\etag$ is a unique feature of NoRP, set
by construction by the weak feedback assumed.
In the cosmological simulations we estimate $\etag\!\simeq\! 1$ and $1.7$ for 
V07 and V19 respectively.
These values of order unity are similar to the results from \citet{dk13}, 
where the outflows from simulated clumps are compared to observed clumps. 
Similar mass-loading values are indicated by observations
\citep{schroetter19,forster19}

\smallskip 
The median stellar exchange parameter, $\etas$, as defined in \equ{etas},
carries large uncertainties, but it is low in all cases except V19. 
In the isolated simulations we estimate
$\etas\! \sim\! 0.1$ and $0.25$ for NoRP and RP respectively, with 
fluctuations on the order of $\pm \!0.2$dex or smaller.
The weaker stellar stripping in NoRP is likely
caused by the clumps being more compact when the feedback is weaker.
In V07 we deduce $\etas \!\sim\! 0.1$ in the range $t/\td \!\sim\! 2\!-\!15$, 
and small negative values outside this range.
In V19 there are actually negative values of $\etas$ during the main stage of
evolution, namely a net gain of stellar mass. 
This is mostly accretion of young stars due to their high density
in the disc of V19 at the high redshifts analyzed.
A word of caution is that the value of $\etas$ may be partly contaminated 
by unbound disc stars in the clump identification, given that the density 
contrast of relatively old stars in the clumps is typically lower than
that of the gas and SFR. 
In summary, the stellar mass exchange in our simulations tends to be small,
and in most cases less important than the other ingredients of the bathtub
model.

\smallskip 
We conclude that the three important model parameters, $\alpha$, $\epsd$ and 
$\etag$, typically fluctuate by only $\pm 0.2$dex or less during the main stage
of evolution, and can therefore be approximated as constants 
during the main phase of the clump lifetime. 
This makes the model a valid approximation in this regime, 
with $\etas$ negligible except in V19. 
The values of these parameters as estimated from the simulations
are in the ball park of the a priori expectations.

\subsection{Simulated clump evolution versus model}

\smallskip 
As a result of the approximate constancy of the basic parameters,
the main phase of clump evolution, after one or a few disc dynamical times,
is characterized by a fairly constant SFR and $\Mg$ and an associated linear 
growth of $\Ms$.
The gas fraction is declining accordingly.  
The total clump mass is initially roughly constant, 
as long as the clump is gas dominated, 
and it is growing with an increasing pace as the stellar mass becomes 
important, approaching a linear growth. 

\smallskip 
Inspecting the evolution of masses in the different simulation types,
we see that in the isolated runs, as the clumps grow from a no-clump uniform 
disc, the clump gas mass is growing from zero until it converges to roughly the 
model-predicted level. In the cosmological runs, as expected, the gas mass 
roughly matches the model predictions starting at an earlier time.
The stellar mass matches the prediction pretty well at all times.
The constancy of the SFR after $\tform$ is apparent in all the simulation 
types, at the level of $\pm\! 0.1$dex.

\smallskip 
The star-formation time $\tsf$, the inverse of the sSFR, 
approaches a linear growth with
time $t$ since clump formation, reflecting the linear growth of $\Ms$ 
and the constancy of the SFR.
Being an overestimate of $t$ at early times,
$\tsf$ becomes a good proxy for $t$ during the main stage of clump evolution.
In most cases, the corrected $\tsfc$ is a better estimate and sometimes a close
lower limit for $t$.
This is true for all the simulation types.

\smallskip 
The depletion time $\tdep$ in units of $\td$, which according to \equ{tdep}
is given by the inverse of $\epsd$, is rather constant. 
During the main stage of evolution (and more so in the initial phase), 
$\tdep$ serves as an upper limit for $t$, while near $\sim\! 10\td$ it becomes 
comparable to $\tsf$ and a fair estimate of $t$.

\smallskip 
At late times, toward the expected asymptotic regime, some of the few 
clumps that survive for more than $20\td$ indeed seem to show marginal
indications for an accelerated growth of stellar mass
at a constant gas fraction and $\tsf$, as expected by the analytic solution.
This, however, is not apparent in the figures that show the medians
of stacked samples of clumps. %



\smallskip  
After evaluating the validity of the model through the constancy of its
parameters, and crudely estimating the model
parameters for each of the simulation types, we also show in
\figs{NoRP} to \ref{fig:V19}
the specific model predictions for the evolution of the medians of the 
simulated quantities $\Mg$, $\Ms$, SFR, $\tsf$ and $\tsfc$ relevant to
each simulation.
The specific values of the model parameters for each simulation type,
as marked in the figures, were estimated from each simulation
(\se{param}, the four upper panels of the figures).
They tend to be in the ball park of the a priori expectations,
though they are not necessarily identical in the best fits of the
different simulations.

\smallskip
The model seems to provide fair qualitative fits to the simulations of all
four types. This is especially true during the main phase of clump evolution,
at $(2\!-\!10)\td$, after the initial settling phase and before the end of
migration, where both the model and simulations show the phase of roughly
constant SFR and $\Mg$ and linear growth of $\Ms$, and demonstrate that
$\tsf$ is a good proxy for time since clump formation.
The NoRP simulation seems to show marginal hints for the late-time asymptotic
phase, which is supposed to occur after the end of migration, where the masses
grow exponentially and $\fg$ and $\tsf$ saturate to a constant value.


\subsection{Time indicators for short-lived clumps}
\label{sec:SL}

While the analysis so far focused on the clumps that survive for more than
$50\Myr$, 
we wish to verify to what extent $\tsf$, $\tsfc$ and $\tdep$ can serve as 
observable indicators of time since formation also for the short-lived clumps.
This will be useful in the attempt to distinguish between the scenarios of
SL and LL clumps.

\smallskip 
The SL clumps typically disrupt before or during the early stages
of the main phase of evolution, by $\sim\, 3 \td$.
However, their evolution during their short lifetimes may be
similar to that of the LL clumps.
In particular, similarly to the LL clumps in their initial phase,
we expect $\tsf$ for SL clumps to be smaller than $\tdep$, as 
$\tsf\!=\!\fsg\tdep$,
and the gas fraction in SL clumps is expected to be above 50\% 
\citep[as in the simulations of][]{oklopcic17}.
After a short initial phase, the $\tsf$ of SL clumps is expected to 
grow linearly with time, much like the LL clumps, but the SL 
clumps are disrupted before their $\tsf$ reaches the levels of $\tdep$.

\smallskip 
In order to crudely test the above predictions, we inspect simulated SL clumps.
For this we appeal to isolated simulations similar to the ones used here
\citep{fensch21}, but where most of the clumps are short-lived.
This was obtained by starting with discs of a low gas
fraction, $\fg\! \sim\! 0.2$, 
and adopting a strong feedback
including thermal supernova feedback, radiation pressure and photoionization.
These simulations typically show 3-6 clumps of baryonic mass 
$\sim\! 10^8\msun$ and larger for at least half the time. 
The key difference from the other simulations used here
is the resulting shorter clump lifetimes.

\smallskip
\Fig{RP}, bottom-right panel, also shows the median and 68\% scatter for 
$\tsf$ in these SL clumps as a function of time from clump formation.
The median for the SL clumps is indeed similar to the median for the LL clumps
from the gas-rich simulations with RP.
After the short initial phase of duration $\sim\! \td$, the star-formation time
$\tsf$ becomes a good proxy for time $t$ also for SL clumps at $(1\!-\!3)\td$.

\begin{figure} 
\centering
\includegraphics[width=0.48\textwidth]
{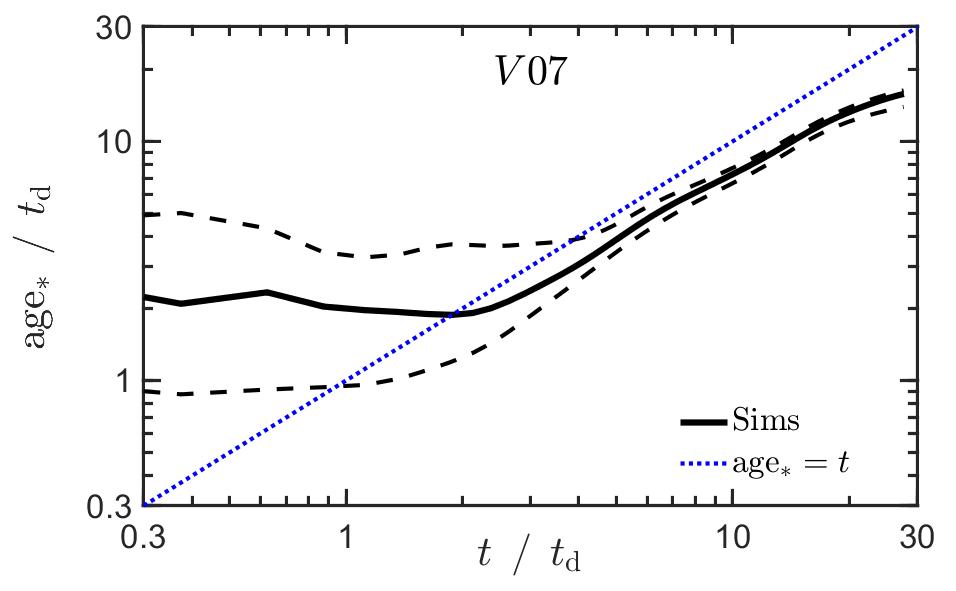} 
\caption{
Age as a proxy for time. Shown are the medians and 68\% scatter
of the average stellar age in clumps versus time since clump formation,
for the clumps in V07.
The age is a proxy for time after the first disc dynamical time.
}
\label{fig:tc_age}
\end{figure}

\smallskip
\begin{figure} 
\centering
\includegraphics[width=0.48\textwidth]
{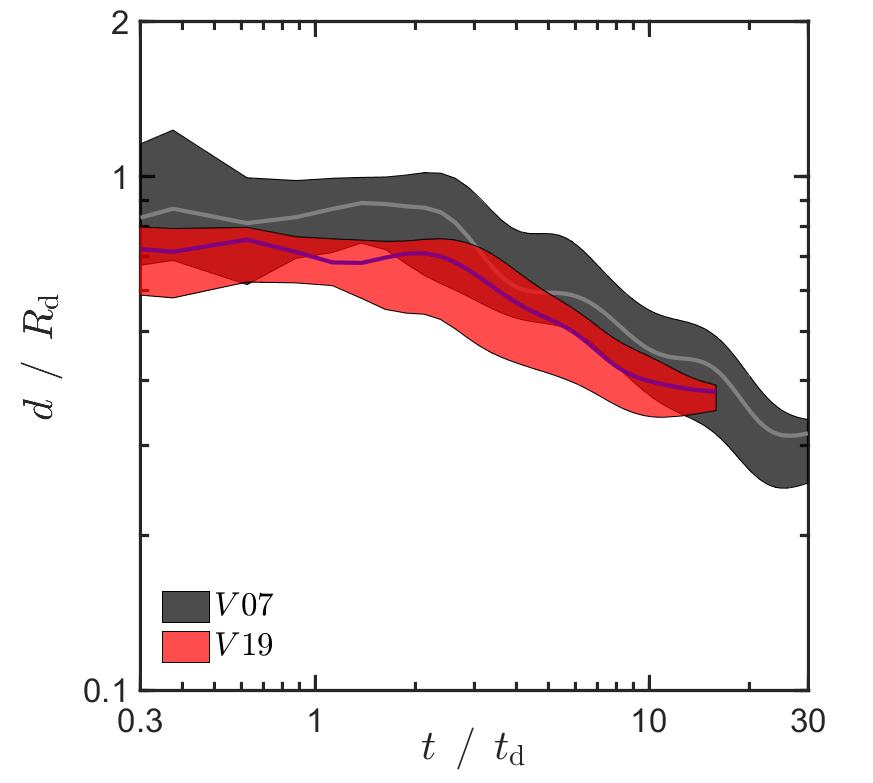}
\caption{
Clump migration.
Shown are the median and 68\% scatter of the clump distance from the galactic
center versus time since clump formation,
for the clumps in the cosmological simulations V07 and V19.
Migration inwards is noticeable after $t\!\sim\! 2\td$,
showing radial migration of $0.45\Rd$ ($0.6\Rd$) 
during $\sim\!10\td$ ($\sim\!20\td$), 
comparable to the migration rate predicted in \equ{tmig}.
}
\label{fig:tc_d}
\end{figure}

\section{Theory versus Observations}
\label{sec:obs}

\subsection{Clumps in CANDELS}

In \citet[][hereafter G18]{guo18}, we have analyzed 3187 clumps
identified in 1269 galaxies from the CANDELS/GOODS-S field 
in the redshift range $z\!=\!0.5\!-\!3$, based on the sample described in 
\citet{guo15}.  The sample, clump detection and estimates of clump properties 
are described in these papers, while we bring only a very brief summary here.

\smallskip 
For each galaxy, the stellar mass and SFR were determined by SED fitting,
while the SFR was also determined directly from the UV flux.
Star-forming galaxies were selected at $0.5 \!<\! z \!<\! 3$
to have global $\Ms \!>\! 10^9 \msun$ and sSFR$\!>\! 10^{-1}\Gyr^{-1}$. 
An apparent magnitude cut of HF160W $< 24.5$ AB has been applied to ensure  
reliable morphology and size measurements. 
In order to be able to identify and measure clumps given the resolution, 
the sample has been limited to galaxies whose effective radii along the galaxy 
semi-major axis (SMA) is larger than 0.2".
To minimize the effect of dust extinction and clump blending, only galaxies 
with axial ratio $q \!>\! 0.5$ were included.
This sample consists of 1655 galaxies. 

\smallskip 
The clumps were detected in rest-frame Near-UV. 
A blob of Near-UV excess is identified as a clump if its UV luminosity is at
least 3\% of the total UV luminosity of the host galaxy.
A total of 3187 clumps have been identified in 1269 galaxies.
In the following first comparison we use a subsample of these galaxies, 
with $\log M_{\rm s,gal}\!=\!9.8\!-\!11.4$ in the redshift-range 
$z\!=\!1.5\!-\!3$ where UV fluxes are available,
a total of 607 clumps in 284 galaxies.

\smallskip 
The flux in each of nine HST bands was measured for each clump using an
aperture of 0.18", and an aperture correction has been applied. 
Background disc subtraction has been applied in each band,
using different levels of correction, ranging from no subtraction to very
aggressive subtraction (G18, Table 2).
The clump properties, including $\Ms$, age and dust reddening,
were derived by fitting their HST SEDs to stellar population synthesis models 
\citep{bruzual_charlot03} with Chabrier IMF \citep{chabrier03},
as described in \citet{guo12}. The SFR used here, for 
galaxies at $z\!>\!1.5$, was estimated directly from the UV flux.
The SFR estimates from UV are independent of the mass, age and SFR estimates 
from SED fitting, so we prefer to use the UV SFR here.

\smallskip
We use below the clump properties as derived after a moderate correction for
disc contamination, the fiducial background subtraction denoted aperi\_v3 
in Table 2 of G18.
The radial gradients turn out to show the same qualitative features for 
the different
levels of correction, from a very defensive subtraction (aper3) to a very 
aggressive subtraction (aperi\_v1),
except for the case of no background subtraction at all (aperi\_v4).

\smallskip
As a note of caution, we learn from \citet{huertas20} that
using SED fitting on aperture as in G18 seems to overestimate the stellar mass 
of clumps by a factor of a few to several. This factor, however, is rather  
constant for all clump masses and for clumps in all galactic radii 
(and therefore likely for clumps of all ages). In our comparisons to the G18
observed clump masses  
(and SFR, which could be subject to a similar overestimate),
this overestimate may thus affect the overall normalization, but it has
little effect on the trends. To accommodate this uncertainty in mass 
normalization, we will normalize our model relations to match the masses at 
young ages.

\begin{figure*} 
\centering
\includegraphics[width=0.44\textwidth]
{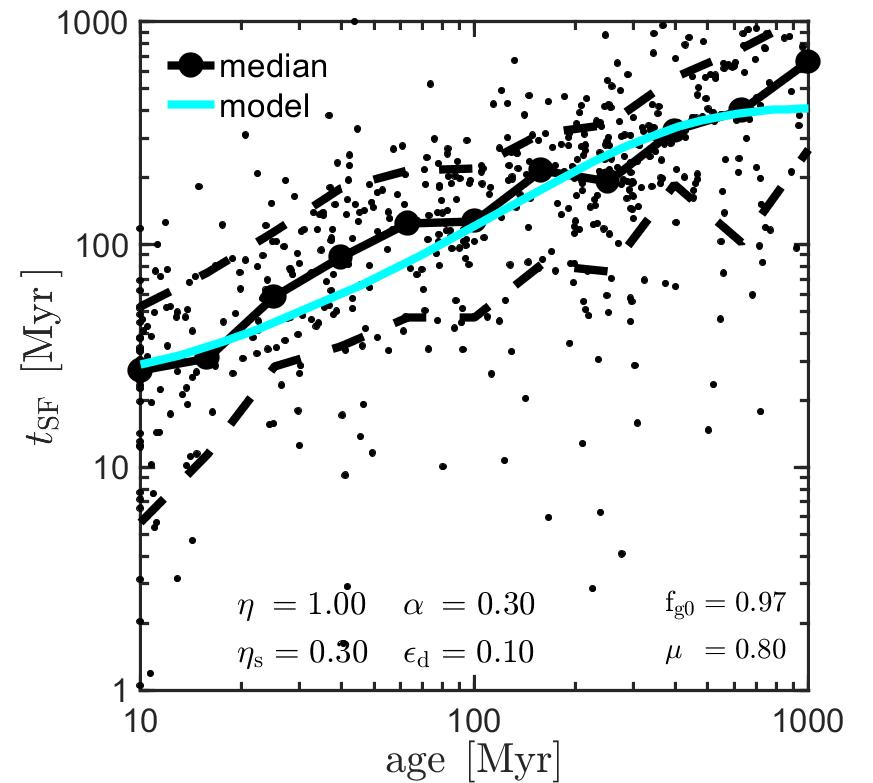}
\includegraphics[width=0.44\textwidth]
{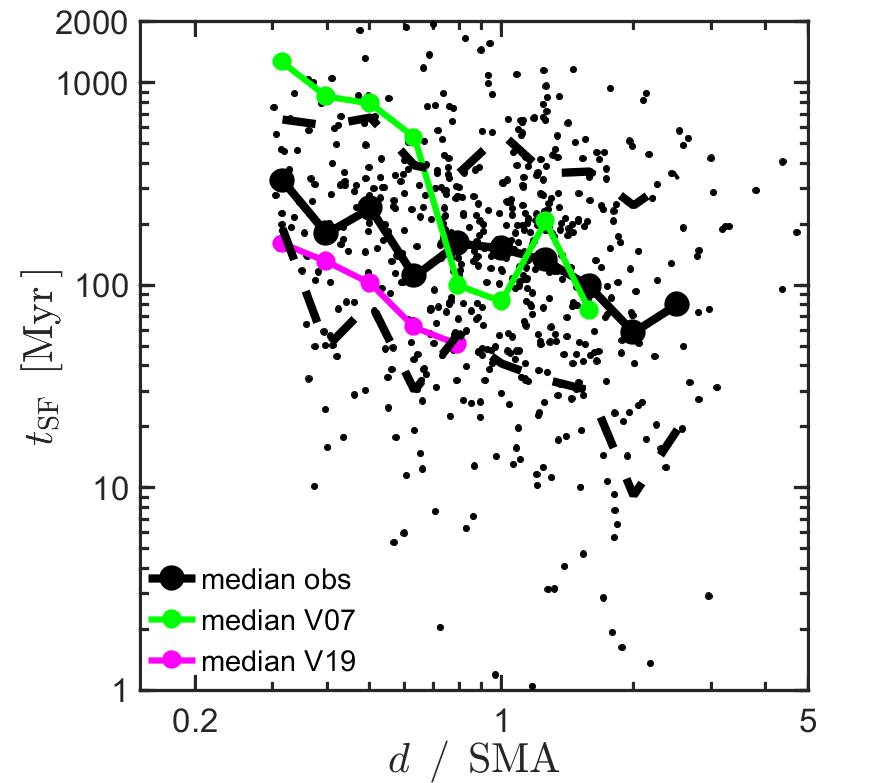}
\includegraphics[width=0.44\textwidth]
{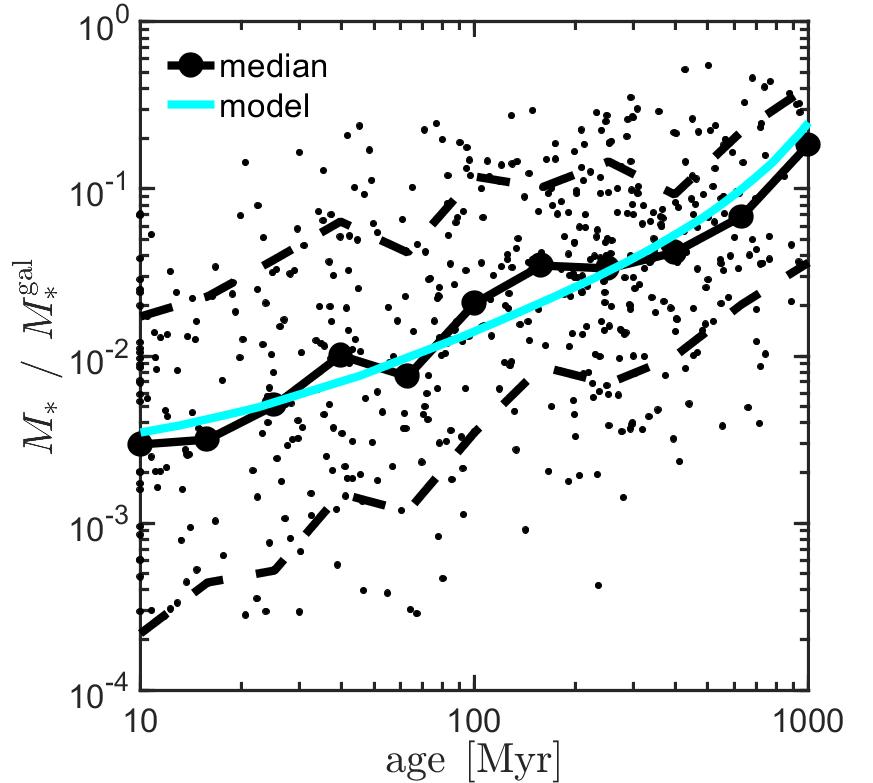}
\includegraphics[width=0.44\textwidth]
{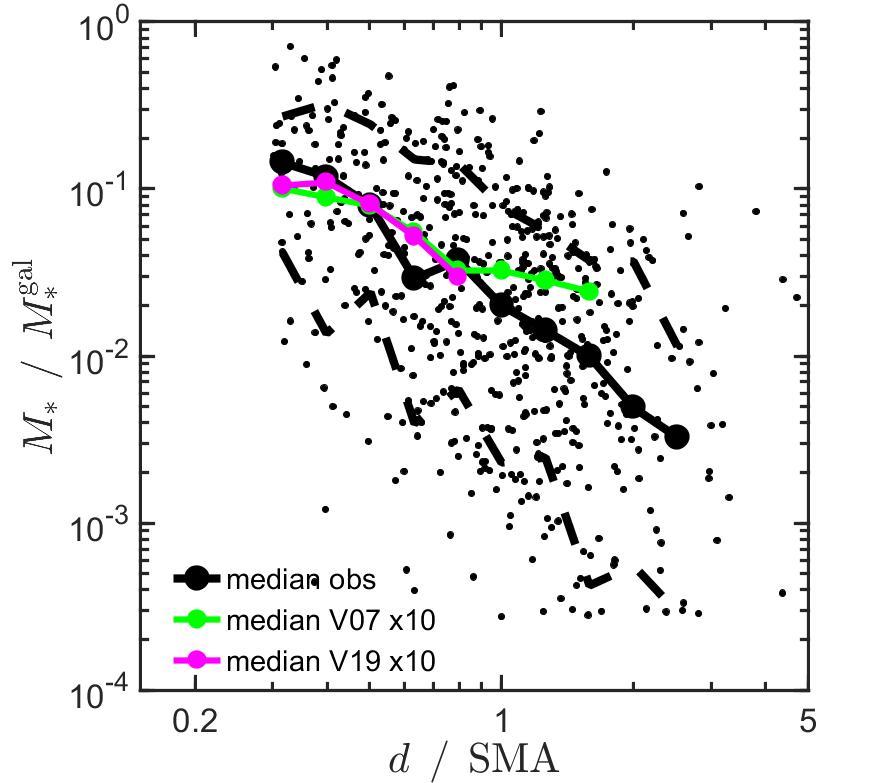}
\includegraphics[width=0.44\textwidth]
{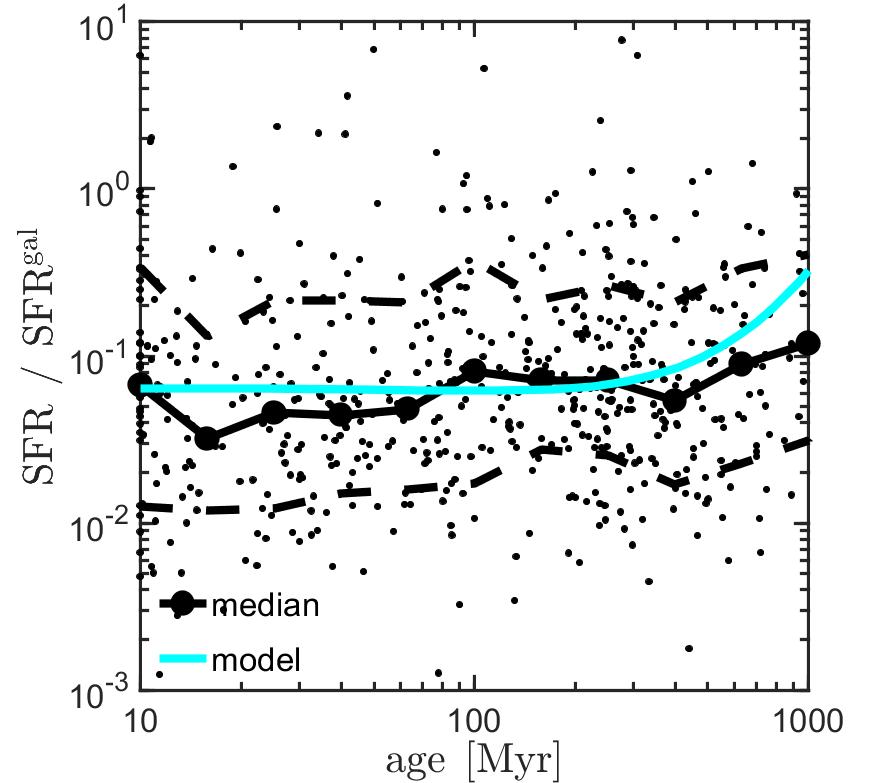}
\includegraphics[width=0.44\textwidth]
{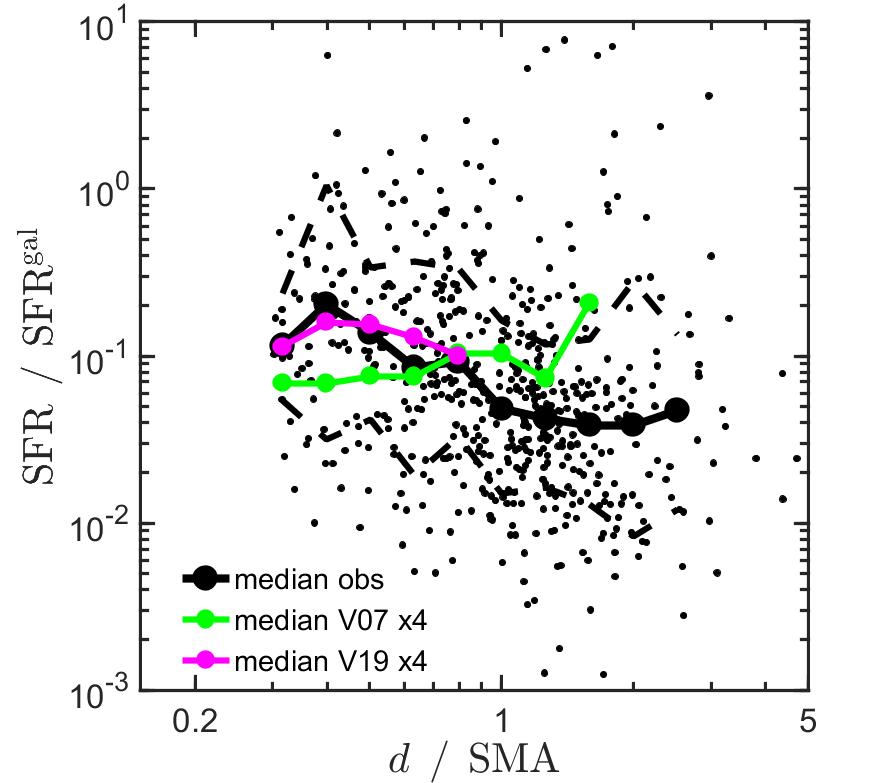}
\caption{
Observed clump properties against clump stellar age (left) and against
galacto-centric distance with respect to the semi-major axis (right).
{\bf Top:} star-formation time $\tsf\!=\!$sSFR$^{-1}$.
{\bf Middle-bottom:} clump stellar mass and SFR with respect to the whole 
galaxy.
The points represent 607 clumps in 284 CANDELS galaxies with
$\log {\Ms}_{\rm gal}\!=\!9.8\!-\!11.4$ at $z\!=\!1.5\!-\!3$,
from the sample analyzed in G18.
The medians and 68\% scatter are shown in log-spaced bins.
A model fit is shown as a function of time, with the parameter choice
$\alpha\!=\!0.3$, $\epsd\!=\!0.1$, $\etag\!=\!1$, $\etas\!=\!0.3$ and
$\fg0\!=\!0.97$,
and with a normalization that matches the properties at young ages.
The median radial gradients from the simulations (M17) 
are shown in comparison to the observed gradients as a function of $d/a$, 
normalized to match the observations when necessary.
}
\label{fig:obs}
\end{figure*}

\subsection{Simulated time vs age and distance}


The available observed quantity for time is the mean stellar age in a clump,
while the main quantity of interest for dynamical clump evolution is the time
since its formation. The age may be larger due to older stars that formed
outside the clump and became part of it at or after clump formation.
These could be a real bound part of the clump, or unbound stars that are
temporarily within the clump boundaries, or foreground/background disc stars 
that are associated with the clump by the 2D observational identification 
method.
In order to find out to what extent the stellar age can serve as a proxy
for time since clump formation, we inspect in \fig{tc_age} the median age as a
function of $t$, both in units of $\td$, for the clumps in the cosmological
simulation V07. 
We learn that after the first disc
dynamical time the age is a good proxy for time since formation,
indicating that the contamination by background stars is small.
In V19 the contamination is more severe.
This could be because at this high redshift
the accreted disc stars are $\lsim 100\Myr$ old 
(see the time range of negative $\etas$ in \fig{V19}) 
while the migration time is only $\sim\! 150\Myr$, making the
clump stellar ages dominated by the young background stars. 

\smallskip
Another observed quantity is the projected distance of the clump from the 
galaxy center relative to the semi-major axis of the galaxy, $d/a$. 
Since the clumps tend to form in the outer disc and migrate inward, 
$d/a$ may also serve as a proxy for time since formation, and gradients of
clump properties as a function of $d/a$ are expected to be related to clump 
evolution in time.
In order to obtain an impression of this, we show in \fig{tc_d} the median of
$d/\Rd$ as a function of $t/\td$ for the clumps in the cosmological
simulations.
We see that after $\sim 3\td$, the clump distance is declining with time, as
expected, reflecting the inward migration.
The effective trend in this regime seems to be roughly $d \!\prop\! t^{-1/2}$ 
for the two cosmological simulations. 
A fair fit is provided by $d/a \!\simeq\! e^{-t/\tmig}$.
This indicates that the translation from
$t$ to $d$ cannot be modeled in detail by assuming a constant radial velocity
during migration.
The migration time is $(10\sdash 20)\td$, in the ball park of the prediction in
\equ{tmig}.

\subsection{Observed clump properties vs age \& distance}

\Fig{obs} puts together the observed CANDELS clumps from G18, 
in the galaxy mass range 
$\log {\Ms}_{\rm gal}\!=\!9.8\!-\!11.4$ and in the redshift range $z=1.5-3$ 
(where UV-based SFR estimates are available),
a total of 607 clumps in 284 galaxies.
Shown are the clump stellar mass and UV-based SFR with respect to the 
whole galaxy, as well as the star-formation time within the clump
$\tsf\!=\!{\rm sSFR}^{-1}$. 
These properties are plotted against clump age (left panels)
and against distance from the galaxy center $d$ with
respect to the semi-major axis $a$ (right panels). 
The medians and 68\% scatter are shown in log-spaced bins.

\smallskip 
The left panels of \fig{obs} show the clump properties as a function of its age.
While the $\tsf$ values derived from the SED-based SFR  
are correlated with the clump ages by construction, 
as they both arise from the same SED fitting, the $\tsf$ derived from 
the UV-based SFR is not expected to be degenerate with the age. 
We notice in the top-left panel  
that $\tsf$ and clump age are still significantly correlated.
If the age is a proxy for time since clump formation, as in V07 (\fig{tc_age}),
we learn from the observed correlation that the star-formation time $\tsf$, 
much like the age, can indeed serve as a proxy for $t$.

\smallskip 
The median clump SFR (bottom-left panel) is rather constant as a function of 
clump age. 
This is consistent with the constant SFR as a function of time predicted by 
the toy model during the main clump evolution phase. 
The median stellar mass is growing roughly linearly with age (middle-left 
panel).
This is as expected from the toy model once the age of the clump stars 
reflects the time since clump formation.

\smallskip
Our bathtub toy-model results for $\tsf$, $\Ms$ and $SFR$
are shown in comparison assuming that stellar age and time since clump
formation are the same.
The parameter choice for a good fit is
$\alpha\!=\!0.3$, $\epsd\!=\!0.1$, $\etag\!=\!1$, $\etas\!=\!0.3$ and 
$\fgi\!=\!0.97$. 
The stellar mass and SFR are normalized to match the observational results at
young ages.
The model reveals remarkable fits to the rates of evolution of the
medians of the observed clump properties as a function of age.

\smallskip 
The right panels of \fig{obs} show the gradients of the same clump properties 
as a function of projected distance from the center $d/a$. 
The top-right panel shows $\tsf$ vs $d/a$.
The anti-correlation between the two is consistent with the expected 
migration inwards and assertion that $\tsf$ is a proxy for time since clump
formation.
As expected, the observed median SFR (bottom-right panel) hints to only a weak 
correlation with distance, consistent with the roughly constant SFR with age,
and the expected constant SFR with time during the main stage of evolution.
The median of the observed $\Ms$ (middle-right panel)
shows a strong anti-correlation with 
distance, reflecting the strong correlation of $\Ms$ with age and
the anti-correlation between age and $d/a$. This is consistent with the 
predicted growth of $\Ms$ in the main stage of clump evolution combined with
the inward migration.

\smallskip 
In principle, one could extend the toy-model predictions to the gradients
as a function of $d/a$ by attempting to model the migration process
and thus translate $t/\td$ to galacto-centric 
distance and to its observable counterpart $d/a$.
This requires accurate modeling of the migration process which we do not
attempt here.
Instead, we display in comparison to the observed radial gradients the medians 
from the radial gradients of the simulated clumps 
(following M17). 
We learn that the slopes of the simulated and observed radial 
gradients are similar, consistent with clump migration.
The normalizations of $\Ms$ and SFR in the simulations and observations 
are different, and were adjusted in the figures as marked in the labels. 
These offsets largely result from the limited observed resolution of more than
$1\kpc$, which causes an overestimate in the masses and SFR assigned to clumps
\citep{huertas20,ginzburg21}. 


\subsection{Other observations}

While we have learned that
the large sample of CANDELS clumps from G18 provides convincing
evidence in favor of long-lived clumps, smaller samples provide
supporting evidence.  
For example, 
several other studies have attempted to estimate clump ages using
stellar population synthesis modeling \citep{elmegreen05,Elmegreen09,
genzel11,forster11b,wuyts12,guo12,zanella15,soto17}. 
These estimates carry large systematic and random uncertainties, partly due
measurement errors and partly due to uncertainties in the SED modeling
including the star-formation histories, dust extinction and
background subtraction.
Given these uncertainties, the estimated ages span a wide range,
from $\sim\! 10\Myr$ to $\lsim\! 1\Gyr$, 
with most estimates indicating average ages of a few hundred $\Myr$,
consistent with the results of G18 
used above and with the predicted LL clumps.

\smallskip 
\citet{cibinel17} used ALMA to observe six clumps in a main-sequence clumpy
galaxy at $z\sim 1.5$ in the HUDF. From the stellar masses and SFRs they
estimate $\tsf \!\sim\! 70\!-\!900\Myr$. 
Using ALMA CO(5-4) observations, they also constrain the gas mass
and derive crude upper limits for the depletion times,   
$\tdep \!<\! (0.1\!-\!2)\Gyr$.
When stacking the six clumps together, they estimate
$\tdep\!\sim\!200\Myr$, comparable to and slightly smaller than the typical
value of $\tsf$ for these clumps, with $\fg \!\sim\! 20-50\%$.  
This argues for long-lived clumps, that had time to consume gas into stars.
Such gas fractions are consistent with the typical gas fractions in simulated
LL clumps as seen in \figs{NoRP} to \ref{fig:V19}
and lower than expected for SL clumps whose gas 
fractions are predicted to be $>\!50\%$ \citep{oklopcic17}.

\smallskip 
A few other studies have examined radial gradients of
clump properties with galacto-centric distance in small samples of clumps
and galaxies \citep{forster11b,guo12,soto17,zanella19}. 
They all find evidence for older, redder, more massive clumps closer to the
disc centre, in qualitative agreement with the finding of G18, 
and the predicted migration-induced radial gradients of LL clumps.

\section{Distinguishing features of LL vs SL}
\label{sec:LLvsSL}

With the theory predictions in hand 
one can attempt to identify the observable properties that may allow a 
distinction between survival versus disruption of massive in-situ clumps.
We refer to long-lived (LL) versus short-lived (SL) in-situ clumps, 
ex-situ clumps (Ex), and the background disc (D).
We base the predictions on the model of the current paper
and the simulation results in M14 and M17.
The distinguishing properties, listed by observables,  
are as follows:

\smallskip\no
\bul {\it Position in the disc:}
The migrating LL are expected at all radii while the SL are expected where
they form, preferentially in the outer disc.
Ex tend to populate the outer disc and they typically migrate less because 
they tend to oscillate aboou the disc.

\smallskip\no
\bul {\it Stellar ages:}
The ages of LL are $\sim\! 200\Myr$ ($0\!-\!500$), 
while for SL they are $\sim\! 50\Myr$ ($<\!200$).
Ex and D have typical stellar ages $\sim\! 10^3\Myr$, in most cases 
$>\!500\Myr$. 

\smallskip\no
\bul {\it Star-formation time:}
The SFR time $\tsf$, the inverse of the sSFR, is a proxy for age for in-situ
clumps (LL and SL), but not necessarily for Ex and D.
The LL were crudely expected in M17 to show a declining radial gradient, 
crudely $\tsf \!\prop\! d^{-1}$,
but no such radial gradient is expected for SL.
A similar radial gradient is expected for Ex and D, with the $\tsf$ of D 
higher by $\sim\! 0.6$dex.

\smallskip\no
\bul {\it Stellar mass:}
For LL typically $\Ms\!>\!10^8\msun$,
while for SL typically $\Ms\!<\!10^{8.5}\msun$, 
on average smaller than LL by $\sim\! 0.5$dex.
For Ex $\Ms\!>\!0.07\Md$, on average larger than LL by 0.5dex.
A declining radial gradient is expected based on M17, for LL crudely 
$\Ms \prop d^{-1}$ and for SL 
$\Ms \prop d^{-0.5}$, 
with the slope for Ex in between.

\smallskip\no
\bul {\it SFR:}
The SFR is expected to be roughly constant with age and distance for the
in-situ clumps (LL and SL).

\smallskip\no
\bul {\it Gas fraction:}
LL are expected to show a range of $\fg$ values, possibly above and below 
$\sim\! 0.5$, while SL would tend to have only high values, $\fg \!>\! 0.5$,
and Ex would tend to show very low values, $\fg \!<\!0.1$.
Based on M17 we expect a radial gradient for LL, crudely 
$\fg \!\prop\! t^{-1}$, namely decreasing with decreasing $d$, 
but roughly constant values for SL and Ex.

\smallskip
Based on the above, the distinguishing properties listed by clump type are 
as follows: 

\smallskip\no\bul
The LL clumps and the SL clumps are different from one another
in age, $\tsf$, $\Ms$, and $\fg$, 
but not in SFR.
In the \vela simulations, they are also distinct by the distribution of
distances from the galaxy center and by the value of $\Ms$.

\smallskip\no\bul
The LL and Ex clumps are distinct by age, $\Ms$, $\fg$, and more weakly by 
$\Ms$ and SFR.

\smallskip\no\bul
However, 
as discussed in appendix \se{contamination},
contamination of the clumps by disc may give false values and 
generate false radial gradients, especially for the intensive quantities such 
as age and $\fg$.

\smallskip\no\bul
The disc contamination of the SFR is expected to be relatively small,
given the high density contrast between clump and disc, so the SFR could be
corrected reliably.
Hopefully, the correction to $\Ms$ can also be manageable, estimated in our
simulations to be at the level of a few tens of percent.
If so, then $\tsf$ can be estimated, and $\tsfc$ can serve as a proxy for
clump age.

\smallskip\no\bul
Most of the ex-situ clumps can be removed based on age ($>\!500\Myr$), 
$\fg$ ($<\!0.1$), and maybe also mass ($>\!0.07\Md$).

\smallskip
Our practical conclusions for distinguishing between LL and SL clumps
are thus as follows:

\smallskip\no\bul
The radial gradient of the age proxy $\tsf$ is the most promising 
distinguishing 
feature among the quantities that can be corrected for disc contamination.

\smallskip\no\bul
The radial gradient of $\Ms$ and its amplitude are potential
distinguishing features that may be corrected for disc contamination,
under the assumption that any gradient that might have been established at 
formation is weaker.

\smallskip\no\bul
Age and $\fg$ are distinguishing features in principle, but they may be 
useful only if the disc contamination is small.

\smallskip  
The observed radial gradients of CANDELS clumps (G18), as summarized 
in \fig{obs}, indicate consistency with the scenario of long-lived, migrating 
clumps,
and inconsistency with the clump population being dominated by short-lived
clumps.
In particular,
the distribution of clump ages, either as estimated from SED fitting or 
from $\tsf$, indicates that most clumps are older than $100\Myr$. 
The clumps span a broad range of distances from the galaxy center,
and both age and $\tsf$ show significant gradients with $d/a$. 
Furthermore, the clump stellar mass shows a strong gradient with $d/a$.

\section{Conclusion}
\label{sec:conc}

Motivated by the desire to understand the nature of the so-prominent
giant clumps in high-$z$ star-forming disc galaxies, 
and in particular attempting to distinguish between 
long-lived and short-lived clumps, we studied the evolution of migrating 
clumps, using an analytic toy model and different types of simulations, 
and confronted theory with CANDELS observations.

\smallskip
Our theoretical results for the migrating long-lived clump evolution can be 
summarized as follows:

\smallskip\no\bul 
The unique feature of the bathtub toy model of clumps, distinguishing it from
the bathtub model for a whole galaxy, is that the gas accretion rate into the
clump scales with the total clump mass, which can vary on the relevant
timescale and is not directly constrained by the cosmological accretion rate. 
This allows an analytic solution specific for clumps,
a sum of exponential growing and decaying modes.

\smallskip\no\bul
The analytic model reveals a main phase of clump evolution, 
in which the SFR and gas mass are constant, and therefore the stellar mass is 
growing linearly in time, such that the inverse of the sSFR is a very useful 
proxy for clump age.
This phase is valid after an initial period of $\sim\!\td$, the disc dynamical 
time, and until after $\sim\!10\,\td$, namely until the clump has likely
completed its inward migration.
This constant SFR is very different from the commonly assumed ``tau-model" 
exponential decay behavior in a closed box, and from the cosmological evolution
of a whole galaxy where the sSFR is systematically decreasing in time in
proportion to $(1+z)^{5/2}$ following the cosmological specific accretion rate.
The tau-model, which is commonly assumed in clump mass estimates, 
is likely to lead to wrong results.

\smallskip\no\bul
Later on, the solution is intended to approach an asymptotic exponential
growth of mass and SFR with constant specific SFR and gas fraction, 
but this phase is unlikely to materialize as the typical migrating clumps 
are expected to disappear at the galaxy center by 
$\sim\!10\!-\!20\td$.

\smallskip\no\bul
The main model parameters are the accretion efficiency per dynamical time
$\alpha \!\sim\! 0.2\!-\!0.4$,
the SFR efficiency in a disc dynamical time $\epsd \!\sim\! 0.1\!-\!0.3$
(corresponding to $\epsf \!\sim\! 0.02$),
and the gas outflow mass-loading factor $\etag \!\sim\! 1\!-\!3$.
The stellar exchange between the clump and the disc has a less-important
role.

\smallskip\no\bul
The evolution of clumps in simulations, both of gas-rich isolated galaxies and
of zoomed-in high-$z$ galaxies in a cosmological setting, with different
feedback strengths (though all at a moderate level), 
is found to be in qualitative agreement with the analytic predictions.
The simulations confirm the validity of the toy model as a sensible
approximation, they yield model parameter values in the expected range, 
and they match the model-predicted evolution patterns. 
The model simplifying assumptions such as the constant disc properties
at different radii and constant model parameters as the clump migrates and
evolves, turn out to be sensible in view of
the qualitative match of the model to the simulations.
The toy model thus provides a very useful tool for further studies of giant
clumps in high-$z$ galaxies.

\smallskip
In view of the above theoretical expectations,
the observational results for massive clumps in a sample of more than 600 
clumps at $z\!=\!1.5\!-\!3$ from CANDLES can be summarized as follows:

\smallskip\no\bul
The observed clumps show a broad distribution of clump ages,
extending from a few to $\sim\!500\Myr$ and beyond.
This is true for ages as estimated from SED fitting or UV
as well as from the star-formation time $\tsf$, the inverse of sSFR, 
which is confirmed to be a useful age proxy, as predicted.
They also show a wide spread of distances from the galaxy center, which are
anti-correlated with the ages, as expected from inward migration.

\smallskip\no\bul
The observed clumps show evolution patterns and radial gradients of clump 
stellar mass and SFR that resemble the theoretical predictions for migrating 
clumps. In particular, the constant SFR over time and radial distance $d$, 
and the corresponding linear rise of stellar mass with age and inversely 
with $d$, resemble the predicted main phase of clump evolution. 

\smallskip\no\bul 
The observations thus favor the scenario where the massive clumps, roughly 
above one per cent of the galaxy stellar mass, largely survive 
feedback, remain long-lived and bound, and migrate inward on a timescale of
a few disc orbital times. The feedback strength as incorporated in the range of
simulations used here, with a mass-loading factor of order unity, is successful
in reproducing the observations, ruling out much stronger, disruptive feedback.

\smallskip\no\bul 
It is likely, though, that lower-mass short-lived clumps co-exist, 
as well as ex-situ clumps that have merged into the disc. We pointed out
features that could help distinguishing them from the in-situ clumps.

\smallskip 
The radial gradient of the age proxy $\tsf$, the inverse of sSFR,
is the most promising distinguishing feature between LL and SL clumps,
among the quantities that can be corrected for disc contamination.
The radial gradient of the clump stellar mass $\Ms$ and its amplitude 
are potential
distinguishing features that may be corrected for disc contamination.
The other estimates of age and gas fraction $\fg$ are distinguishing features 
in principle, but they may be useful only in cases where the disc 
contamination is small.

\smallskip 
A potential caveat is contamination of the signal of migration by gradients in
clump properties that were implanted at clump formation by radial variations
in disc properties. Our simulations indicate that the initial radial gradients
of clump properties are weaker than the gradients induced by clump migration
and the strong gradients observed, but this requires further investigation.

\smallskip 
Preliminary kinematical analysis of observed $z\!=\!1\!-\!2$ star-forming discs 
(Genzel et al., private communication)
indicates detections of radial velocities inward of gas emitting
H$\alpha$ and CO, as residual velocities after subtracting a rotating disc and
a dispersive bulge. The indicated inward velocities seem to be in the ball park
of the predicted migration rate, yet to be properly observed and analyzed.
If these residual velocities are largely associated with clumps, they may
provide direct evidence for clump migration, thus supporting their long-lived 
nature.  

\smallskip 
The need for survival of massive clumps does not permit a strong 
ejective and disruptive feedback in these clumps. 
On the other hand, the indicated low overall
stellar-to-halo mass ratio, based on abundance matching of observed
galaxies with simulated $\Lambda$CDM haloes 
\citep[e.g.,][]{rodriguez17,moster18,behroozi19}, 
requires strong preventive feedback that would suppress the total stellar mass.
The subgrid models for supernova and stellar feedback, as incorporated in 
current simulations, fail to simultaneously fulfill these two conflicting 
requirements.  Thus, understanding the complex nature of feedback presents 
a non-trivial challenge.

\section*{Acknowledgments}

We thank Andi Burkert, Natascha Forster-Schreiber,  
Reinhard Genzel, Mark Krumholz and Linda Tacconi for stimulating discussions.
AD has been supported by the grants ISF 861/20 and DIP 030-9111.
NM has been supported by the Moore Foundation through grant GBMF739, 
by NSF PHY-1748958.
FB has been supported by an ERC grant StG-257720 and by the CosmoComp ITN.
The isolated simulations used in this work were
performed on Curie at CEA/TGCC on GENCI resources (allocations 2016-04-2192
and 2017-04-2192).
We thank Florent Renaud for optimizing the radiative feedback scheme in the
RAMSES code for the isolated simulations.
DC is a Ramon-Cajal fellow and is supported by the grant PGC2018-094975-C21.
The cosmological simulations were performed at the National
Energy Research Scientific Computing centre (NERSC), Lawrence Berkeley National
Laboratory, and at NASA Advanced Supercomputing (NAS) at NASA Ames Research
Centre.
The analysis was performed on the Astric cluster at the Hebrew University.

\section*{DATA AVAILABILITY}

Data and results underlying this article will be shared on reasonable request 
to the corresponding author.

\bibliographystyle{mn2e}
\bibliography{mass}

\begin{thebibliography}{99}
\expandafter\ifx\csname natexlab\endcsname\relax\def\natexlab#1{#1}\fi

\bibitem[{{Agertz}, {Teyssier} \& {Moore}(2009){Agertz}, {Teyssier}, \&
  {Moore}}]{agertz09}
{Agertz} O., {Teyssier} R., {Moore} B., 2009, \mnras, 397, L64

\bibitem[{{Behrendt}, {Burkert} \& {Schartmann}(2015){Behrendt}, {Burkert}, \&
  {Schartmann}}]{behrendt15}
{Behrendt} M., {Burkert} A., {Schartmann} M., 2015, \mnras, 448, 1007

\bibitem[{{Behroozi} {et~al}\mbox{.}(2019){Behroozi}, {Wechsler}, {Hearin}, \&
  {Conroy}}]{behroozi19}
{Behroozi} P., {Wechsler} R.~H., {Hearin} A.~P., {Conroy} C., 2019, \mnras,
  488, 3143

\bibitem[{{Bouch{\'e}} {et~al}\mbox{.}(2010){Bouch{\'e}}, {Dekel}, {Genzel},
  {Genel}, {Cresci}, {F{\"o}rster Schreiber}, {Shapiro}, {Davies}, \&
  {Tacconi}}]{bouche10}
{Bouch{\'e}} N. {et~al.}, 2010, \apj, 718, 1001

\bibitem[{{Bournaud}, {Elmegreen} \& {Elmegreen}(2007){Bournaud}, {Elmegreen},
  \& {Elmegreen}}]{bournaud07c}
{Bournaud} F., {Elmegreen} B.~G., {Elmegreen} D.~M., 2007, \apj, 670, 237

\bibitem[{{Bournaud} {et~al}\mbox{.}(2014){Bournaud}, {Perret}, {Renaud},
  {Dekel}, {Elmegreen}, {Elmegreen}, {Teyssier}, {Amram}, {Daddi}, {Duc},
  {Elbaz}, {Epinat}, {Gabor}, {Juneau}, {Kraljic}, \& {Le Floch'}}]{bournaud14}
{Bournaud} F. {et~al.}, 2014, \apj, 780, 57

\bibitem[{{Bruzual} \& {Charlot}(2003)}]{bruzual_charlot03}
{Bruzual} G., {Charlot} S., 2003, \mnras, 344, 1000

\bibitem[{{Cacciato}, {Dekel} \& {Genel}(2012){Cacciato}, {Dekel}, \&
  {Genel}}]{cacciato12}
{Cacciato} M., {Dekel} A., {Genel} S., 2012, \mnras, 421, 818

\bibitem[{{Cava} {et~al}\mbox{.}(2018){Cava}, {Schaerer}, {Richard},
  {P{\'e}rez-Gonz{\'a}lez}, {Dessauges-Zavadsky}, {Mayer}, \&
  {Tamburello}}]{cava18}
{Cava} A., {Schaerer} D., {Richard} J., {P{\'e}rez-Gonz{\'a}lez} P.~G.,
  {Dessauges-Zavadsky} M., {Mayer} L., {Tamburello} V., 2018, Nature Astronomy,
  2, 76

\bibitem[{{Ceverino} {et~al}\mbox{.}(2016{\natexlab{a}}){Ceverino}, {Arribas},
  {Colina}, {Rodr{\'{\i}}guez Del Pino}, {Dekel}, \&
  {Primack}}]{ceverino16_outflow}
{Ceverino} D., {Arribas} S., {Colina} L., {Rodr{\'{\i}}guez Del Pino} B.,
  {Dekel} A., {Primack} J., 2016{\natexlab{a}}, \mnras, 460, 2731

\bibitem[{{Ceverino}, {Dekel} \& {Bournaud}(2010){Ceverino}, {Dekel}, \&
  {Bournaud}}]{cdb10}
{Ceverino} D., {Dekel} A., {Bournaud} F., 2010, \mnras, 404, 2151

\bibitem[{{Ceverino} {et~al}\mbox{.}(2012){Ceverino}, {Dekel}, {Mandelker},
  {Bournaud}, {Burkert}, {Genzel}, \& {Primack}}]{ceverino12}
{Ceverino} D., {Dekel} A., {Mandelker} N., {Bournaud} F., {Burkert} A.,
  {Genzel} R., {Primack} J., 2012, \mnras,

\bibitem[{{Ceverino} \& {Klypin}(2009)}]{ceverino09}
{Ceverino} D., {Klypin} A., 2009, \apj, 695, 292

\bibitem[{{Ceverino} {et~al}\mbox{.}(2014){Ceverino}, {Klypin}, {Klimek},
  {Trujillo-Gomez}, {Churchill}, {Primack}, \& {Dekel}}]{ceverino14}
{Ceverino} D., {Klypin} A., {Klimek} E.~S., {Trujillo-Gomez} S., {Churchill}
  C.~W., {Primack} J., {Dekel} A., 2014, \mnras, 442, 1545

\bibitem[{{Ceverino}, {Primack} \& {Dekel}(2015){Ceverino}, {Primack}, \&
  {Dekel}}]{ceverino15_shape}
{Ceverino} D., {Primack} J., {Dekel} A., 2015, \mnras, 453, 408

\bibitem[{{Ceverino} {et~al}\mbox{.}(2016{\natexlab{b}}){Ceverino},
  {S{\'a}nchez Almeida}, {Mu{\~n}oz Tu{\~n}{\'o}n}, {Dekel}, {Elmegreen},
  {Elmegreen}, \& {Primack}}]{ceverino16_drops}
{Ceverino} D., {S{\'a}nchez Almeida} J., {Mu{\~n}oz Tu{\~n}{\'o}n} C., {Dekel}
  A., {Elmegreen} B.~G., {Elmegreen} D.~M., {Primack} J., 2016{\natexlab{b}},
  \mnras, 457, 2605

\bibitem[{{Chabrier}(2003)}]{chabrier03}
{Chabrier} G., 2003, \pasp, 115, 763

\bibitem[{{Cibinel} {et~al}\mbox{.}(2017){Cibinel}, {Daddi}, {Bournaud},
  {Sargent}, {le Floc'h}, {Magdis}, {Pannella}, {Rujopakarn}, {Juneau},
  {Zanella}, {Duc}, {Oesch}, {Elbaz}, {Jagannathan}, {Nyland}, \&
  {Wang}}]{cibinel17}
{Cibinel} A. {et~al.}, 2017, \mnras, 469, 4683

\bibitem[{{Daddi} {et~al}\mbox{.}(2010){Daddi}, {Bournaud}, {Walter},
  {Dannerbauer}, {Carilli}, {Dickinson}, {Elbaz}, {Morrison}, \& {et
  al.}}]{daddi10}
{Daddi} E. {et~al.}, 2010, \apj, 713, 686

\bibitem[{{Dav{\'e}}, {Finlator} \& {Oppenheimer}(2012){Dav{\'e}}, {Finlator},
  \& {Oppenheimer}}]{dave12}
{Dav{\'e}} R., {Finlator} K., {Oppenheimer} B.~D., 2012, \mnras, 421, 98

\bibitem[{{Dekel} \& {Burkert}(2014)}]{db14}
{Dekel} A., {Burkert} A., 2014, \mnras, 438, 1870

\bibitem[{{Dekel} {et~al}\mbox{.}(2020{\natexlab{a}}){Dekel}, {Ginzburg},
  {Jiang}, {Freundlich}, {Lapiner}, {Ceverino}, \& {Primack}}]{dekel20_flip}
{Dekel} A., {Ginzburg} O., {Jiang} F., {Freundlich} J., {Lapiner} S.,
  {Ceverino} D., {Primack} J., 2020{\natexlab{a}}, \mnras, 493, 4126

\bibitem[{{Dekel} \& {Krumholz}(2013)}]{dk13}
{Dekel} A., {Krumholz} M.~R., 2013, \mnras, 432, 455

\bibitem[{{Dekel} {et~al}\mbox{.}(2020{\natexlab{b}}){Dekel}, {Lapiner},
  {Ginzburg}, {Freundlich}, {Jiang}, {Finish}, {Kretschmer}, {Lin}, {Ceverino},
  {Primack}, {Giavalisco}, \& {Ji}}]{dekel20_ring}
{Dekel} A. {et~al.}, 2020{\natexlab{b}}, \mnras, 496, 5372

\bibitem[{{Dekel} \& {Mandelker}(2014)}]{dm14}
{Dekel} A., {Mandelker} N., 2014, \mnras, 444, 2071

\bibitem[{{Dekel}, {Sari} \& {Ceverino}(2009){Dekel}, {Sari}, \&
  {Ceverino}}]{dsc09}
{Dekel} A., {Sari} R., {Ceverino} D., 2009, \apj, 703, 785

\bibitem[{{Dekel} \& {Silk}(1986)}]{ds86}
{Dekel} A., {Silk} J., 1986, \apj, 303, 39

\bibitem[{{Dekel} {et~al}\mbox{.}(2013){Dekel}, {Zolotov}, {Tweed}, {Cacciato},
  {Ceverino}, \& {Primack}}]{dekel13}
{Dekel} A., {Zolotov} A., {Tweed} D., {Cacciato} M., {Ceverino} D., {Primack}
  J.~R., 2013, \mnras, 435, 999

\bibitem[{{Dessauges-Zavadsky} {et~al}\mbox{.}(2019){Dessauges-Zavadsky},
  {Richard}, {Combes}, {Schaerer}, {Rujopakarn}, {Mayer}, {Cava}, {Boone},
  {Egami}, {Kneib}, {P{\'e}rez-Gonz{\'a}lez}, {Pfenniger}, {Rawle}, {Teyssier},
  \& {van der Werf}}]{dessauges19}
{Dessauges-Zavadsky} M. {et~al.}, 2019, Nature Astronomy, 3, 1115

\bibitem[{{Elmegreen} \& {Elmegreen}(2005)}]{elmegreen05}
{Elmegreen} B.~G., {Elmegreen} D.~M., 2005, \apj, 627, 632

\bibitem[{{Elmegreen} {et~al}\mbox{.}(2009){Elmegreen}, {Elmegreen}, {Ximena
  Fernandez}, \& {Lemonias}}]{Elmegreen09}
{Elmegreen} B.~G., {Elmegreen} D.~M., {Ximena Fernandez} M., {Lemonias} J.~J.,
  2009, \apj, 692, 12

\bibitem[{{Faure} {et~al}\mbox{.}(2021){Faure}, {Bournaud}, {Fensch}, {Daddi},
  {Behrendt}, {Burkert}, \& {Richard}}]{faure21}
{Faure} B., {Bournaud} F., {Fensch} J., {Daddi} E., {Behrendt} M., {Burkert}
  A., {Richard} J., 2021, \mnras, 502, 4641

\bibitem[{{Fensch} \& {Bournaud}(2021)}]{fensch21}
{Fensch} J., {Bournaud} F., 2021, \mnras, 505, 3579

\bibitem[{{Fisher} {et~al}\mbox{.}(2019){Fisher}, {Bolatto}, {White},
  {Glazebrook}, {Abraham}, \& {Obreschkow}}]{fisher19}
{Fisher} D.~B., {Bolatto} A.~D., {White} H., {Glazebrook} K., {Abraham} R.~G.,
  {Obreschkow} D., 2019, \apj, 870, 46

\bibitem[{{Fisher} {et~al}\mbox{.}(2017){Fisher}, {Glazebrook}, {Abraham},
  {Damjanov}, {White}, {Obreschkow}, {Basset}, {Bekiaris}, {Wisnioski},
  {Green}, \& {Bolatto}}]{fisher17}
{Fisher} D.~B. {et~al.}, 2017, \apjl, 839, L5

\bibitem[{{Forbes}, {Krumholz} \& {Burkert}(2012){Forbes}, {Krumholz}, \&
  {Burkert}}]{forbes12}
{Forbes} J., {Krumholz} M., {Burkert} A., 2012, \apj, 754, 48

\bibitem[{{Forbes} {et~al}\mbox{.}(2014){Forbes}, {Krumholz}, {Burkert}, \&
  {Dekel}}]{forbes14a}
{Forbes} J.~C., {Krumholz} M.~R., {Burkert} A., {Dekel} A., 2014, \mnras, 438,
  1552

\bibitem[{{F{\"o}rster Schreiber} {et~al}\mbox{.}(2006){F{\"o}rster Schreiber},
  {Genzel}, {Lehnert}, {Bouch{\'e}}, {Verma}, {Erb}, {Shapley}, {Steidel},
  {Davies}, {Lutz}, {Nesvadba}, {Tacconi}, {Eisenhauer}, {Abuter}, {Gilbert},
  {Gillessen}, \& {Sternberg}}]{forster06}
{F{\"o}rster Schreiber} N.~M. {et~al.}, 2006, \apj, 645, 1062

\bibitem[{{F{\"o}rster Schreiber} {et~al}\mbox{.}(2018){F{\"o}rster Schreiber},
  {Renzini}, {Mancini}, {Genzel}, {Bouch{\'e}}, {Cresci}, {Hicks}, \& {et
  al.}}]{forster18}
{F{\"o}rster Schreiber} N.~M., {Renzini} A., {Mancini} C., {Genzel} R.,
  {Bouch{\'e}} N., {Cresci} G., {Hicks} E.~K.~S., {et al.}, 2018, \apjs, 238,
  21

\bibitem[{{F{\"o}rster Schreiber} {et~al}\mbox{.}(2011){F{\"o}rster Schreiber},
  {Shapley}, {Genzel}, {Bouch{\'e}}, {Cresci}, {Davies}, {Erb}, {Genel},
  {Lutz}, {Newman}, {Shapiro}, {Steidel}, {Sternberg}, \&
  {Tacconi}}]{forster11b}
{F{\"o}rster Schreiber} N.~M. {et~al.}, 2011, \apj, 739, 45

\bibitem[{{F{\"o}rster Schreiber} {et~al}\mbox{.}(2019){F{\"o}rster Schreiber},
  {{\"U}bler}, {Davies}, {Genzel}, {Wisnioski}, {Belli}, {Shimizu}, {Lutz},
  {Fossati}, {Herrera-Camus}, {Mendel}, {Tacconi}, {Wilman}, {Beifiori},
  {Brammer}, {Burkert}, {Carollo}, {Davies}, {Eisenhauer}, {Fabricius},
  {Lilly}, {Momcheva}, {Naab}, {Nelson}, {Price}, {Renzini}, {Saglia},
  {Sternberg}, {van Dokkum}, \& {Wuyts}}]{forster19}
{F{\"o}rster Schreiber} N.~M. {et~al.}, 2019, \apj, 875, 21

\bibitem[{{Freundlich} {et~al}\mbox{.}(2013){Freundlich}, {Combes}, {Tacconi},
  {Cooper}, {Genzel}, {Neri}, {Bolatto}, {Bournaud}, {Burkert}, {Cox}, {Davis},
  {F{\"o}rster Schreiber}, {Garcia-Burillo}, {Gracia-Carpio}, {Lutz}, {Naab},
  {Newman}, {Sternberg}, \& {Weiner}}]{freundlich13}
{Freundlich} J. {et~al.}, 2013, \aap, 553, A130

\bibitem[{{Genel} {et~al}\mbox{.}(2012){Genel}, {Naab}, {Genzel}, {F{\"o}rster
  Schreiber}, {Sternberg}, {Oser}, {Johansson}, {Dav{\'e}}, {Oppenheimer}, \&
  {Burkert}}]{genel12}
{Genel} S. {et~al.}, 2012, \apj, 745, 11

\bibitem[{{Gentry} {et~al}\mbox{.}(2017){Gentry}, {Krumholz}, {Dekel}, \&
  {Madau}}]{Gentry17}
{Gentry} E.~S., {Krumholz} M.~R., {Dekel} A., {Madau} P., 2017, \mnras, 465,
  2471

\bibitem[{{Genzel} {et~al}\mbox{.}(2008){Genzel}, {Burkert}, {Bouch{\'e}},
  {Cresci}, {F{\"o}rster Schreiber}, {Shapley}, {Shapiro}, {Tacconi}, \& {et
  al.,}}]{genzel08}
{Genzel} R. {et~al.}, 2008, \apj, 687, 59

\bibitem[{{Genzel} {et~al}\mbox{.}(2014){Genzel}, {F{\"o}rster Schreiber},
  {Lang}, {Tacchella}, {Tacconi}, {Wuyts}, \& {et al.}}]{genzel14_rings}
{Genzel} R., {F{\"o}rster Schreiber} N.~M., {Lang} P., {Tacchella} S.,
  {Tacconi} L.~J., {Wuyts} S., {et al.}, 2014, \apj, 785, 75

\bibitem[{{Genzel} {et~al}\mbox{.}(2011){Genzel}, {Newman}, {Jones},
  {F{\"o}rster Schreiber}, {Shapiro}, {Genel}, {Lilly}, \& {et al.}}]{genzel11}
{Genzel} R., {Newman} S., {Jones} T., {F{\"o}rster Schreiber} N.~M., {Shapiro}
  K., {Genel} S., {Lilly} S.~J., {et al.}, 2011, \apj, 733, 101

\bibitem[{{Genzel} {et~al}\mbox{.}(2006){Genzel}, {Tacconi}, {Eisenhauer},
  {F{\"o}rster Schreiber}, {Cimatti}, {Daddi}, {Bouch{\'e}}, \& {et
  al.}}]{genzel06}
{Genzel} R., {Tacconi} L.~J., {Eisenhauer} F., {F{\"o}rster Schreiber} N.~M.,
  {Cimatti} A., {Daddi} E., {Bouch{\'e}} N., {et al.}, 2006, \nat, 442, 786

\bibitem[{{Ginzburg} {et~al}\mbox{.}(2021){Ginzburg}, {Huertas-Company},
  {Dekel}, {Mandelker}, {Snyder}, {Ceverino}, \& {Primack}}]{ginzburg21}
{Ginzburg} O., {Huertas-Company} M., {Dekel} A., {Mandelker} N., {Snyder} G.,
  {Ceverino} D., {Primack} J., 2021, \mnras, 501, 730

\bibitem[{{Goldreich} \& {Lynden-Bell}(1965)}]{goldreich65_thick}
{Goldreich} P., {Lynden-Bell} D., 1965, \mnras, 130, 97

\bibitem[{{Guo} {et~al}\mbox{.}(2015){Guo}, {Ferguson}, {Bell}, {Koo},
  {Conselice}, {Giavalisco}, {Kassin}, {Lu}, {Lucas}, {Mandelker}, {McIntosh},
  {Primack}, {Ravindranath}, {Barro}, {Ceverino}, {Dekel}, {Faber}, {Fang},
  {Koekemoer}, {Noeske}, {Rafelski}, \& {Straughn}}]{guo15}
{Guo} Y. {et~al.}, 2015, \apj, 800, 39

\bibitem[{{Guo} {et~al}\mbox{.}(2012){Guo}, {Giavalisco}, {Ferguson},
  {Cassata}, \& {Koekemoer}}]{guo12}
{Guo} Y., {Giavalisco} M., {Ferguson} H.~C., {Cassata} P., {Koekemoer} A.~M.,
  2012, \apj, 757, 120

\bibitem[{{Guo} {et~al}\mbox{.}(2018){Guo}, {Rafelski}, {Bell}, {Conselice},
  {Dekel}, {Faber}, {Giavalisco}, {Koekemoer}, {Koo}, {Lu}, {Mandelker},
  {Primack}, {Ceverino}, {de Mello}, {Ferguson}, {Hathi}, {Kocevski}, {Lucas},
  {P{\'e}rez-Gonz{\'a}lez}, {Ravindranath}, {Soto}, {Straughn}, \&
  {Wang}}]{guo18}
{Guo} Y. {et~al.}, 2018, \apj, 853, 108

\bibitem[{{Hopkins}, {Quataert} \& {Murray}(2012){Hopkins}, {Quataert}, \&
  {Murray}}]{Hopkins12b}
{Hopkins} P.~F., {Quataert} E., {Murray} N., 2012, \mnras, 421, 3522

\bibitem[{{Huertas-Company} {et~al}\mbox{.}(2020){Huertas-Company}, {Guo},
  {Ginzburg}, {Lee}, {Mandelker}, {Metter}, {Primack}, {Dekel}, {Ceverino},
  {Faber}, {Koo}, {Koekemoer}, {Snyder}, {Giavalisco}, \& {Zhang}}]{huertas20}
{Huertas-Company} M. {et~al.}, 2020, \mnras, 499, 814

\bibitem[{{Immeli} {et~al}\mbox{.}(2004{\natexlab{a}}){Immeli}, {Samland},
  {Gerhard}, \& {Westera}}]{Immeli04_b}
{Immeli} A., {Samland} M., {Gerhard} O., {Westera} P., 2004{\natexlab{a}},
  \aap, 413, 547

\bibitem[{{Immeli} {et~al}\mbox{.}(2004{\natexlab{b}}){Immeli}, {Samland},
  {Westera}, \& {Gerhard}}]{immeli04_a}
{Immeli} A., {Samland} M., {Westera} P., {Gerhard} O., 2004{\natexlab{b}},
  \apj, 611, 20

\bibitem[{{Inoue} {et~al}\mbox{.}(2016){Inoue}, {Dekel}, {Mandelker},
  {Ceverino}, {Bournaud}, \& {Primack}}]{inoue16}
{Inoue} S., {Dekel} A., {Mandelker} N., {Ceverino} D., {Bournaud} F., {Primack}
  J., 2016, \mnras, 456, 2052

\bibitem[{{Kravtsov}(2003)}]{krav03}
{Kravtsov} A.~V., 2003, \apjl, 590, L1

\bibitem[{{Kravtsov}, {Klypin} \& {Khokhlov}(1997){Kravtsov}, {Klypin}, \&
  {Khokhlov}}]{krav97}
{Kravtsov} A.~V., {Klypin} A.~A., {Khokhlov} A.~M., 1997, \apjs, 111, 73

\bibitem[{{Krumholz} \& {Burkert}(2010)}]{krum_burkert10}
{Krumholz} M.~R., {Burkert} A., 2010, \apj, 724, 895

\bibitem[{{Krumholz} {et~al}\mbox{.}(2018){Krumholz}, {Burkhart}, {Forbes}, \&
  {Crocker}}]{krum18}
{Krumholz} M.~R., {Burkhart} B., {Forbes} J.~C., {Crocker} R.~M., 2018, \mnras,
  477, 2716

\bibitem[{{Krumholz} \& {Dekel}(2010)}]{kd10}
{Krumholz} M.~R., {Dekel} A., 2010, \mnras, 406, 112

\bibitem[{{Krumholz} \& {Dekel}(2012)}]{kd12}
{Krumholz} M.~R., {Dekel} A., 2012, \apj, 753, 16

\bibitem[{{Krumholz}, {Dekel} \& {McKee}(2012){Krumholz}, {Dekel}, \&
  {McKee}}]{kdm12}
{Krumholz} M.~R., {Dekel} A., {McKee} C.~F., 2012, \apj, 745, 69

\bibitem[{{Krumholz} \& {Thompson}(2012)}]{krum_thom12}
{Krumholz} M.~R., {Thompson} T.~A., 2012, \apj, 760, 155

\bibitem[{{Krumholz} \& {Thompson}(2013)}]{krum_thom13}
{Krumholz} M.~R., {Thompson} T.~A., 2013, \mnras, 434, 2329

\bibitem[{{Lenki{\'c}} {et~al}\mbox{.}(2021){Lenki{\'c}}, {Bolatto}, {Fisher},
  {Glazebrook}, {Obreschkow}, {Abraham}, \& {Ambachew}}]{lenkic21}
{Lenki{\'c}} L., {Bolatto} A.~D., {Fisher} D.~B., {Glazebrook} K., {Obreschkow}
  D., {Abraham} R., {Ambachew} L., 2021, \mnras

\bibitem[{{Lilly} {et~al}\mbox{.}(2013){Lilly}, {Carollo}, {Pipino}, {Renzini},
  \& {Peng}}]{lilly13}
{Lilly} S.~J., {Carollo} C.~M., {Pipino} A., {Renzini} A., {Peng} Y., 2013,
  \apj, 772, 119

\bibitem[{{Mandelker} {et~al}\mbox{.}(2017){Mandelker}, {Dekel}, {Ceverino},
  {DeGraf}, {Guo}, \& {Primack}}]{mandelker17}
{Mandelker} N., {Dekel} A., {Ceverino} D., {DeGraf} C., {Guo} Y., {Primack} J.,
  2017, \mnras, 464, 635

\bibitem[{{Mandelker} {et~al}\mbox{.}(2014){Mandelker}, {Dekel}, {Ceverino},
  {Tweed}, {Moody}, \& {Primack}}]{mandelker14}
{Mandelker} N., {Dekel} A., {Ceverino} D., {Tweed} D., {Moody} C.~E., {Primack}
  J., 2014, \mnras, 443, 3675

\bibitem[{{Mayer} {et~al}\mbox{.}(2016){Mayer}, {Tamburello}, {Lupi}, {Keller},
  {Wadsley}, \& {Madau}}]{mayer16}
{Mayer} L., {Tamburello} V., {Lupi} A., {Keller} B., {Wadsley} J., {Madau} P.,
  2016, \apjl, 830, L13

\bibitem[{{Moody} {et~al}\mbox{.}(2014){Moody}, {Guo}, {Mandelker}, {Ceverino},
  {Mozena}, {Koo}, {Dekel}, \& {Primack}}]{moody14}
{Moody} C.~E., {Guo} Y., {Mandelker} N., {Ceverino} D., {Mozena} M., {Koo}
  D.~C., {Dekel} A., {Primack} J., 2014, \mnras, 444, 1389

\bibitem[{{Moster}, {Naab} \& {White}(2018){Moster}, {Naab}, \&
  {White}}]{moster18}
{Moster} B.~P., {Naab} T., {White} S. D.~M., 2018, \mnras, 477, 1822

\bibitem[{{Murray}, {Quataert} \& {Thompson}(2010){Murray}, {Quataert}, \&
  {Thompson}}]{murray10}
{Murray} N., {Quataert} E., {Thompson} T.~A., 2010, \apj, 709, 191

\bibitem[{{Newman} {et~al}\mbox{.}(2012){Newman}, {Shapiro Griffin}, {Genzel},
  {Davies}, {F{\"o}rster-Schreiber}, {Tacconi}, {Kurk}, {Wuyts}, {Genel},
  {Lilly}, {Renzini}, {Bouch{\'e}}, {Burkert}, {Cresci}, {Buschkamp},
  {Carollo}, {Eisenhauer}, {Hicks}, {Lutz}, {Mancini}, {Naab}, {Peng}, \&
  {Vergani}}]{newman12}
{Newman} S.~F. {et~al.}, 2012, \apj, 752, 111

\bibitem[{{Noguchi}(1999)}]{noguchi99}
{Noguchi} M., 1999, \apj, 514, 77

\bibitem[{{Oklop{\v c}i{\'c}} {et~al}\mbox{.}(2017){Oklop{\v c}i{\'c}},
  {Hopkins}, {Feldmann}, {Kere{\v s}}, {Faucher-Gigu{\`e}re}, \&
  {Murray}}]{oklopcic17}
{Oklop{\v c}i{\'c}} A., {Hopkins} P.~F., {Feldmann} R., {Kere{\v s}} D.,
  {Faucher-Gigu{\`e}re} C.-A., {Murray} N., 2017, \mnras, 465, 952

\bibitem[{{Palla} \& {Stahler}(1999)}]{palla99}
{Palla} F., {Stahler} S.~W., 1999, \apj, 525, 772

\bibitem[{{Perret} {et~al}\mbox{.}(2014){Perret}, {Renauld}, {Epinat}, {Amram},
  {Bournaud}, {Contini}, {Teyssier}, \& {Lambert}}]{perret14}
{Perret} V., {Renauld} F., {Epinat} B., {Amram} P., {Bournaud} F., {Contini}
  T., {Teyssier} R., {Lambert} J.-C., 2014, \aap, 562, A1

\bibitem[{{Renaud} {et~al}\mbox{.}(2013){Renaud}, {Bournaud}, {Emsellem},
  {Elmegreen}, {Teyssier}, {Alves}, {Chapon}, {Combes}, {Dekel}, {Gabor},
  {Hennebelle}, \& {Kraljic}}]{renaud13}
{Renaud} F. {et~al.}, 2013, \mnras, 436, 1836

\bibitem[{{Renaud}, {Romeo} \& {Agertz}(2021){Renaud}, {Romeo}, \&
  {Agertz}}]{renaud21}
{Renaud} F., {Romeo} A.~B., {Agertz} O., 2021, arXiv e-prints, arXiv:2106.00020

\bibitem[{{Rodr{\'{\i}}guez-Puebla}
  {et~al}\mbox{.}(2017){Rodr{\'{\i}}guez-Puebla}, {Primack}, {Avila-Reese}, \&
  {Faber}}]{rodriguez17}
{Rodr{\'{\i}}guez-Puebla} A., {Primack} J.~R., {Avila-Reese} V., {Faber} S.~M.,
  2017, \mnras, 470, 651

\bibitem[{{Rujopakarn} {et~al}\mbox{.}(2019){Rujopakarn}, {Daddi}, {Rieke},
  {Puglisi}, {Schramm}, {P{\'e}rez-Gonz{\'a}lez}, {Magdis}, {Alberts},
  {Bournaud}, {Elbaz}, {Franco}, {Kawinwanichakij}, {Kohno}, {Narayanan},
  {Silverman}, {Wang}, \& {Williams}}]{rujopakarn19}
{Rujopakarn} W. {et~al.}, 2019, \apj, 882, 107

\bibitem[{{Schroetter} {et~al}\mbox{.}(2019){Schroetter}, {Bouch{\'e}}, {Zabl},
  {Contini}, {Wendt}, {Schaye}, {Mitchell}, {Muzahid}, {Marino}, {Bacon},
  {Lilly}, {Richard}, \& {Wisotzki}}]{schroetter19}
{Schroetter} I. {et~al.}, 2019, \mnras, 490, 4368

\bibitem[{{Soto} {et~al}\mbox{.}(2017){Soto}, {de Mello}, {Rafelski},
  {Gardner}, {Teplitz}, {Koekemoer}, {Ravindranath}, {Grogin}, {Scarlata},
  {Kurczynski}, \& {Gawiser}}]{soto17}
{Soto} E. {et~al.}, 2017, \apj, 837, 6

\bibitem[{{Tacchella} {et~al}\mbox{.}(2016{\natexlab{a}}){Tacchella}, {Dekel},
  {Carollo}, {Ceverino}, {DeGraf}, {Lapiner}, {Mandelker}, \&
  {Primack}}]{tacchella16_prof}
{Tacchella} S., {Dekel} A., {Carollo} C.~M., {Ceverino} D., {DeGraf} C.,
  {Lapiner} S., {Mandelker} N., {Primack} J.~R., 2016{\natexlab{a}}, \mnras,
  458, 242

\bibitem[{{Tacchella} {et~al}\mbox{.}(2016{\natexlab{b}}){Tacchella}, {Dekel},
  {Carollo}, {Ceverino}, {DeGraf}, {Lapiner}, {Mandelker}, \& {Primack
  Joel}}]{tacchella16_ms}
{Tacchella} S., {Dekel} A., {Carollo} C.~M., {Ceverino} D., {DeGraf} C.,
  {Lapiner} S., {Mandelker} N., {Primack Joel} R., 2016{\natexlab{b}}, \mnras,
  457, 2790

\bibitem[{{Tacconi} {et~al}\mbox{.}(2010){Tacconi}, {Genzel}, {Neri}, {Cox},
  {Cooper}, {Shapiro}, {Bolatto}, {Bouch{\'e}}, \& {et al.}}]{tacconi10}
{Tacconi} L.~J. {et~al.}, 2010, \nat, 463, 781

\bibitem[{{Tacconi} {et~al}\mbox{.}(2018){Tacconi}, {Genzel}, {Saintonge},
  {Combes}, {Garc{\'\i}a-Burillo}, {Neri}, {Bolatto}, {Contini}, {F{\"o}rster
  Schreiber}, {Lilly}, {Lutz}, {Wuyts}, {Accurso}, {Boissier}, {Boone},
  {Bouch{\'e}}, {Bournaud}, {Burkert}, {Carollo}, {Cooper}, {Cox}, {Feruglio},
  {Freundlich}, {Herrera-Camus}, {Juneau}, {Lippa}, {Naab}, {Renzini},
  {Salome}, {Sternberg}, {Tadaki}, {{\"U}bler}, {Walter}, {Weiner}, \&
  {Weiss}}]{tacconi18}
{Tacconi} L.~J. {et~al.}, 2018, \apj, 853, 179

\bibitem[{{Tacconi} {et~al}\mbox{.}(2013){Tacconi}, {Neri}, {Genzel}, {Combes},
  {Bolatto}, {Cooper}, {Wuyts}, \& {et al.}}]{tacconi13}
{Tacconi} L.~J., {Neri} R., {Genzel} R., {Combes} F., {Bolatto} A., {Cooper}
  M.~C., {Wuyts} S., {et al.}, 2013, \apj, 768, 74

\bibitem[{{Tamburello} {et~al}\mbox{.}(2015){Tamburello}, {Mayer}, {Shen}, \&
  {Wadsley}}]{tamburello15}
{Tamburello} V., {Mayer} L., {Shen} S., {Wadsley} J., 2015, \mnras, 453, 2490

\bibitem[{{Teyssier}(2002)}]{teyssier02}
{Teyssier} R., 2002, \aap, 385, 337

\bibitem[{{Tomassetti} {et~al}\mbox{.}(2016){Tomassetti}, {Dekel}, {Mandelker},
  {Ceverino}, {Lapiner}, {Faber}, {Kneller}, {Primack}, \&
  {Sai}}]{tomassetti16}
{Tomassetti} M. {et~al.}, 2016, \mnras, 458, 4477

\bibitem[{{Toomre}(1964)}]{toomre64}
{Toomre} A., 1964, \apj, 139, 1217

\bibitem[{{Wuyts} {et~al}\mbox{.}(2012){Wuyts}, {F{\"o}rster Schreiber},
  {Genzel}, {Guo}, {Barro}, {Bell}, {Dekel}, {Faber}, {Ferguson}, {Giavalisco},
  {Grogin}, {Hathi}, {Huang}, {Kocevski}, {Koekemoer}, {Koo}, \& {et
  al.,}}]{wuyts12}
{Wuyts} S. {et~al.}, 2012, \apj, 753, 114

\bibitem[{{Zanella} {et~al}\mbox{.}(2015){Zanella}, {Daddi}, {Le Floc'h},
  {Bournaud}, {Gobat}, {Valentino}, {Strazzullo}, {Cibinel}, {Onodera},
  {Perret}, {Renaud}, \& {Vignali}}]{zanella15}
{Zanella} A. {et~al.}, 2015, \nat, 521, 54

\bibitem[{{Zanella} {et~al}\mbox{.}(2019){Zanella}, {Le Floc'h}, {Harrison},
  {Daddi}, {Bernhard}, {Gobat}, {Strazzullo}, {Valentino}, {Cibinel},
  {S{\'a}nchez Almeida}, {Kohandel}, {Fensch}, {Behrendt}, {Burkert},
  {Onodera}, {Bournaud}, \& {Scholtz}}]{zanella19}
{Zanella} A. {et~al.}, 2019, \mnras, 489, 2792

\bibitem[{{Zolotov} {et~al}\mbox{.}(2015){Zolotov}, {Dekel}, {Mandelker},
  {Tweed}, {Inoue}, {DeGraf}, {Ceverino}, {Primack}, {Barro}, \&
  {Faber}}]{zolotov15}
{Zolotov} A. {et~al.}, 2015, \mnras, 450, 2327

\end{thebibliography}
%
%

\appendix
\section{Contamination by disc}
\label{sec:contamination}


One practical difficulty in distinguishing between the scenarios is the
possible contamination of the clumps by disc stars.
These could be either unbound stars that are temporarily within the clump
boundaries, or stars in the foreground and background of the clump along the
line of sight once the clump size is significantly smaller than the disc
height.
The contamination may systematically affect the measured clump properties,
and if the level of contamination varies with distance from the galaxy center
it may generate false radial gradients even for short-lived clumps that do not
migrate.
%
This was demonstrated in a simulation of a rather compact galaxy
where SL clumps appear to show weak but non-negligible
age and mass radial gradients and the contaminated stellar ages range up to
$300\Myr$ in the inner disc \citep{oklopcic17}.

\smallskip 
Extensive quantities, such as stellar mass and SFR, can be crudely
``de-contaminated" by subtracting the estimated disc contribution within the 2D
clump boundaries based on the mean surface density of that quantity in the
disc, as evaluated either in a ring around the clump or in a similar distance
from the galaxy center.
Naturally, the correction is more manageable in cases where the contrast between
clump and disc is higher, which implies that the corrected SFR would be more
reliable than the corrected $\Ms$
\citep{forster11b,wuyts12,mandelker14,cibinel17}.
Such a correction is not straightforward in the case of non-extensive
quantities such as age.
Thus, the star-formation time $\tsf\!=\!$ sSFR$^{-1}$, which is a ratio of two
extensive quantities that could in principle be decontaminated,
may serve as a more reliable
estimator of the clump dynamical age for the purpose of distinguishing between
LL and SL clumps.

\smallskip 
For the CANDELS clumps, the disc subtraction is applied separately to the flux
in each of the nine HST bands before the SED fitting that derives the mass, SFR
and age. This is repeated using different levels of correction, ranging from no
subtraction to very aggressive subtraction. The variation in the clump
properties and their radial gradients under the different levels of correction
can help us estimate the validity of these quantities as distinguishing
features between the scenarios.
We compare the results with a weak correction to the fiducial case
with a moderate correction and to the case with the most aggressive correction.
We find that the distributions of ages have medians
consistent with the LL scenario and in conflict with the SL
scenario independent of the correction applied.
We find, surprisingly, that the radial gradients in $\Ms$, age, $\tsf$ and SFR
are declining steeper for a stronger correction,
meaning that the correction is stronger at larger radii.
This is different from what we see in our various simulations,
where the contamination is rather independent of distance, and is opposite to
the finding of
\citet{oklopcic17}
in their simulated galaxy where the contamination is stronger at small radii.
We find in the G18 data that there is a strong mass radial gradient independent
of the correction level, indicating that the mass radial gradient
is a valid distinguishing feature between the scenarios.
A significant age radial gradient exists at all levels of correction,
but it is smeared out by contamination when no correction is applied.
The age radial gradient thus
become a valid distinguishing feature once some correction for contamination is
applied, even at weak level.

\smallskip 
Another way to estimate the effect of disc contamination is by investigating it
in our simulations. We find in the isolated simulations a typical contamination
of 35\% in stellar mass, mostly by unbound stars within the clump boundaries,
and naturally a weaker contamination in SFR.
In the cosmological simulation V07 at $z\!=\!2.5\!-\!1$ the median
contamination in stellar mass is at the level of only 10\%, while in V19 at
$z\!=\!5\!-\!3$ it is about 60\%.
These estimates were derived by comparing the stellar mass within the clump
boundaries in 3D to the projected mass in 2D including the stars above and
below the clump.
We also find that the relative contamination is independent of distance from
the galactic center. This indicates that in our simulated galaxies, even when
the contamination is significant and may generate a systematic bias, as in V19,
it does not generate false radial gradients.



\label{lastpage}
\end{document}